\documentclass[nofootinbib,showkeys,floatfix,onecolumn,times]{revtex4-2}

\usepackage{bm}		
\usepackage{graphicx,amssymb}
\usepackage{latexsym,color}
\usepackage{hyperref}
\usepackage[utf8]{inputenc}
\usepackage{natbib}

\usepackage{graphicx}	
\usepackage{amsmath}	

\begin{document}
\count\footins = 1000
\title{Constraining photon  trajectories in   black hole shadows}
\author{D. Pugliese\&Z. Stuchl\'{\i}k}
\email{daniela.pugliese@physics.slu.cz}
\affiliation{
Research Centre for Theoretical Physics and Astrophysics, Institute of Physics,
  Silesian University in Opava,
 Bezru\v{c}ovo n\'{a}m\v{e}st\'{i} 13, CZ-74601 Opava, Czech Republic
}

\begin{abstract}
We examine the  shadow  cast  by a Kerr     black hole, focusing on constraints  on photons corresponding to different shadow  boundaries. The photons are  related to   different orbital ranges and  impact parameter values, creating a   map  of the shadow boundaries.  Our analysis  fixes  also the  conditions under which  it is possible to observe an ``imprint"  of the black hole (outer) ergosurface and (outer) ergoregion on the  Kerr black hole shadow boundary.
The counter-rotating case resulted  strongly constrained with respect to the co-rotating case, constituting a remarkable and significant difference  where the counter-rotating component associated with the  shadow boundary  is strongly distinct from the co-rotating one.  However,  in this framework, even   the co-rotating photons  imply restrictions  on conditions on the spins and planes, which are  bounded by limiting values.
We believe the results found here, being a tracer for the central black hole,   can constitute new templates for the ongoing observations. 
\end{abstract}

\date{\today}



\def\be{\begin{equation}}
\def\ee{\end{equation}}
\def\bea{\begin{eqnarray}}
\newcommand{\cc}{\mathrm{C}}
\newcommand{\jj}{\mathrm{J}}
\def\eea{\end{eqnarray}}
\newcommand{\tb}[1]{\textbf{{#1}}}
\newcommand{\ttb}[1]{\textbf{#1}}
\newcommand{\rtb}[1]{\textcolor[rgb]{1.00,0.00,0.00}{\tb{#1}}}
\newcommand{\btb}[1]{\textcolor[rgb]{0.00,0.00,1.00}{\tb{#1}}}
\newcommand{\otb}[1]{\textcolor[rgb]{1.00,0.47,0.22}{\tb{#1}}}
\newcommand{\ptb}[1]{\textcolor[rgb]{0.57,0.14,0.58}{\tb{#1}}}
\newcommand{\gtb}[1]{\textcolor[rgb]{0.00,0.47,0.24}{\tb{#1}}}
\newcommand{\pp}{\textbf{()}}
\newcommand{\non}[1]{{\LARGE{\not}}{#1}}
\newcommand{\mso}{\mathrm{mso}}
\newcommand{\Qa}{\mathcal{Q}}
\newcommand{\mbo}{\mathrm{mbo}}
\newcommand{\il}{~}
\newcommand{\rc}{\rho_{\ti{C}}}
\newcommand{\dd}{\mathcal{D}}
\newcommand{\Sie}{\mathcal{S}}
\newcommand{\Sa}{\mathcal{S}}
\newcommand{\mnras}{MNRAS}
\newcommand{\aap}{A\&A}

\newcommand{\actaa}{Acta Astron.}
\newcommand{\Fa}{\mathcal{F}}
\newcommand{\Mie}{\mathcal{M}}
\newcommand{\Em}{\mathcal{E}}
\newcommand{\La}{\mathcal{L}}
\newcommand{\Ta}{{\mbox{\scriptsize  \textbf{\textsf{T}}}}}

    \newcommand{\oo}{\mathrm{O}}

\maketitle

\section{Introduction}
The first image
of the black hole  (\textbf{BH}) shadow was released  by the Event Horizon Telescope (\textbf{EHT})  Collaboration\footnote{The Event Horizon Telescope: (https://eventhorizontelescope.org): a virtual Earth-sized  large telescope array consisting of a global network of radio telescopes  using  very--long--baseline interferometry (\textbf{VLBI})  at   millimeter and   sub-millimeter wavelengths--see also \cite{EHT2}, for  further information on the instrument.} in 2019,  first results concerning the detection of an event horizon of a super--massive
black hole  (\textbf{SMBH})
 at the center of a neighboring elliptical  Messier \textbf{87} (\textbf{M87})
galaxy were announced in  \cite{EHT1,EHT2,EHT2c,EHT2d,EHT2e,EHT2f}.    The corresponding linear-polarimetric \textbf{EHT} images of the center of \textbf{M87}  were presented  in \cite{EHT21a}.
In \cite{EHT21b} the resolved polarization structures (and Atacama Large Millimeter/submillimeter Array observation (\textbf{ALMA}))  have been  compared   to  theoretical models.  (A new sharper image of the \textbf{M87} \textbf{BH} has been released, in 2023 created with PRIMO algorithm \cite{Medeiros}).
In 2022 the \textbf{EHT} Collaboration has been   able to observe the shadow cast by
the \textbf{SMBH}   Sagittarius \textbf{A*} (\textbf{SgrA*})  in the center of our galaxy\footnote{The \textbf{EHT} images showed  rings of synchrotron emissions.
The rings appeared  asymmetric,   consistently with  synchrotron emission from a hot plasma orbiting  a (Kerr)  \textbf{BH} (and subjected to  gravitational lensing). In particular,  the ring morphological characteristics  reflect the \textbf{BH} parameters (mass and spin).
In these models, magnetized
\textbf{ADAF}  (advection-dominated accretion flow) appears to emerge as the underlying best model  explaining the observed synchrotron emission.
Models of magnetized, radiatively inefficient
accretion flow/\textbf{ADAFs}  have been constructed
  for  \textbf{SgrA*}  fitting  the observed
spectral energy density (characteristic  low luminosity of \textbf{SgrA*},
relative  to the Eddington limit,  appears to suggest
 matter falling  onto  the  \textbf{BH}   as
radiatively inefficient/\textbf{ADAF}).
Analytical modelization  aims therefore   at the description of the
emitted synchrotron radiation sources (which could also consist in
a  magnetized torus, for example, a geometrically thick
accretion flow emitting  thermal synchrotron) and
a  jet  with  thermal and non--thermal synchrotron radiation.
Semi--analytic models of synchrotron emission from relativistic jet,  general relativistic magnetohydrodynamical (\textbf{GRMHD}) simulations of \textbf{ADAFs},  distinguish in general  two different modes:  the standard and normal evolution model (SANE), (magnetic fields are turbulent and midplane magnetic field pressure is less than the gas pressure), and the magnetically arrested disk (MAD) (strong magnetic fields)--\cite{EHT22a,EHT22b,EHT22c,EHT22d,EHT22e,EHT22f}.} \cite{EHT22a,EHT22b,EHT22c,EHT22d,EHT22e,EHT22f}.
 The \textbf{EHT} observations, providing   immediate evidences of an event horizon presence, constitute a consistent  advance in  astronomy  having  made also   possible to focus on physical phenomena in the   close   proximity of the  \textbf{BH}. (The  \textbf{BH} shadow  could be related to complex  astrophysical phenomena, typically accretion disks and related jets orbiting close  to the horizon\footnote{The \textbf{EHT} reported also observations of Blazar 3C 279 (4C--05.55, NRAO413, PKS1253--05),  high resolution images of the jet produced by the \textbf{SMBH} sitting at the center of Centaurus A  and the distant Blazar J1924-2914 (the  radio-loud quasar is also known as (PKS 1921--293, OV--236))--see also \cite{Kim,Issaoun,Janssen,2023ApJ...944...55P,Curd23}, and
 the observations of the  flat--spectrum radio quasar  NRAO 530 (1730-130, J1733--1304)\cite{Jorstad}.}.)

We expect that it will be possible to observe even more refined and clearer details from the astronomical observations in near future (see  \cite{Lu23}).  There is a continuous improvement (and updating) of the    \textbf{SMBH} \textbf{M87*} image, through new analyses and  simulations.
Advancements  and new data analysis techniques can  allow a more clear observation and description of the \textbf{BH} environment and  more  refined analyses will  be able to clearly  discern  the presence and structure of  different  photon  orbits, as predicted by the theory.
In \cite{ccdpp}, for example,  the  \textbf{M87*} images from \textbf{EHT} collaboration  has been  re-analyzed, using  a series of kinetic plasma simulations,
 predicting   the  images during the   outbursts  characterizing \textbf{M87*}.
It was  found that, following magnetic-field instabilities, radio-wave hot spots  (connected to closed magnetic structures--plasmoids--of  the \textbf{BH} magnetosphere) may appear as  rotating around the \textbf{BH} shadow on an orbit   three times larger than the \textbf{BH} size.
In \cite{Avery},   revisiting  the past observations, it has been  claimed to have spotted  the sharp light ring   around  \textbf{M87*}--see also \cite{Medeiros,PalumboWong, Avery2,Johnson,2022MNRAS.517.2462L, 2023ApJ...944...55P}. (The ring was predicted
 by general relativistic magnetohydrodynamical  (\textbf{GRMHD}) simulations in  the near horizon  region of  \textbf{M87}*). This  new analysis  also provides evidence of a rotating jet ejected from the \textbf{BH} region
--\cite{2022MNRAS.517.2462L} and   \cite{Medeiros,PalumboWong,Avery, Avery2,W22SafAstar,2022ApJ...940..182T,Tamburini,2022MNRAS.517.2462L}.
A recent analysis of \textbf{EHT} data sharpened the  view of the glowing gas around the \textbf{BH}.
The new image was  reconstructed using  the PRIMO algorithm, \citep{Medeiros};
new observations, obtained combined data from   radio telescopes
Global Millimetre \textbf{VLBI} Array, \textbf{ALMA}, and the Greenland Telescope,  have shown the  ring-like accretion structure in \textbf{M87} connecting the  \textbf{BH}  plasma jet  to the  \textbf{BH} and the accretion matter\citep{Lu23}.

\medskip

\textbf{BHs} shadows have been studied analytically in   numerous studies  \citep{Bardeen1973,Synge,Luminet,Chandra}.
The standard notion of the  \textbf{BH}   shadow defines rather the silhouette (shadow boundary)
than shadow due to definition of the dark region in radiating screen  in situation where the  \textbf{BH}   is located before the screen and the distant observer \citep{1972ApJ...178..347B,Bardeen1973}.
The boundary of the shadow is   constituted by  unstable photon (spherical and circular) orbits
\citep{1972ApJ...178..347B,Bardeen1973,Hioki,ZS,SZ,
SZZ,Johannsen,Ghasemi-Li-Bambi,Rezzolla}, whereas the shadow itself  is the dark  region bounded by
 shadow boundary. From methodological view--point, in general the \textbf{BH} shadow boundary
is found    considering  photons
coming  from all the  points at infinity.   The observer is   located at infinity, and the \textbf{BH} has a well defined and fixed
 inclination angle\footnote{The
angle between the \textbf{BH}  rotation axis and the observer line of
sight.}. Photons on unstable orbits   can  reach  the far away  observer after perturbation.  The  photons  moving  towards the  \textbf{BH}  could  be  trapped in
unstable spherical orbits, or  they can  be captured by  \textbf{BH} gravitational field  producing the  central dark region (shadow), or they could escape to infinity, being   detected by the distant observer  if   in line of sight. However, the observed  shadow  boundary can be substantially modified  by properties of the orbiting matter and even  by properties of photons radiated by the orbiting matter,  depending  on properties   of the region of the  light
distribution  and its  source.
Possible source of the light can be
the accretion disk  orbiting the central  \textbf{BH}.
In these conditions,  the photons could   interact with  the accreting  plasma  affecting   therefore  the  photons distribution and the intensity.
The shadow in this case has been also
extensively  studied  for example in
\cite{FMA00,T04,BD05,BL06,BN06,HCS07,Doku}.

The shape and size  of  a \textbf{BH} shadow boundary  depend   in particular on  the \textbf{BH} parameters. Hence observations of the shadow boundary
not only provide compelling \textbf{BH} evidence, but also
an  estimation on the \textbf{BH} parameters. In order to fit astronomical observations, several
observables were constructed using special points on the
shadow boundary in the celestial coordinates (see  \cite{Tamburini}).
These   new observables should carry information on the central spinning  \textbf{BH}  and different orbiting structures.
 {In this work we study  some aspects of the
 \textbf{BH} shadows boundary analyzing  photon   spherical orbits in orbital  ranges   associated  to accretion process.
In our analysis we solve the equations for the null geodesics constituting the  shadow boundary, fixing constraints on the photons impact parameter $\ell$ or the radius $r=$constant. In this way we individuate points on the \textbf{BHs} shadow profiles correspondent to the fixed constraints on $\ell$ or   $r$,  investigated  for all values of the \textbf{BH} spin parameter.
 We consider photons co-rotating and counter-rotating motion (determined by  $\ell$). 
We also  explore the effects of frame dragging on the shadow boundary,   investigating   the possibility to observe  an  ``imprint" of the outer  ergoregion and the outer  ergosurface on the shadows boundary, by examining the    photon spherical orbits in the  outer ergoregion and on the  outer ergosurface.}

The structure of this article is as follows.
 In Sec.\il(\ref{Sec:quaconsta}) we introduce the spacetime metric. Constants of motion and geodesics equations are summarized  in Sec.\il(\ref{Sec:carter-equa}).
Black hole shadows are introduced  in Sec.\il(\ref{Sec:shadows}).
Shadows analysis  is in Sec.\il(\ref{Sec:first-lun}).
We investigate  shadows  for specific impact parameter  in    Sec.\il(\ref{Sec:allell}),
while in
  Sec.\il(\ref{Sec:radiii-shadows}) solutions are analysed  for fixed $r$ related to particular spherical orbital  ranges.
   In Sec.\il(\ref{Sec:from-ergo})  photons spherically orbiting on  the outer ergosurface, and in the outer ergoregion are investigated.
  Concluding remarks are in Sec.\il(\ref{Sec:conclu-Rem}). {In Appendix\il(\ref{Sec:co-rota}) there are further notes on the Kerr spacetime geodesic structure}.
In Appendix\il(\ref{App:allell})   various aspects of the analysis have been further detailed.
\section{The spacetime metric}\label{Sec:quaconsta}
In the Boyer-Lindquist (BL)  coordinates
\( \{t,r,\theta ,\phi \}\)\footnote{We adopt the
geometrical  units $c=1=G$ and  the $(-,+,+,+)$ signature, Latin indices run in $\{0,1,2,3\}$.  The radius $r$ has unit of
mass $[M]$, and the angular momentum  units of $[M]^2$, the velocities  $[u^t]=[u^r]=1$
and $[u^{\phi}]=[u^{\theta}]=[M]^{-1}$ with $[u^{\phi}/u^{t}]=[M]^{-1}$,
$[u_{\phi}/u_{t}]=[M]$ and an angular momentum per
unit of mass $[L]/[M]=[M]$.},
the   Kerr  spacetime line element   reads
%
\bea \label{alai}&& ds^2=-\left(1-\frac{2Mr}{\Sigma}\right)dt^2+\frac{\Sigma}{\Delta}dr^2+\Sigma
d\theta^2+\left[(r^2+a^2)+\frac{2M r a^2}{\Sigma}\sin^2\theta\right]\sin^2\theta
d\phi^2
-\frac{4rMa}{\Sigma} \sin^2\theta  dt d\phi,
\\&&\label{Eq:delta}
\mbox{where}\quad
\Delta\equiv a^2+r^2-2 rM\quad\mbox{and}\quad \Sigma\equiv a^2 (1-\sin^2\theta)+r^2.
\eea
 Parameter  $a=J/M\geq0$ is the metric spin, where  total angular momentum is   $J$  and  the  gravitational mass parameter is $M$.
  A Kerr \textbf{BH} is defined  by the condition $a\in[0,M]$. The extreme Kerr \textbf{BH}  has dimensionless spin $a/M=1$. The non-rotating   case $a=0$ is the   Schwarzschild \textbf{BH} solution.
  (The Kerr naked singularities (\textbf{NSs}) have  $a>M$.)

    The \textbf{BH}   horizons are
\bea
   r_-\leq r_+,\quad \mbox{where} \quad r_{\pm}\equiv M\pm\sqrt{M^2-a^2}.
\eea
The outer ergoregion of the spacetime is  $]r_+,r_{\epsilon}^+
]$, where the  outer  ergosurface   $r_{\epsilon}^+$ is
\bea\label{Eq:sigma-erg}
r_{\epsilon}^{+}\equiv M+\sqrt{M^2- a^2 (1-\sigma)}\quad\mbox{with}\quad \sigma\equiv \sin^2\theta\in [0,1],
\eea
and  there is
  $r_{\epsilon}^+=2M$  in the equatorial plane  ($\sigma=1$),
and  $r_+<r_{\epsilon}^+$ on   $\theta\neq0$.

In the following, we will  use  dimensionless    units with $M=1$ (where $r\rightarrow r/M$  and $a\rightarrow a/M$).

\subsection{Geodesics equations and constants of motion}\label{Sec:carter-equa}
{The equatorial plane is the symmetry plane for the metric, and the constant $r$
orbits on this plane are circular.}
The  geodesic equations  in Kerr  spacetime are fully separable. There are    four constants of motion. For convenience  we summarize  the  (Carter) equations  of motion as follows (see \cite{Carter}):
\bea&&\label{Eq:eqCarter-full}
 \dot{t}=\frac{1}{\Sigma}\left[\frac{P \left(a^2+r^2\right)}{\Delta}-{a \left[a \Em\sigma-\La\right]}\right],\quad
\dot{r}=\pm \frac{\sqrt{R}}{\Sigma};\quad \dot{\theta}=\pm \frac{\sqrt{T}}{\Sigma},\quad\dot{\phi}=\frac{1}{\Sigma}\left[\frac{a P}{\Delta}-\left[{a \Em-\frac{\La}{\sigma}}\right]\right];
\eea
{for $p^a=dx^a/d \tau\equiv u^a\equiv\{ \dot{t},\dot{r},\dot{\theta},\dot{\phi}\}$,   the geodesic  tangent four-vector, where $\tau$ is an affine parameter, normalized so  that $p^ap_a=-\mu^2$,
and  $\mu$ is the rest mass  of the test particle where  a null geodesics has $\mu=0$ \footnote{For the seek of simplicity  we adopted    notation $\dot{q}$ or $u^a$ for  photons and particles, the context should avoid any possible misunderstanding.}, and
\bea\label{Eq:eich}
& P\equiv \Em \left(a^2+r^2\right)-a \La,\quad & R\equiv P^2-\Delta \left[(\La-a \Em)^2+\mu^2 r^2+\Qa\right],\quad T\equiv \Qa-
(\cos\theta)^2 \left[a^2 \left(\mu^2-\Em^2\right)+\left(\frac{\La^2}{\sigma}\right)\right], 
\eea
where $\Qa$ is  the Carter constant of motion. Quantities $(\Em, \La)$ are constants of geodesic  motions
\bea&&\label{Eq:EmLdef}
\Em=-(g_{t\phi} \dot{\phi}+g_{tt} \dot{t}),\quad \La=g_{\phi\phi} \dot{\phi}+g_{t\phi} \dot{t}, %
\mu^2,
\eea
defined  from   the Kerr geometry  rotational  Killing field   $\xi_{\phi}\equiv \partial_{\phi}$,
     and  the Killing field  $\xi_{t}\equiv \partial_{t}$
representing the stationarity of the  background.

 The constant $\La$ in Eq.\il(\ref{Eq:EmLdef}) may be interpreted       as the axial component of the angular momentum  of a test    particle following
timelike geodesics. Constant  $\Em$  represents the  energy of the test particle
 related to the static observers at infinity\footnote{Note, we assume $\Em>0$ (and $\dot{t}>0$). This condition  for co-rotating fluids in the ergoregion has to be discussed further. In the ergoregion  particles can also have $\La=0$ (i.e. $\ell=0$). However this condition characterizing the ergoregion  is not associated to geodesic  circular motion in the \textbf{BH} spacetimes. There are no solutions in general for $
\dot{t}\geq 0$, $\Em<0$ $T\geq0$ and $ \ell<0$ (for $r>r_+,a\in [0,1], \sigma\in [0,1]$). We assume the so--called positive root states $\dot{t}>0$--for details see \cite{MTW,B1,B2}}.
%
%
We  introduce also  the specific  angular momentum (called also impact parameter)
 \bea&&\label{Eq:flo-adding}
\ell\equiv\frac{\La}{\Em}=-\frac{g_{\phi\phi}u^\phi  +g_{\phi t} u^t }{g_{tt} u^t +g_{\phi t} u^\phi},  
\eea
%

  %
With  $a>0$,  the  fluids and particles  counter-rotation  (co-rotation) is defined by $\ell a<0$ ($\ell a>0$).

 Static  observers, having   four-velocity   $\dot{\theta}=\dot{r}=\dot{\phi}=0$,
cannot exist inside the ergoregion. Whereas, trajectories   $\dot{r}\geq0$, including photons  crossing the outer ergosurface  and escaping outside
in the region $r\geq r_{\epsilon}^+$ are possible.
\section{Shadows}\label{Sec:shadows}
In this section we discuss  the concept of \textbf{BH} shadows, introducing    the quantities $(q,\ell)$ (constants of motion), the celestial coordinates $(\alpha,\beta)$  and we  discuss the constraints adopted  in our analysis.

\medskip

 Focusing on   the null  geodesic
equations (\ref{Eq:eqCarter-full}), the boundary of the
\textbf{BH} shadow is  determined by the (unstable) photon orbits  defined by:
\bea\label{Eq:radial-condition}
R=\partial_r R=0,\quad \partial_r^2R>0,
\eea
 {hereafter refereed as set $(\mathfrak{R})$ of equations,} within  constraints provided by the  set\footnote{In this work we also consider solution $ \partial_r^2R=0$. Note $R$ does not depend explicitly on $\sigma$. (In particular we shall consider also  the condition $T\geq 0$). There are no solutions for $
\dot{t}\geq0$, $\Em<0$, $T\geq0$, $
\ell>0$, $R=0$, $R'=0$.} of Eqs\il(\ref{Eq:eqCarter-full}).

We introduce   the motion constants  $q$  and $\ell$  (impact parameter)
\bea\label{Eq:lq-constants}
\ell\equiv \frac{\La}{\Em},\quad q\equiv \frac{\Qa}{\Em^2}.
\eea
 Following \cite{Bardeen1973} we  assume  the observer is located at infinity (i.e, at very large distance in practical calculation),
introducing the celestial coordinates\footnote{Note  we can define $\alpha=-{\ell}/{\sin\theta}$ {where} $\alpha=-\ell$ and $ \beta=\pm\sqrt{q}$ (where $q\geq0$)  for  $\theta={\pi}/{2}$. Definition
$\alpha=-{\ell}/\sqrt{\sigma}$ is  equivalent to consider $\theta\in [0,\pi]$.}
$\alpha$ and $\beta$
%
 for the  null geodesics 
\bea&&\label{Eq:beta-d-defin}
\alpha=-\frac{\ell}{\sqrt{\sigma}},\quad
\beta=\pm\sqrt{q-q_c}\
\\
&&\nonumber  \mbox{for}\quad
\sigma\neq0\quad\mbox{and}\quad  q\geq q_c,\quad\mbox{where}
\quad
 q_{c}\equiv\frac{(1-\sigma ) \left(\ell^2-\ell_c^2 \right)}{\sigma },\quad \ell_c^2\equiv a^2 \sigma.
 \eea

The  shadow boundary is, at fixed  spin $a$ and angle  $\sigma$, a closed curve in  the plane  $\alpha-\beta$,  and its  points correspond to  constant   $r$,   (solutions of $(\mathfrak{R})$),  $q$ and  parameter\footnote{For example, conditions
($R=0$,$R'=0$) for  $a=0$ and  $q=0$ correspond to  $(r=3, \ell=\pm 3\sqrt{3})$,  photon orbit on the equatorial plane of the Schwarzschild spacetime (where there is
$R''>0$),
and point $(\alpha,\beta)=(\mp 3\sqrt{3},0)$  of the Schwarzschild  \textbf{BH} shadow boundary
.}  $\ell$.

In this work we study the solutions $(r,q,\ell)$, corresponding  to  the points $(\alpha,\beta)$ of the shadow boundary, then analysing the variation of
$(r,q,\ell)$  with  $(a,\sigma)$.
In order to do this, $r$ and  $\ell$ are  set in  ranges   defined by some  boundary values.
We then   proceed by solving system   $(\mathfrak{R})$  for the  boundary $r$ and  $\ell$ values, obtaining, for fixed radius $r$, the solutions  $(\alpha,\beta,\ell,q)$ and, for fixed parameter
  $\ell$, solutions $(\alpha,\beta,r,q)$, then analyzed for all values of  $a$ and  $\sigma$. Ultimately, we obtain  the parts of the shadow boundary associated to the  fixed constraints, and  therefore  a  shadow boundary map (in dependence on the selected ranges of values for $\ell$ or $r$ for  all possible values of   $(a,\sigma)$).
Below we discuss   the  limit values for $r$ and $\ell$.

In Sec.\il(\ref{Sec:from-ergo}) we solve  system  $\mathfrak{(R)}$  for  fixed  $r\in ]r_+,r_{\epsilon}^+]$. Solutions are unstable  spherical  photon orbits located in the outer ergoregion and on the outer ergosurface.
The  frame--dragging effects on  the \textbf{BH} shadow profile are explored   also
studying   the solutions of the equations $\mathfrak{(R)}$ for $\dot{\phi}=0$, i.e. for   null  geodesics, with $\ell<0$, on
 an inversion surface--\citep{retro-inversion}.
An  inversion surface  is a  closed   surface,  defined by condition  $\dot{\phi}=0$ on the  geodesics,  embedding a  Kerr \textbf{BH}, and located out of the  Kerr \textbf{BH} outer ergosurface, which can be seen as effect of the Kerr spacetime frame--dragging--Figs\il(\ref{Fig:Plotrte4a}).
We use the notation $\mathrm{Q}_\Ta$  for any quantity $\mathrm{Q}$ considered at the inversion point, where  within the condition $\dot{\phi}=0$,  there is
 $\ell=\ell_\Ta\equiv \left.-\frac{g_{t\phi}}{g_{tt}}\right|_\Ta$. %

For the other  limiting values,  we will  consider radii
$\{r_\gamma^\pm,r_{mbo}^{\pm},r_{mso}^{\pm}\}$ and,  for the parameter $\ell$,
 the functions $\{\ell_{mso}^\pm,\ell_{mbo}^\pm,\ell_{\gamma}^\pm\}$,
defined in Appendix\il(\ref{Sec:co-rota})  for co-rotating $(-)$ and counter-rotating $(+)$ motion. (There is  $\mathrm{Q}_\bullet$ for any quantity $\mathrm{Q}$ evaluated at $r_\bullet$.)
Radii  $r_\gamma^\pm(a)$ are  photon  circular orbits on the equatorial plane\footnote{For the definition in relation to the photon sphere see for example \cite{2021GReGr..53...10T,Perlick,Johnson,Yang,2018EPJC...78..879C}.}.
They are solutions of  $\mathfrak{(R)}$ for $\sigma=1$ ($q=0$) and $\ell=\ell_\gamma^\pm(a)$ respectively.
Radii  $r_{mbo}^{\pm}$ are the marginal  circular bound orbits,   and radii $r_{mso}^{\pm}$ are marginal stable circular orbits
for time--like particles on the equatorial  plane,  having  $\ell= \ell_{mbo}^\pm$  and $\ell= \ell_{mso}^\pm$ respectively \citep{1972ApJ...178..347B}.
Therefore, radii
$\{r_{mbo}^{\pm},r_{mso}^{\pm}\}$  cannot be solutions of $\mathfrak{(R)}$ with   $\ell^\pm\in \{\ell_{mso}^\pm,\ell_{mbo}^\pm\}$ respectively, $q=0$ and  $\sigma=1$.
However, here we solve  $\mathfrak{(R)}$   on the \emph{spherical surfaces} defined by  radii $\{r_{mbo}^{\pm},r_{mso}^{\pm},r_\gamma^\pm\}$,    examining the  unstable photons circular orbits (solutions of  $\mathfrak{(R)}$)  on  $r=\{r_{mbo}^{\pm},r_{mso}^{\pm},r_\gamma^\pm\} $ for a \emph{general} angle  $\sigma$ and constants\footnote{That is, considering for example the radius $r_{mbo}^{\pm}$, defining a {spherical surface} embedding the  \textbf{BH},  a solution of  $\mathfrak{(R)}$   can exist on this surface on an angle $\sigma\neq 1$, a constant $q\neq 0$ and $\ell\neq \ell_{mbo}^\pm$. The corresponding solutions $(\sigma,q,\ell)$ are then detailed studied in Sec.\il(\ref{Sec:radiii-shadows}), with the corresponded point $(\alpha,\beta)$ on the shadow boundary and studied at the variation of $a$.}  $(\ell,q)$.
  Similarly, we find   solutions of  $\mathfrak{(R)}$  for spherical  photon orbits with parameter $\ell^\pm\in \{\ell_{mso}^\pm,\ell_{mbo}^\pm,\ell_{\gamma}^\pm\}$
  at a \emph{general} radius %
  $r$, a general  angle $\sigma$  and constant\footnote{In fact,   a solution of  $\mathfrak{(R)}$ with   $\ell_{mbo}^{\pm}$, for example,   can exist  on an orbit  $r\neq r_{mbo}^\pm$,  for   $\sigma\neq 1$ and a constant $q\neq 0$. The  solutions $(r,q,\sigma)$ are then detailed studied in Sec.\il(\ref{Sec:allell}),  with the corresponded point $(\alpha,\beta)$ on the shadow boundary, and studied at the variation of $a$.}  $q$.

The spherical surfaces $\{r_{mbo}^{\pm},r_{mso}^{\pm},r_\gamma^\pm\}$ can provide an immediate  astrophysical context in relation to the  structures orbiting around the \textbf{BH}\footnote{There is
$
\mp\ell_{mso}^\pm\leq \mp\ell_{mbo}^\pm\leq \mp\ell_{\gamma}^\pm
$    and $
r_{mso}^\pm>r_{mbo}^\pm>r_{\gamma}^\pm
$, being valid separately for the upper and lower signs. However, it is important to stress that  the  relation between the $(\pm)$ limiting impact parameters, and  the relative location of  the $(\pm)$  orbital spherical shells, defined by the   radii $(r_{mso}^\pm,r_{mbo}^\pm,r_{\gamma}^\pm
,r_\epsilon^+(a),r_\Ta(a,\ell_\Ta))$,  depend not trivially on the \textbf{BH}  spin, as it  can be easy seen by plotting the functions  with respect to the \textbf{BH} spin-- see Figs\il(\ref{Fig:Plotscadenztoaglrl})  and   \cite{retro-inversion}.}. In fact, the accretion disk inner edge (cusp)  orbiting on  the \textbf{BH} equatorial plane is located in region     $r\in]r_{mbo}^\pm,r_{mso}^\pm]$\footnote{This assumption is widely adopted (and  well grounded)  in \textbf{BH} Astrophysics, and in the following analysis we   will   use this  assumption  {independently} of  other specific details  of the  accretion disk models.} of the equatorial plane, and therefore, close to  $\sigma=1$, the inner region of an axially symmetric counter--rotating or co--rotating accretion torus is in the  spherical shell  $]r_{mbo}^\pm,r_{mso}^\pm]$ respectively.
The inner orbital  shell,  $[r_{\gamma}^\pm, r_{mbo}^\pm]$,
covers  on the equatorial plane the cusps of the
 open (hydrodynamical) toroidal structures  corresponding to  matter funnels along the \textbf{BH} rotational axis, which are  generally  associated to  geometrically    thick accretion  tori,-- \citep{KJA78,AJS78,Sadowski:2015jaa,Lasota:2015bii,proto-jet,Lyutikov(2009),Madau(1988),Sikora(1981)}.

In geometrically thin disks, the  flows, freely falling from the  cusps can be considered leaving the marginally stable circular orbit with  $\ell=\ell_{mso}^\pm$,  and  more in general  with $\mp\ell^\pm\in ]\mp\ell_{mbo}^\pm,\mp\ell_{mso}^\pm]$  from $]r_{mbo}^\pm,r_{mso}^\pm]$ (or
$\mp\ell^\pm\in [\mp\ell_{mbo}^\pm,\mp\ell_{\gamma}^\pm[$ according to the cusp location in $]r_{\gamma}^\pm, r_{mbo}^\pm]$ for the open configurations).
These momenta also fix the range of location of the accretion disk  center, defined as the  maximum pressure  point in the disks--see Appendix\il(\ref{Sec:co-rota}).

\begin{figure*}
\centering
    \includegraphics[width=8cm]{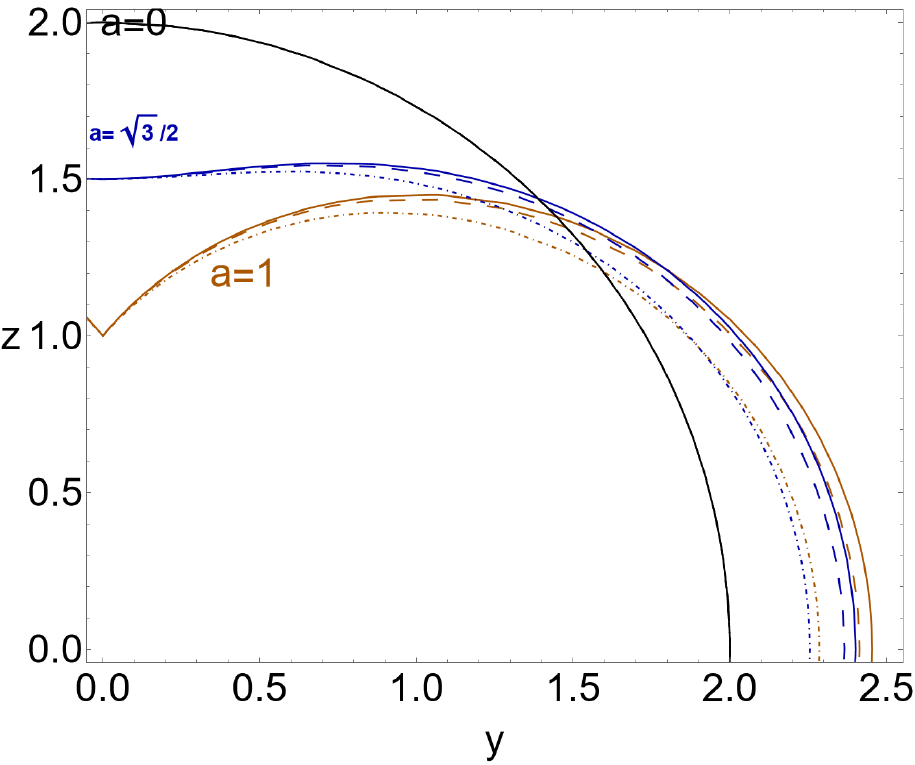}
    \includegraphics[width=8cm]{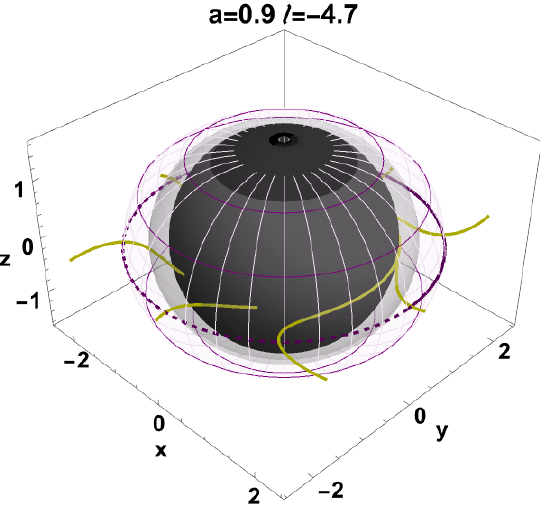}
   \caption{Left panel:  inversion surfaces sections, for   $\ell_{mso}^+$ (plain) and $\ell_{mbo}^+$  (dashed), $\ell_{\gamma}^+$  (dotted-dashed). It is $r= \sqrt{y^2+z^2}$ and $\theta=\arccos({z}/r)$.   Right panel:  counter-rotating  photon orbits  (yellow curves) to the   \textbf{BH} (black region). Inversion point $r_{\Ta}$  is the deep-purple curve. It is  $\{z=r \cos\theta,y=r \sin \theta \sin\phi,x=r \sin\theta \cos\phi\}$.
    Gray region is the outer ergosurface, light-purple  region is  $r< r_\Ta(\sigma_\Ta)$.}\label{Fig:Plotrte4a}
\end{figure*}
Hence,   in   Sec.\il(\ref{Sec:allell}) we investigate the  solutions  of $(\mathfrak{R})$,  for
$\ell\in\{\ell_{mso}^\pm,\ell_{mbo}^\pm,\ell_{\gamma}^\pm,\ell_\Ta\}$,   in terms of orbit $r$, constant $q$, angle $\sigma$, and coordinates $(\alpha,\beta)$.
In
  Sec.\il(\ref{Sec:radiii-shadows})   we examine  the  solutions    of $(\mathfrak{R})$,
for  $r=\{r_{\gamma}^\pm,r_{mbo}^\pm,r_{mso}^\pm\}$, and  in the spherical shells
$r\in[r_{mbo}^\pm,r_{mso}^\pm]$, $r\in[r_{\gamma}^\pm,r_{mbo}^\pm]$, in terms of parameter $\ell$ and $(q,\sigma,\alpha,\beta)$.

\section{Shadows analysis}\label{Sec:first-lun}
\subsection{Parameters $\ell\in\{\ell_{mso}^\pm,\ell_{mbo}^\pm,\ell_{\gamma}^\pm,\ell_{\Ta}\}$}\label{Sec:allell}
{In general for \textbf{BHs} with spin\footnote{Spin  $a_g$ is expression of the \textbf{BH} horizons in terms of the (horizons) specific angular momentum $\ell$ and,  can be expressed in terms of the $\alpha$ celestial coordinate and the angle $\sigma$.-- see Figs\il(\ref{Fig:Plotfondoagl}). The \textbf{BH}  horizons  angular momentum $\ell_H^\pm\equiv 1/\omega_H^\pm$, where $\omega_H^\pm(a)\equiv {r_\mp}/{2a}$ are the \textbf{BH} horizons frequencies (relativistic angular velocity), are related  to parameter  $\ell$--\citep{bundle-EPJC-complete}.}
$a\in ]0,a_g[$,  spherical  photon  orbits, solutions of   system  $\mathfrak{(R)}$   with    $\ell\in\{\ell_{mso}^\pm,\ell_{mbo}^\pm\}$,  are  on  radius $r=r_\lambda\neq\{r_{mso}^\pm,r_{mbo}^\pm\}$ with motion  constant $q=q_\lambda\geq 0$},  where
\bea
&&\label{Eq:qlambda}
a_g(\ell)\equiv \frac{4 \ell }{\ell ^2+4}=-\frac{4 \alpha  \sqrt{\sigma }}{\alpha ^2 \sigma +4}\quad\mbox{and}\quad
q_\lambda\equiv \frac{r \left[a^2 (r+2)-4 a \ell +r^3-(r-2) \ell ^2\right]}{\Delta},\quad
r_{\lambda }\equiv\frac{3-a (a+\ell )}{\sqrt[3]{3} \sqrt[3]{\lambda}}+\frac{\sqrt[3]{\lambda }}{3^{2/3}}+1 \eea
where $\lambda \equiv \frac{1}{6} \sqrt{\left[54(1- a^2)\right]^2+108 [a (a+\ell )-3]^3}+9(1- a^2)$.
%
{Functions $(q_\lambda,r_\lambda)$ will be further constrained when evaluated  for $\ell\in\{\ell_{mso}^\pm,\ell_{mbo}^\pm\}$, as we  discuss  in details below}.
In Appendix\il(\ref{App:allell}),  various aspects of the analysis have been further detailed.

First, by using  the condition $T\geq0$ (and $\dot{t}>0$),
we find that  for
 $a\in ]0,1]$ there is $\sigma\in [\sigma_\lambda,1]$ and,
  for the Schwarzschild \textbf{BH},  solutions exist  for     $\sigma\in [\sigma_\ell,1]$,   where $\sigma_\ell$ and $\sigma_\lambda$ are
\bea
&&\label{Eq:sigmalambda}
\sigma_{\ell}\equiv \frac{\ell ^2}{q+\ell ^2},\quad
\sigma _{\lambda }\equiv \frac{q+\ell ^2-a^2 \left[\sqrt{\frac{\left(q+\ell ^2\right)^2}{a^4}+\frac{2 \left(q-\ell ^2\right)}{a^2}+1}-1\right]}{2 a^2}.
\eea
Therefore,
$\sigma$ is bottom bounded by a value $\sigma_\ell$ or $\sigma_\lambda$, depending on the \textbf{BH} spin  and the Carter constant.
The limiting   angle $\sigma_\lambda$ is shown in Figs\il(\ref{Fig:Plotsigmaglr}) for $\ell\in \{\ell_{mso}^\pm, \ell_{mbo}^\pm\}$ for \textbf{BHs} with spin $a\in[0,1]$. It is clear how   $\sigma_\lambda$  changes for the co-rotating and counter-rotating case, and for slowly spinning and fast spinning  \textbf{BHs}  for value $\ell=\ell_{mso}^\pm$.
{Solutions $r_\lambda$, for $\ell\in \{\ell_{mso}^\pm,\ell_{mbo}^\pm,\ell_{\gamma}^\pm,\ell_\Ta\}$, are  shown in Figs\il(\ref{Fig:Plotscadenztoaglrl}) together with  radii $ \{r_{mso}^\pm,r_{mbo}^\pm,r_{\gamma}^\pm,r_\Ta\}$.
Solutions $(\sigma_\lambda,q_\lambda)$, relative to the cases $\ell\in \{\ell_{mso}^\pm,\ell_{mbo}^\pm,\ell_{\gamma}^\pm,\ell_\Ta\}$, are  shown in Figs\il(\ref{Fig:Plotfondoaglrsigmapbcasdu},\ref{Fig:Plotsigmaglr}).  In Fig.\il(\ref{Fig:Plotfondoaglrsigmapbcasdu})--left panel
 we also show  the  celestial coordinate $\alpha$ as function of the \textbf{BH}  dimensionless  spin $a$,  for  $\sigma=1$, evaluated on the parameter $\ell\in \{\ell_{mso}^\pm,\ell_{mbo}^\pm,\ell_{\gamma}^\pm\}$.}

\begin{figure}
\centering
\includegraphics[width=5.6cm]{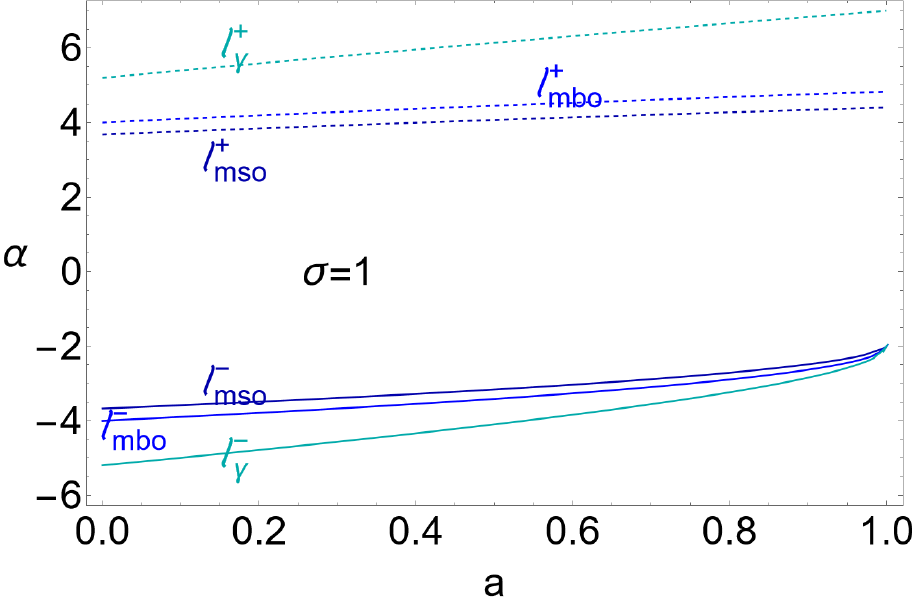}
\includegraphics[width=5.6cm]{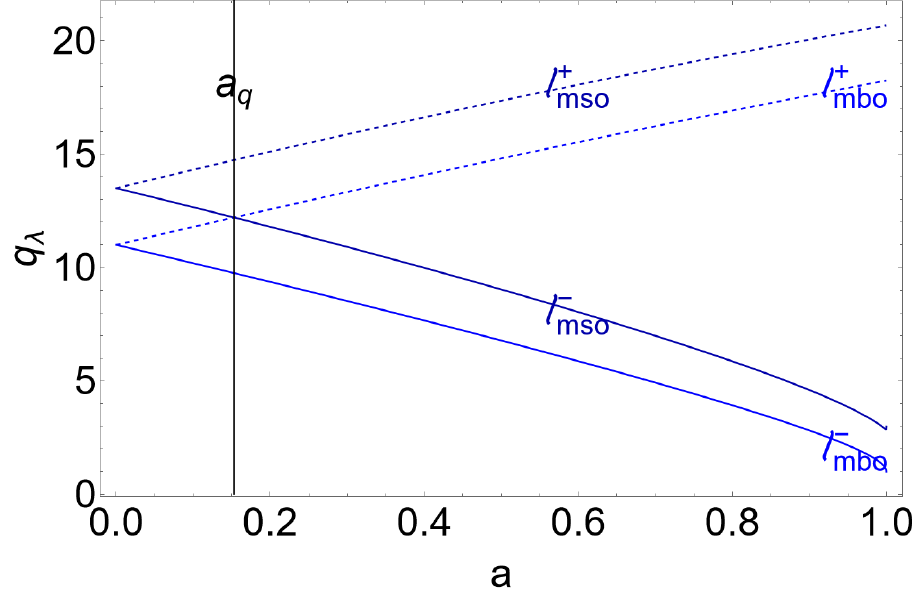}
\includegraphics[width=5.6cm]{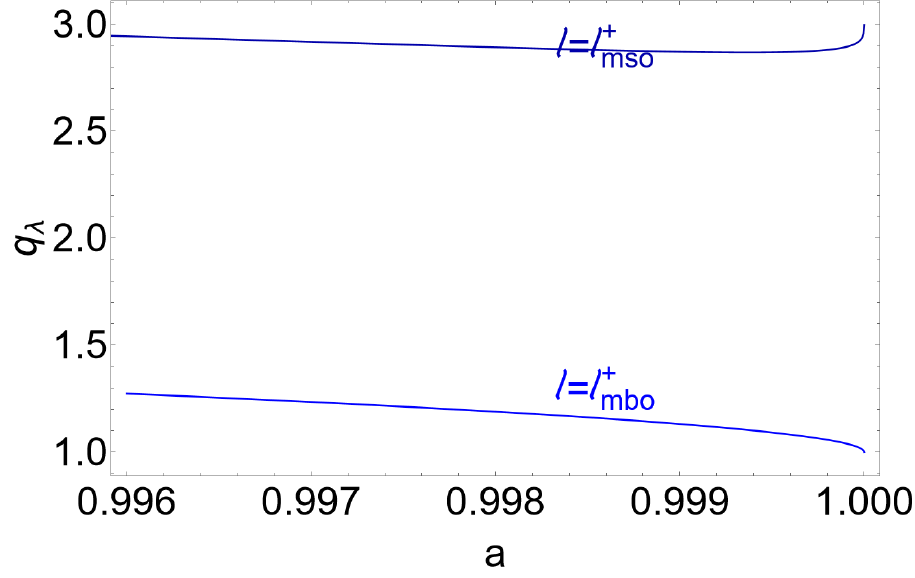}
\caption{For $\ell\gtrless 0$ there is $\alpha\lessgtr 0$ respectively, and solutions $(-\alpha)$ are not represented.
Quantity  $q_\lambda$ is in  Eq.\il(\ref{Eq:qlambda}), spin $a_q$ is in Eqs\il(\ref {Eq:aq-definition}). Right panel is a close up view  of the center panel. }\label{Fig:Plotfondoaglrsigmapbcasdu}
\end{figure}
\begin{figure}
\centering
\includegraphics[width=5.25cm]{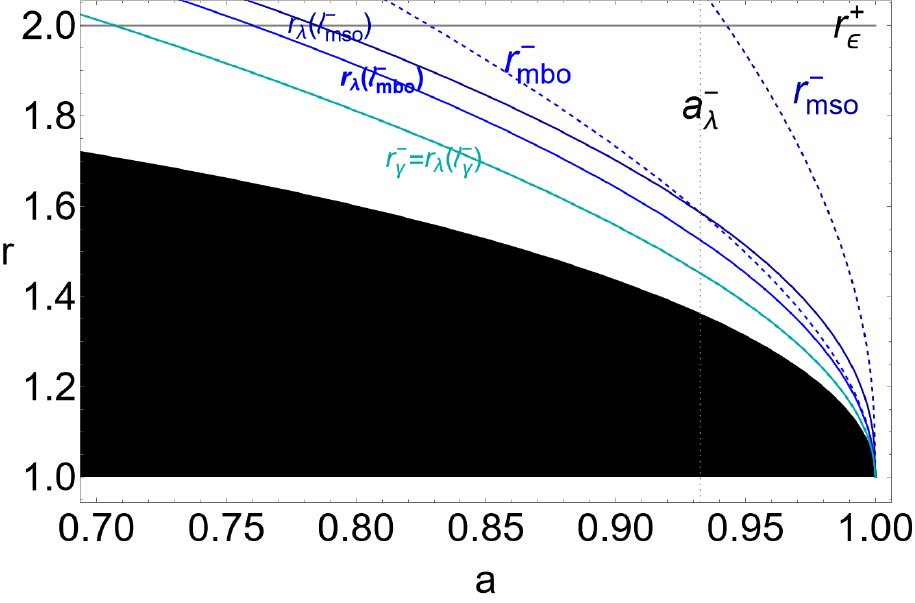}
\includegraphics[width=5.25cm]{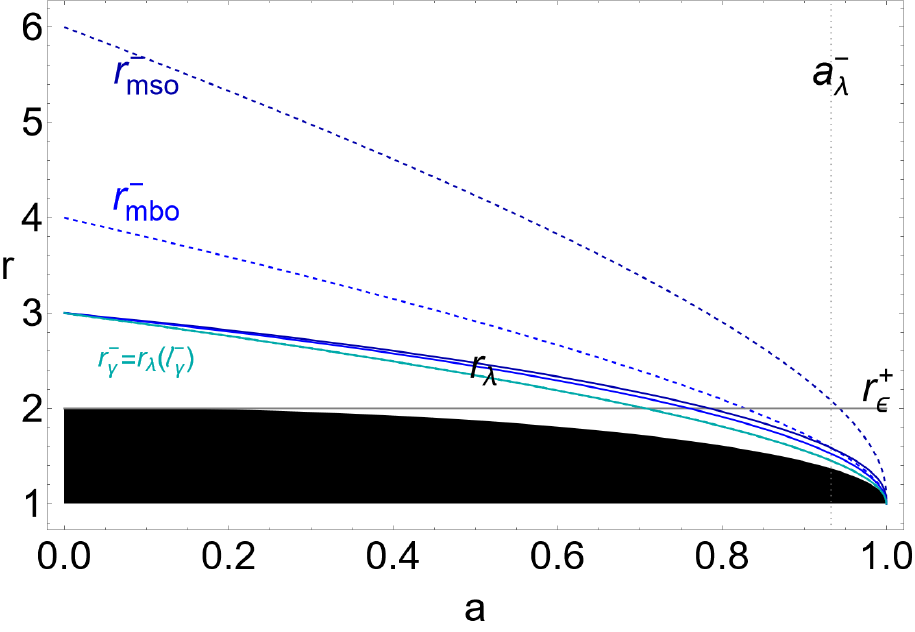}
\includegraphics[width=5.25cm]{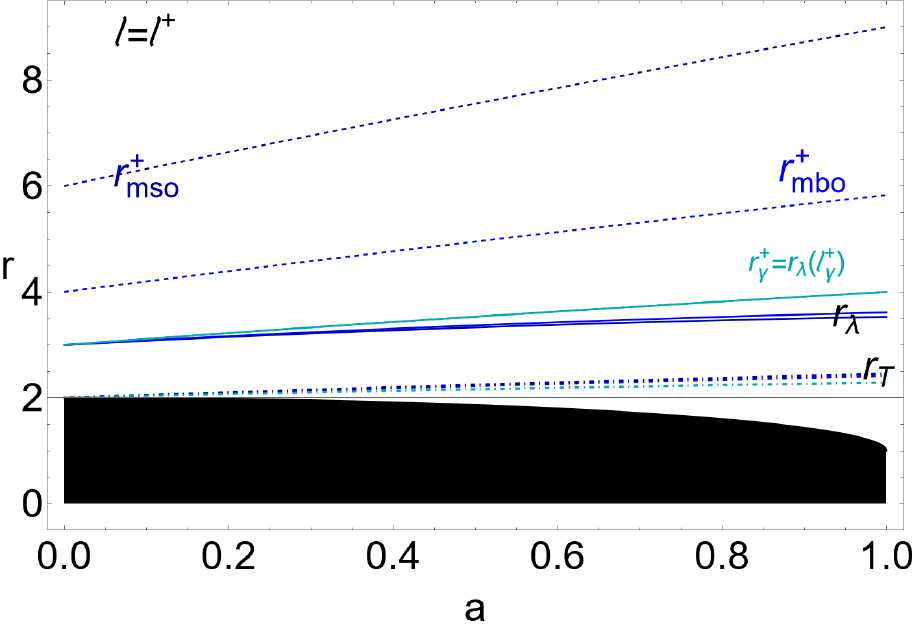}
\caption{%
  Radius $r_\lambda$ (solid curves) and $r_\Ta$ (right panel: dotted--dashed curves)  are plotted for
   $\ell\in \{\ell_\gamma^\pm,\ell_{mbo}^\pm,\ell_{mso}^\pm\}$ (cyan, blue,darker-blue) together with the radii
   $\{r_\gamma^\pm,r_{mbo}^\pm,r_{mso}^\pm\}$ (dashed curves).  Spin $a_{\lambda}^-$ is in Eq.\il(\ref{Eqs:amsom}). Left  panel is a close-up view of the center panel.}\label{Fig:Plotscadenztoaglrl}
\end{figure}
\begin{figure}
\centering
\includegraphics[width=6cm]{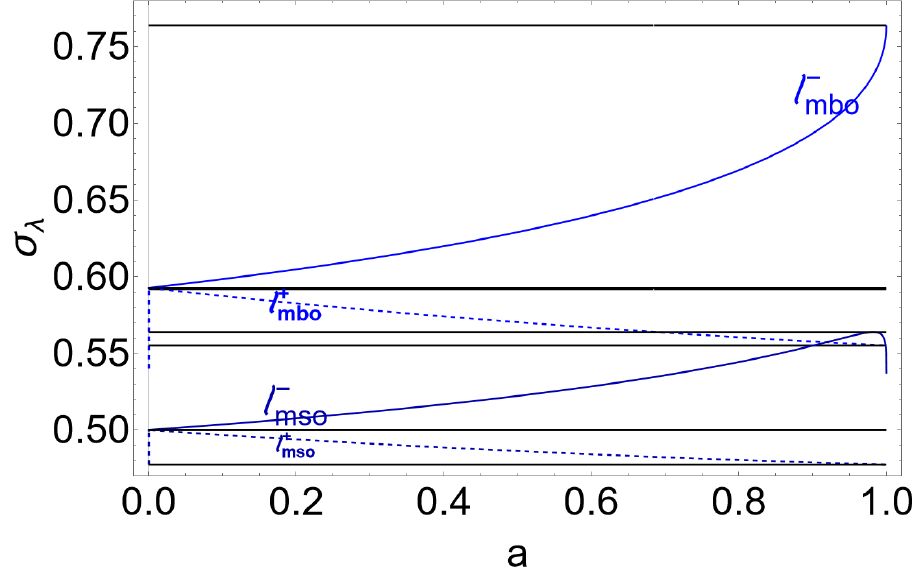}
\includegraphics[width=5.6cm]{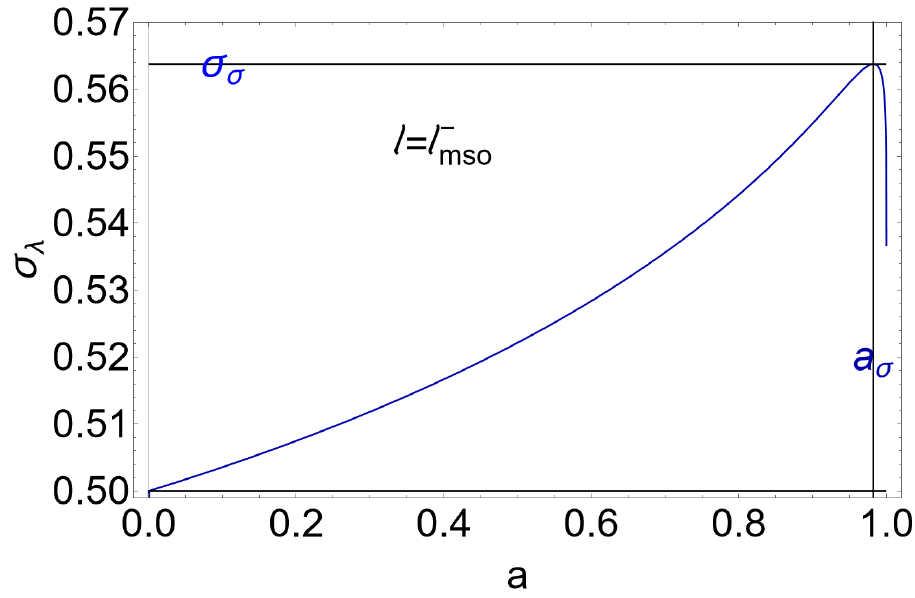}
\caption{Quantity  $\sigma_\lambda$ of Eq.\il(\ref{Eq:sigmalambda}) for  $\ell\in\{\ell_{mso}^\pm,\ell_{mbo}^\pm\}$. Left panel is a close--up view of the right panel. Maximum value $(a_\sigma, \sigma_\sigma)$  is in Eq.\il(\ref{Eqs:amsom}).}\label{Fig:Plotsigmaglr}
\end{figure}

\medskip

\textbf{{Parameter $\ell=\ell_{mso}^-$}}

\medskip

The case  $\ell=\ell_{mso}^-$ is illustrated in   Figs\il(\ref{Fig:Plotbetrodon1lmsom}),
where we show the first results of the analysis for this constraint  by  giving the constrained  $\beta$ as a function of the spin $a$ for \emph{all} possible $\sigma$ (left  upper panel), and as function of the angle  $\sigma$ for \emph{each} \textbf{BH} spin $a$ (central upper panel), and finally as function of the celestial coordinate $\alpha$ for $\sigma\in [0,1]$ and for different  $ a$, as signed on the curves. Each point of a curve is for a different $\sigma$. These panels provide immediate  information on the parts of  shadow boundary   correspondent to  null geodesics with $\ell=\ell_{mso}^-$  for different $(\sigma, a)$.
Lower panels
summarize the results for this constraint,  by relating the coordinates $(\beta,\alpha)$ to $(r,q)$, in the entire range of possible $a$ and $\sigma$.

Thus, as clear also from Figs\il(\ref{Fig:Plotbetrodon1lmsom}), photons with $\ell=\ell_{mso}^-$  can be observed for   large latitudes, i.e. $\sigma\gtrapprox0.53$. The  coordinate $\beta$ (in magnitude) increases with $\sigma$ and decreases with $a$ for $\sigma<\sigma_\sigma$ and $a<a_\sigma$ (where $a_{\sigma}\equiv 0.98217$, and $\sigma _{\sigma }\equiv 0.563773$--see Figs\il(\ref{Fig:Plotsigmaglr})).

Considering  $\beta$ for different  $a$ and $\sigma$, it  is clear that  difference appears for  large latitude  angles, i.e. $\sigma>\sigma_\sigma$,  and small  $\sigma\in [0.53,\sigma_\sigma]$. Angle  $\sigma_\sigma$ (and spin  $a=a_\sigma$)  is a critical value where  there can be  $\beta=0$.
With  $\sigma>\sigma_\sigma$ and $ a>a_\sigma$,  solutions   appear for small values of $(\beta,\alpha)$ in magnitude. The external regions  of the plane $\alpha-\beta$, correspondent to the  larger  values of $(\beta,\alpha)$ in magnitude, distinguish  slower from faster  spinning \textbf{BHs}, and the smaller $(\sigma\approx 0.56)$ from larger latitudinal angles.

\begin{figure}
\centering
\includegraphics[width=5.6cm]{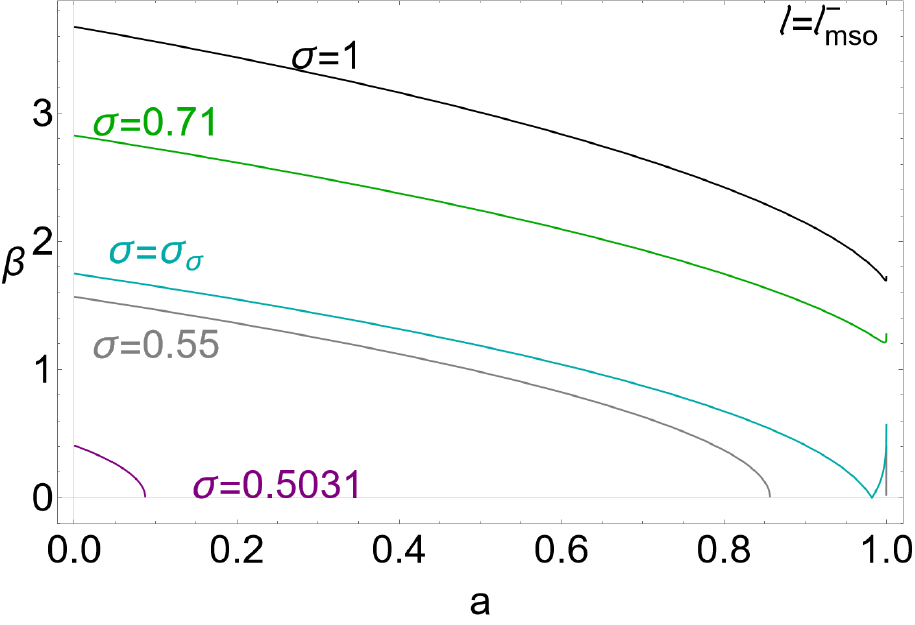}
\includegraphics[width=5.6cm]{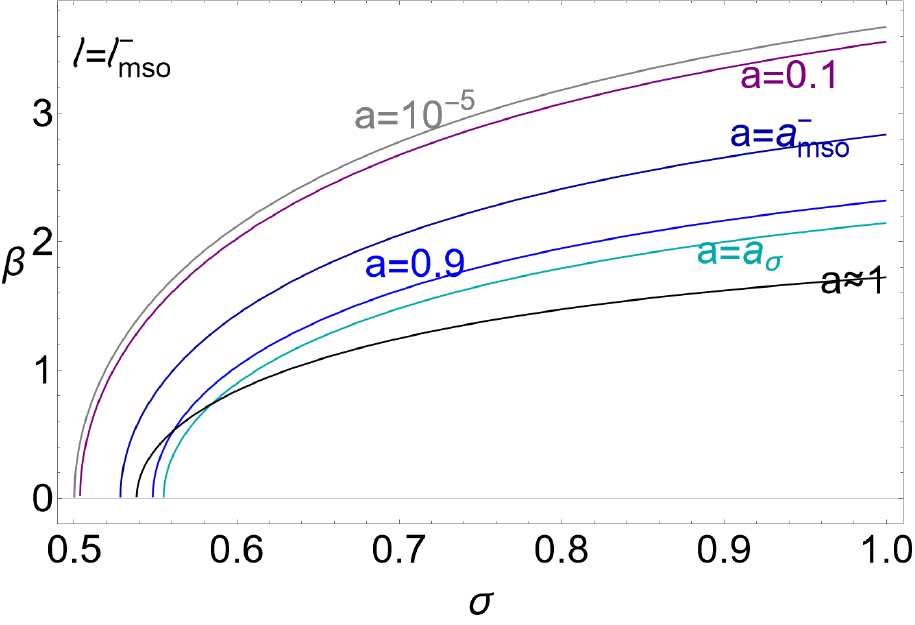}
\includegraphics[width=5.6cm]{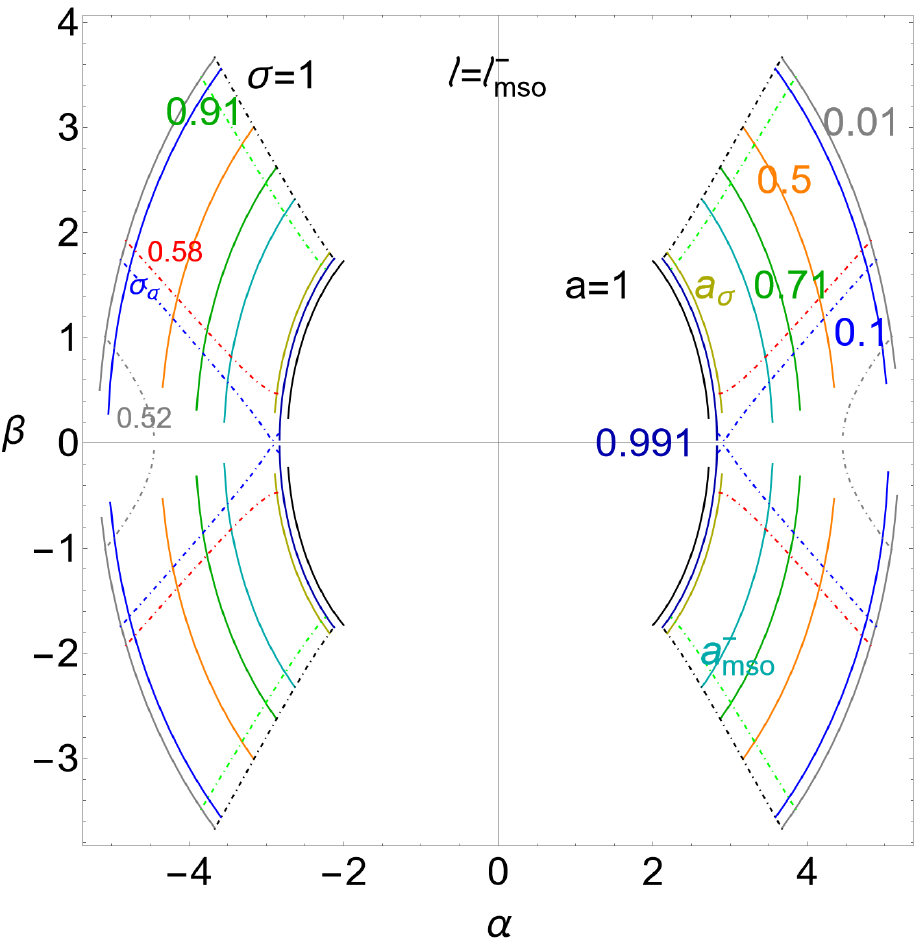}
\includegraphics[width=4.25cm]{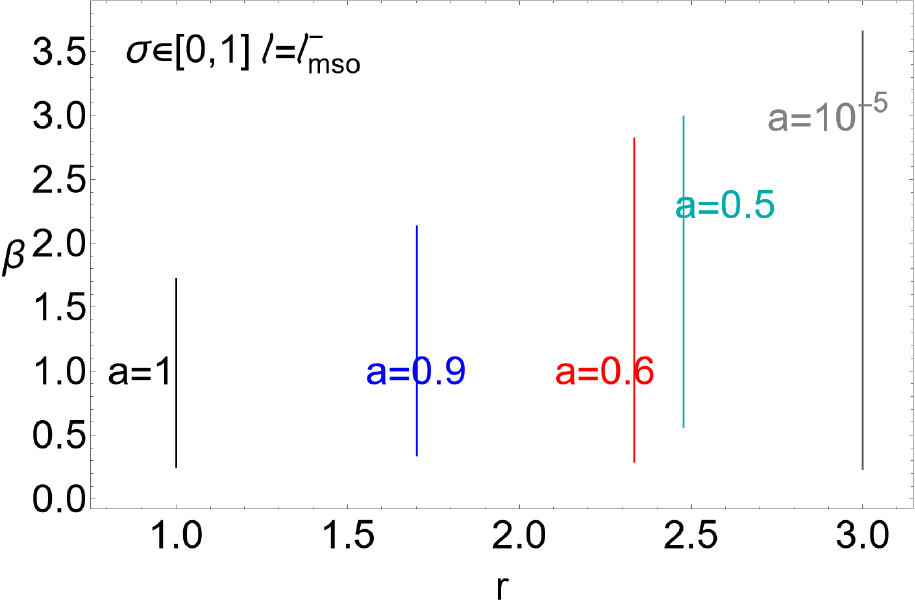}
\includegraphics[width=4.25cm]{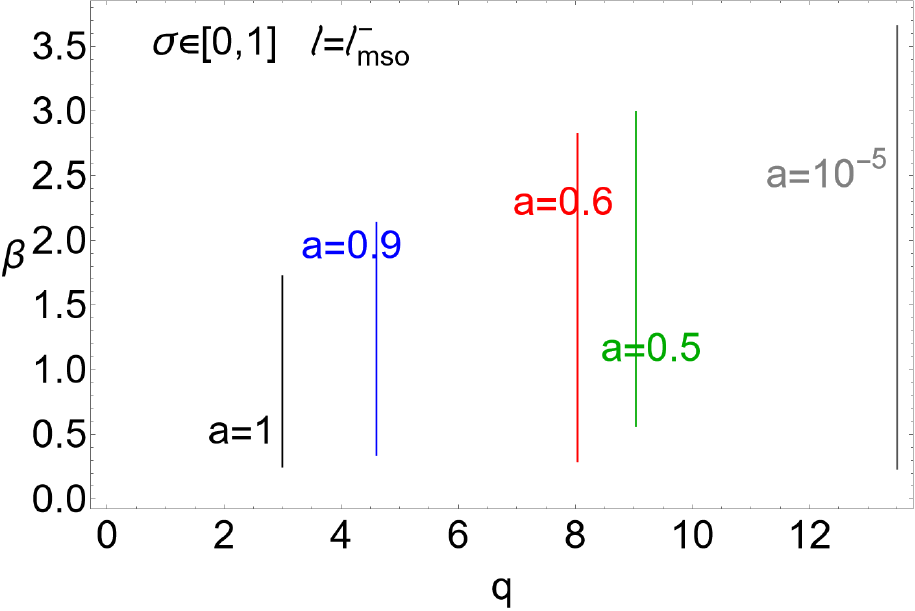}
\includegraphics[width=4.25cm]{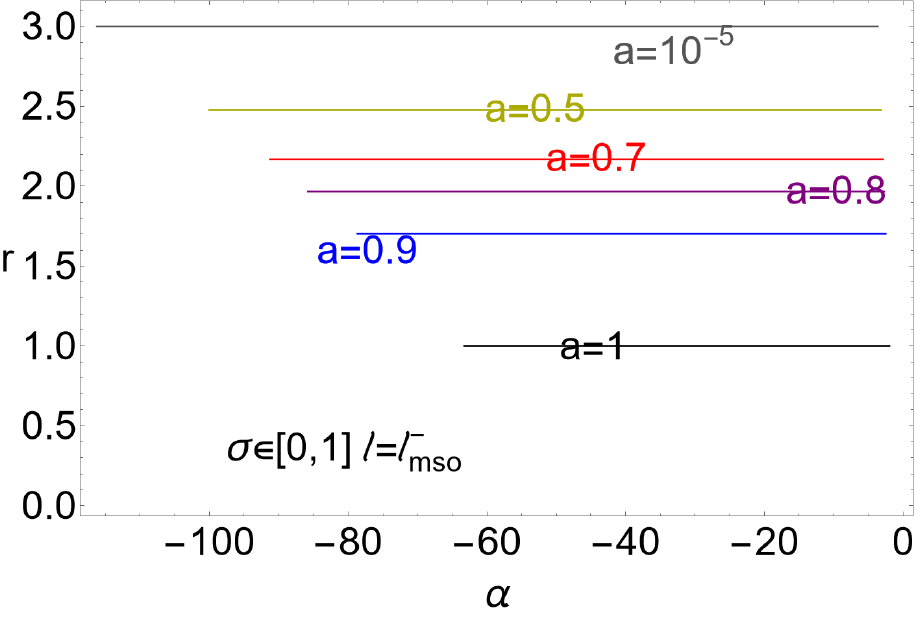}
\includegraphics[width=4.25cm]{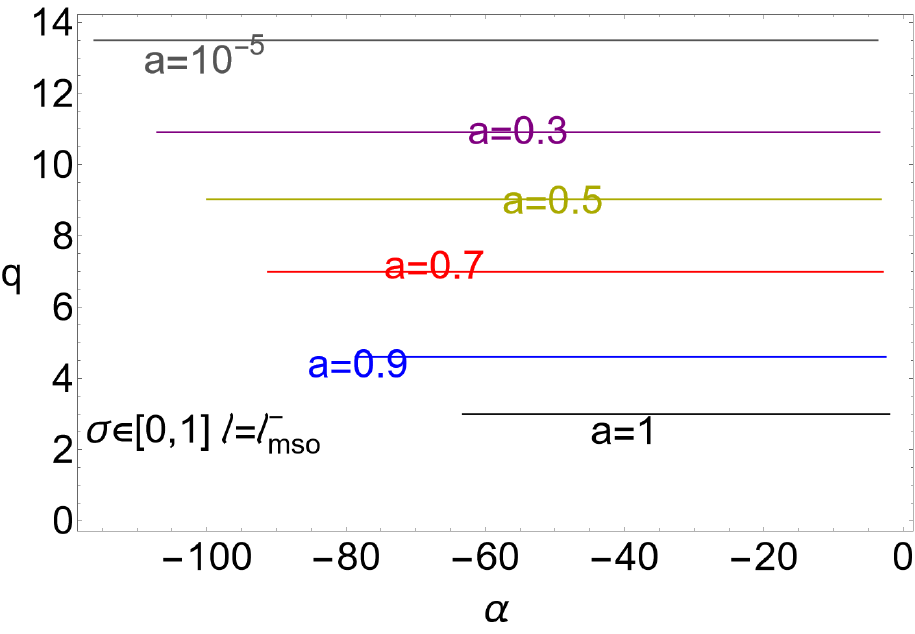}
\caption{Case $\ell=\ell_{mso}^-$.  Upper panels: each point of a  solid curve is for a different $\sigma$.
Dotted-dashed lines are for different  $\sigma$ signed on the curves. Each point of a  dotted-dashed curve is for a different spin $a$.  Bottom panels:  each point of a curve is for a different   $\sigma$.}.\label{Fig:Plotbetrodon1lmsom}
\end{figure}

From Figs\il(\ref{Fig:Plotfondoaglrsigmapbcasdu},\ref{Fig:Plotscadenztoaglrl})   we see that  $r_\lambda$ and $q_\lambda$ are constant for fixed $a$ and different $\alpha$ (and confirmed also by Figs\il(\ref{Fig:Plotbetrodon1lmsom})). The radius $r\in[1,3]$  and parameter $q$ decrease with the \textbf{BH}  dimensionless  spin and the  $\beta$ in magnitude and $q$ ranges decrease with the \textbf{BH} spin.
The celestial coordinate $\alpha$ is negative, since $\ell>0$. As shown in Figs\il(\ref{Fig:Plotbetrodon1lmsom}) and Fig.\il(\ref{Fig:Plotfondoaglrsigmapbcasdu}) the  magnitude $\alpha$ decreases with $a$ and $q$, increases with $r$ and the range of $\alpha$ values  increases with $r$ and decreases with $a$.

\medskip

\textbf{{Parameter $\ell=\ell_{mso}^+$}}

\medskip

Photons with  $\ell=\ell_{mso}^+$ are considered in  Figs\il(\ref{Fig:Plotverinimsop}). It is clear how   $|\beta|$ increases with the \textbf{BH} spin and  the angle $\sigma>0.479$. For $\sigma>0.55$ there is $|\beta|>0$.
For this constraint, small values of $(\alpha,\beta) $ (in magnitude), i.e. the inner regions of the $\alpha-\beta$ plane,  characterize  slowly spinning  \textbf{BHs}.
From Figs\il(\ref{Fig:Plotbetrodon1lmsopbm})--lower panels (and Fig.\il(\ref{Fig:Plotfondoaglrsigmapbcasdu}) it is clear as $r$ and $q$ are constant for different $(\sigma,\beta,\alpha)$.
The radius $r_\lambda$ and the quantity $q_\lambda$ are larger  then in for $\ell=\ell_{mso}^-$, increasing  with the  \textbf{BH} spin.  The range of values of the celestial coordinate  $\beta$  increases with $(r,a)$. On the other hand, the range of values for the  $\alpha$  coordinate increases with ($r,a,q$).

\begin{figure}
\centering
\includegraphics[width=5.6cm]{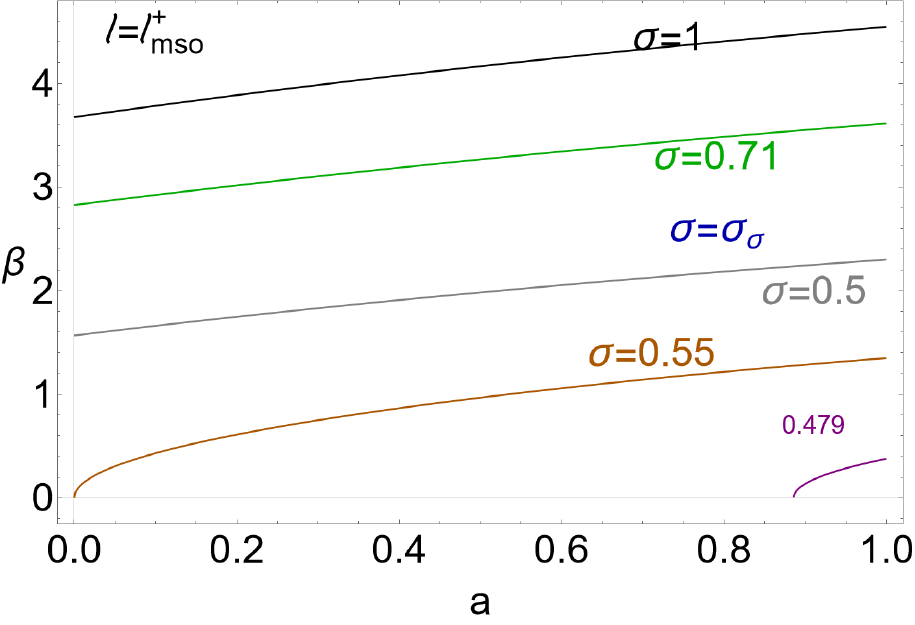}
\includegraphics[width=5.6cm]{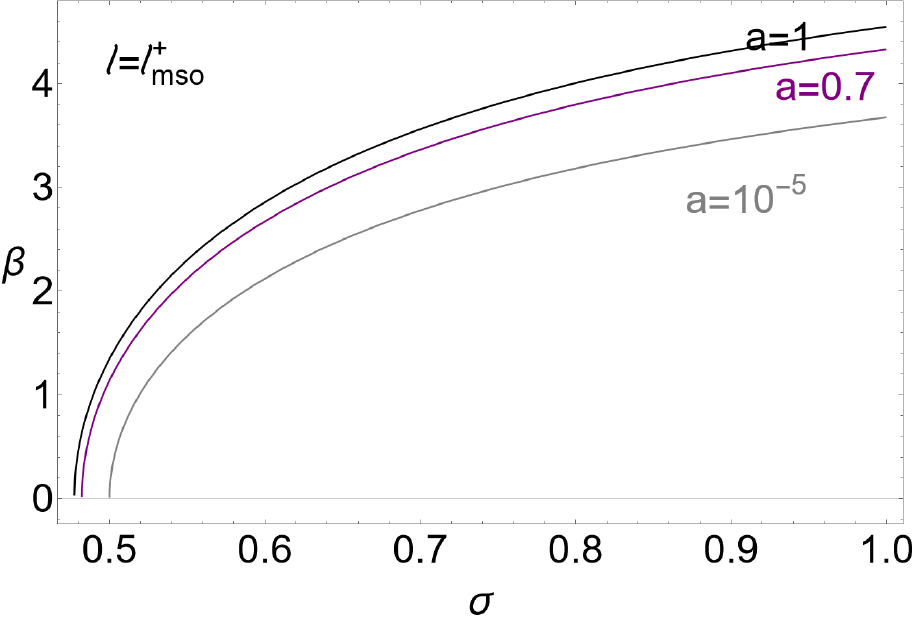}
\includegraphics[width=5.6cm]{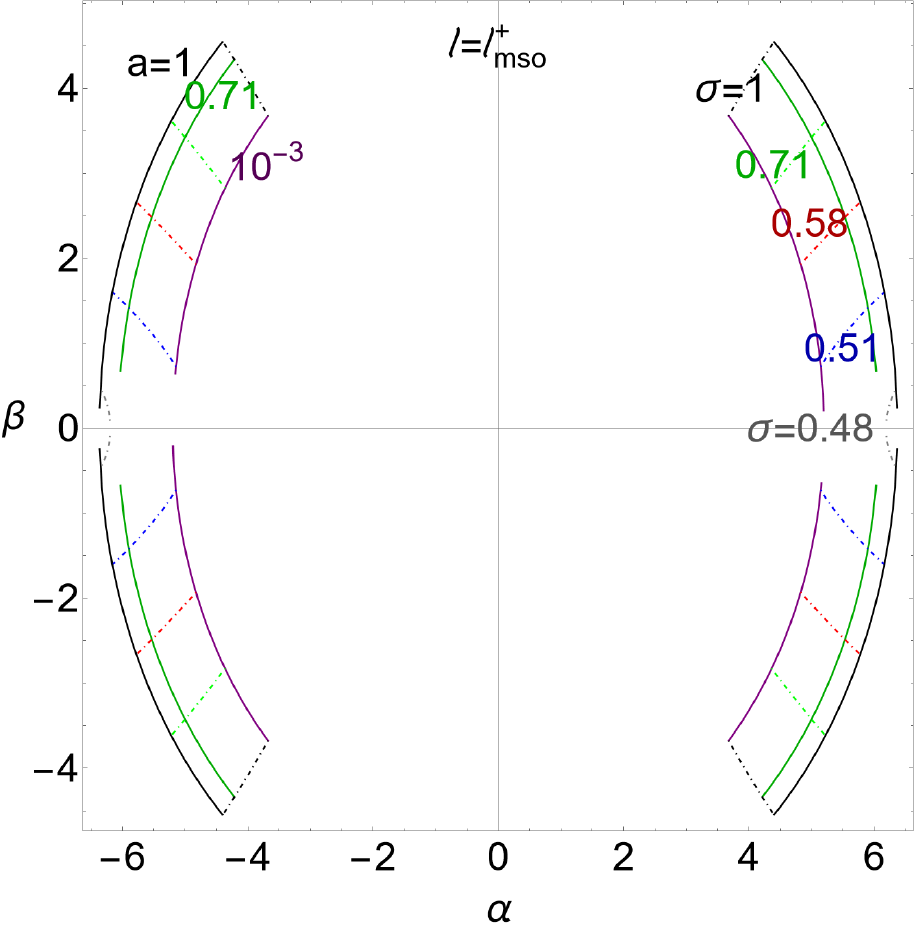}
\includegraphics[width=4.25cm]{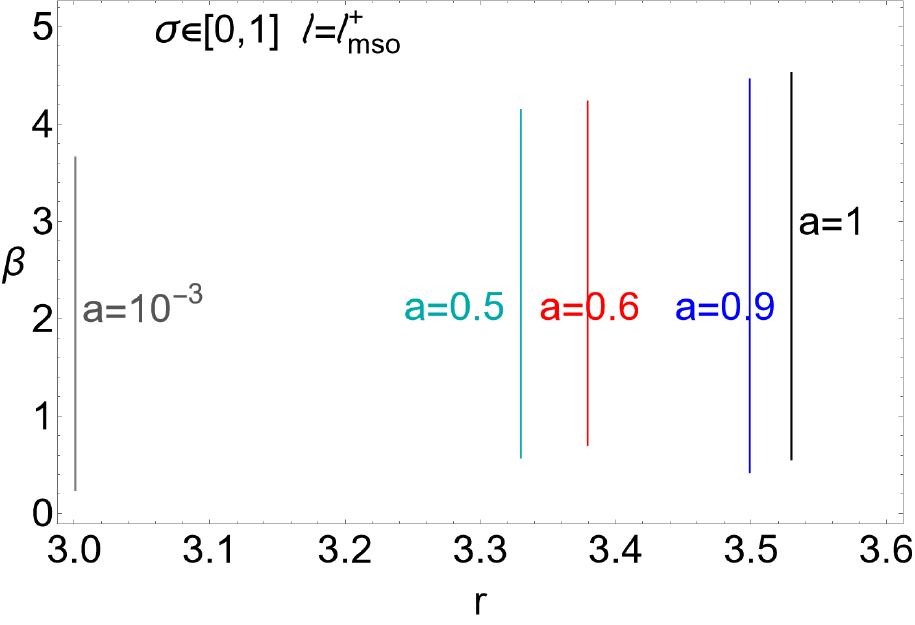}
\includegraphics[width=4.25cm]{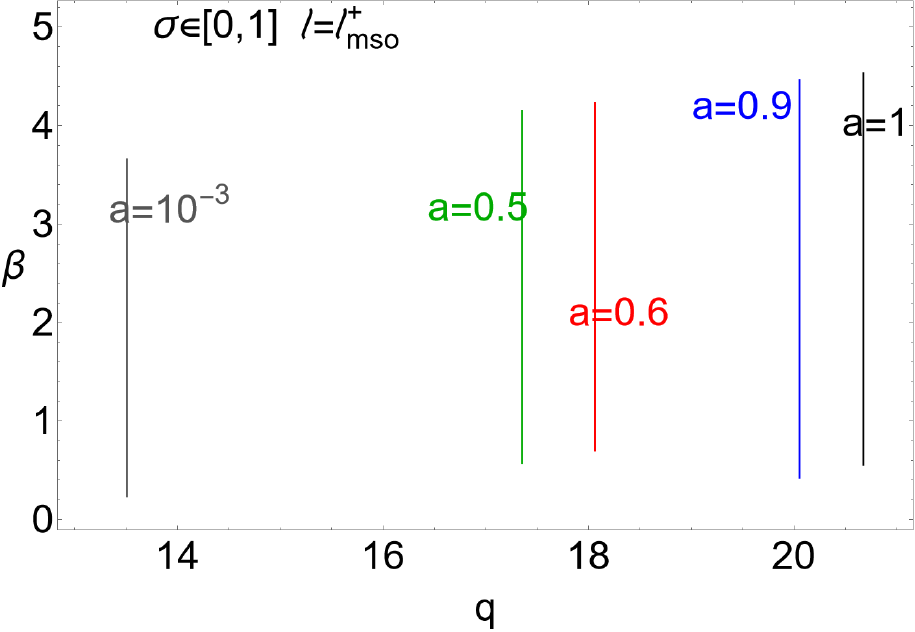}
\includegraphics[width=4.25cm]{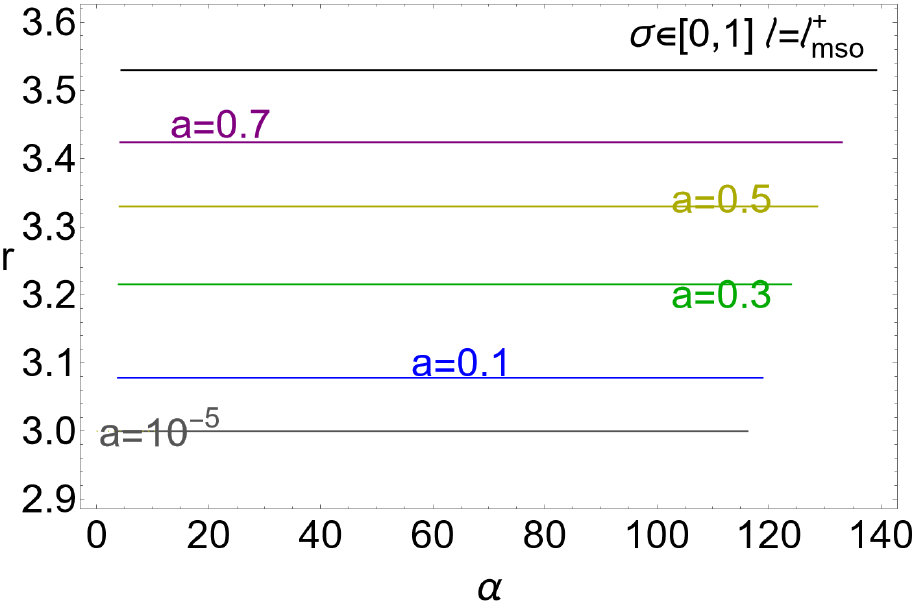}
\includegraphics[width=4.25cm]{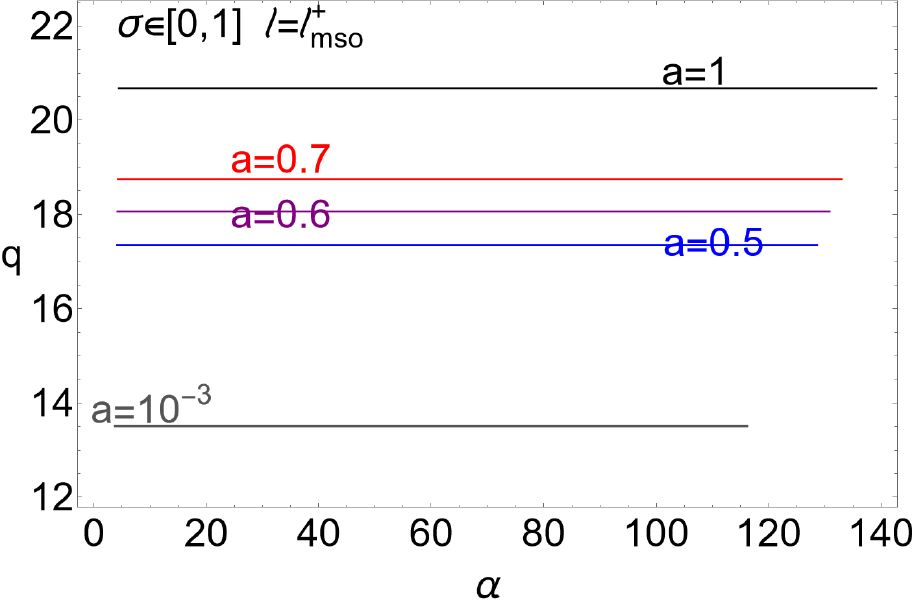}
\caption{Analysis of  the $\ell=\ell_{mso}^+$ case. {(For further details see Figs\il(\ref{Fig:Plotbetrodon1lmsom}).) %
}}\label{Fig:Plotverinimsop}
\end{figure}
\begin{figure}
\centering
\includegraphics[width=5.6cm]{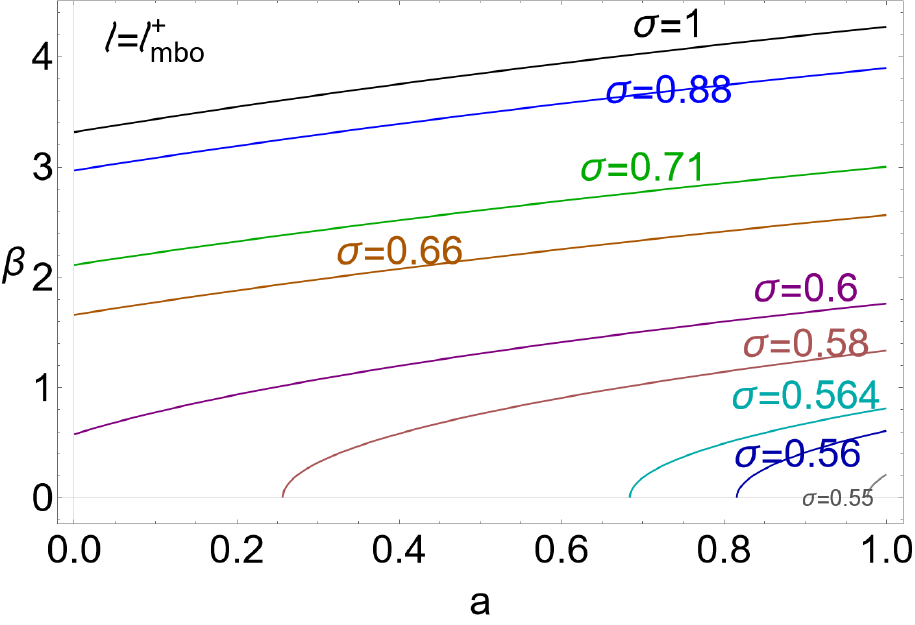}
\includegraphics[width=5.6cm]{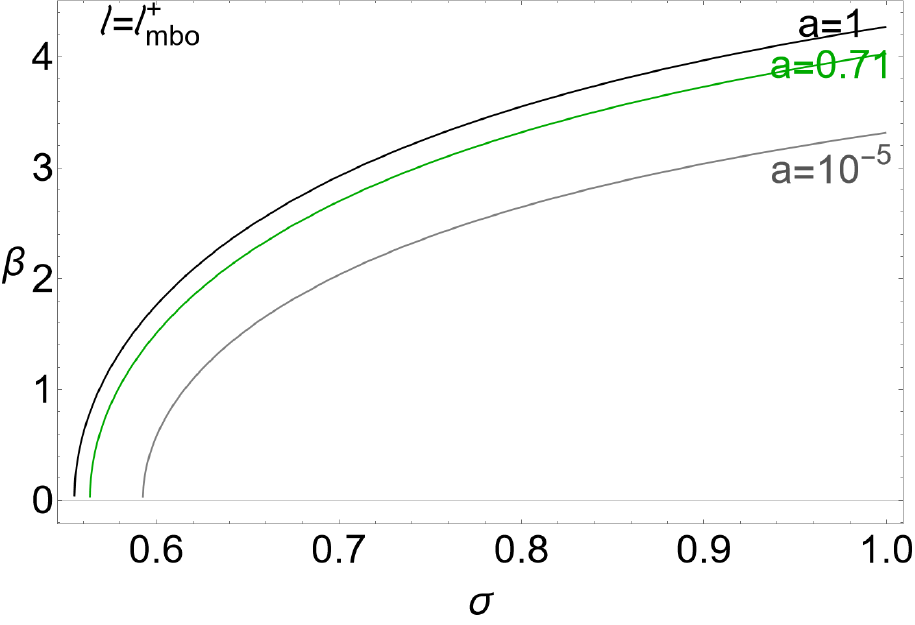}
\includegraphics[width=5.6cm]{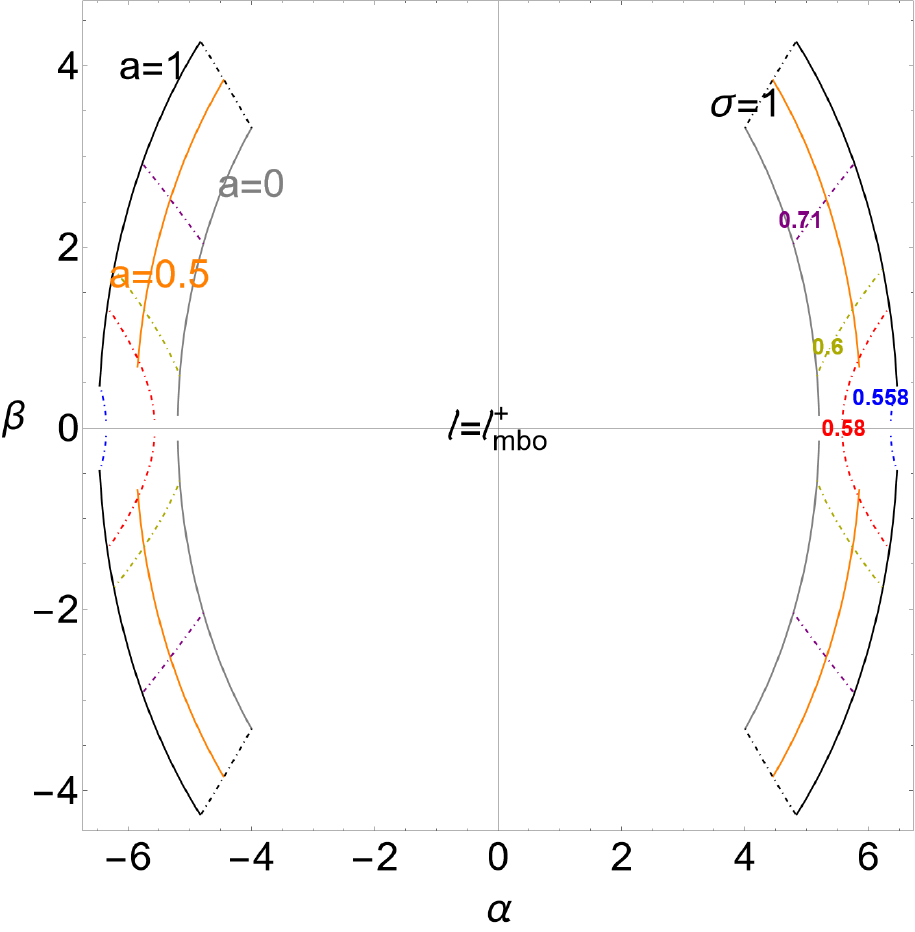}
\includegraphics[width=4.25cm]{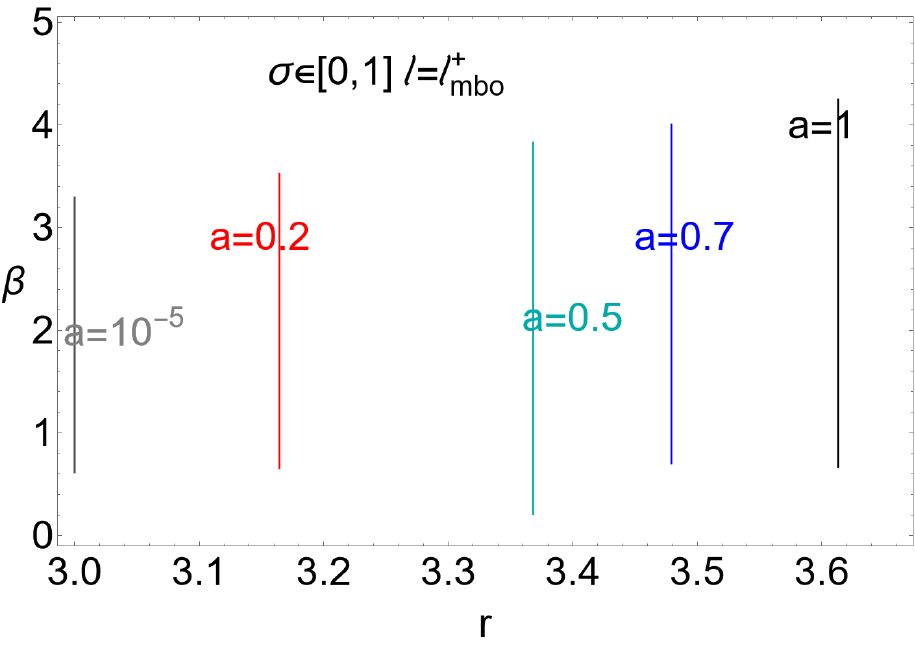}
\includegraphics[width=4.25cm]{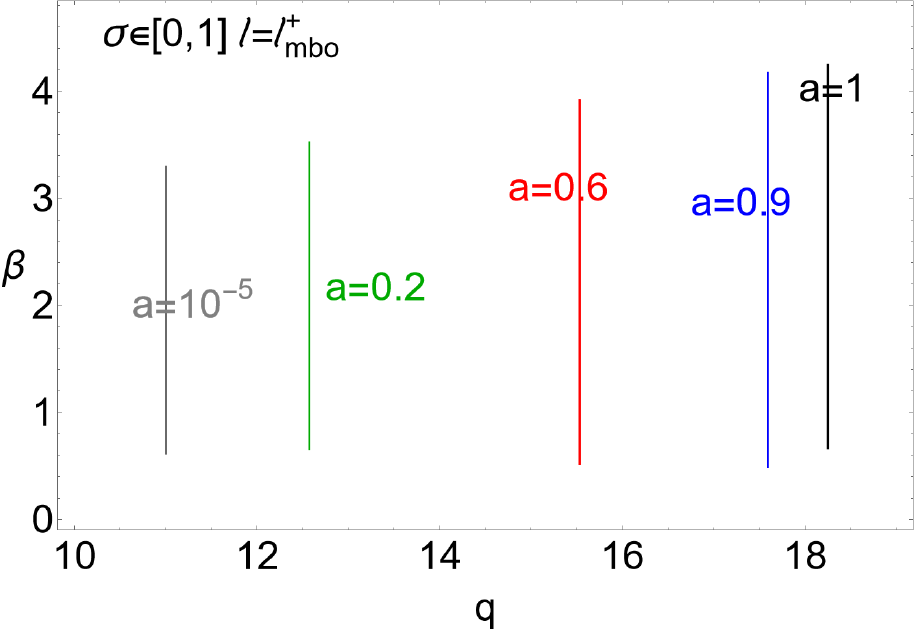}
\includegraphics[width=4.25cm]{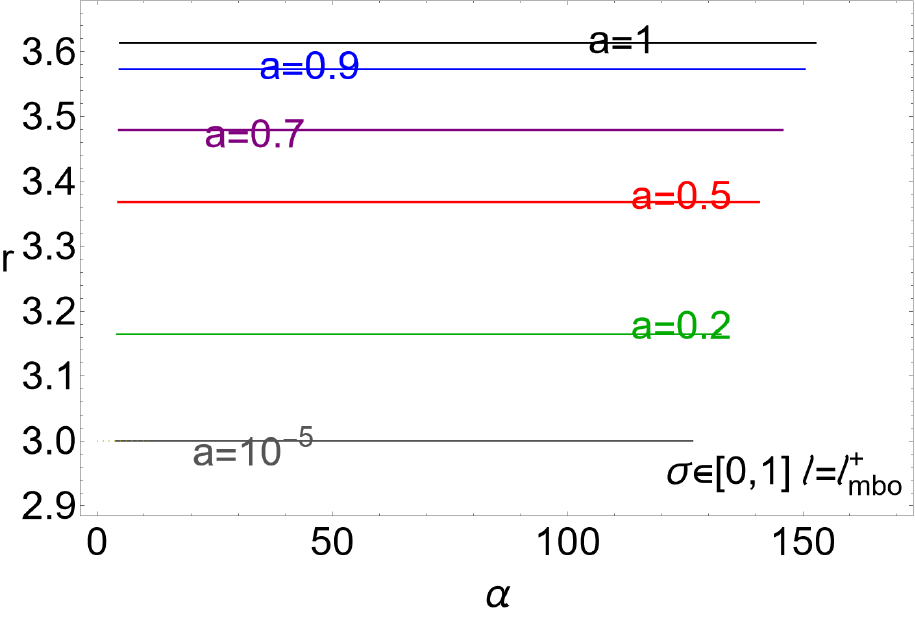}
\includegraphics[width=4.25cm]{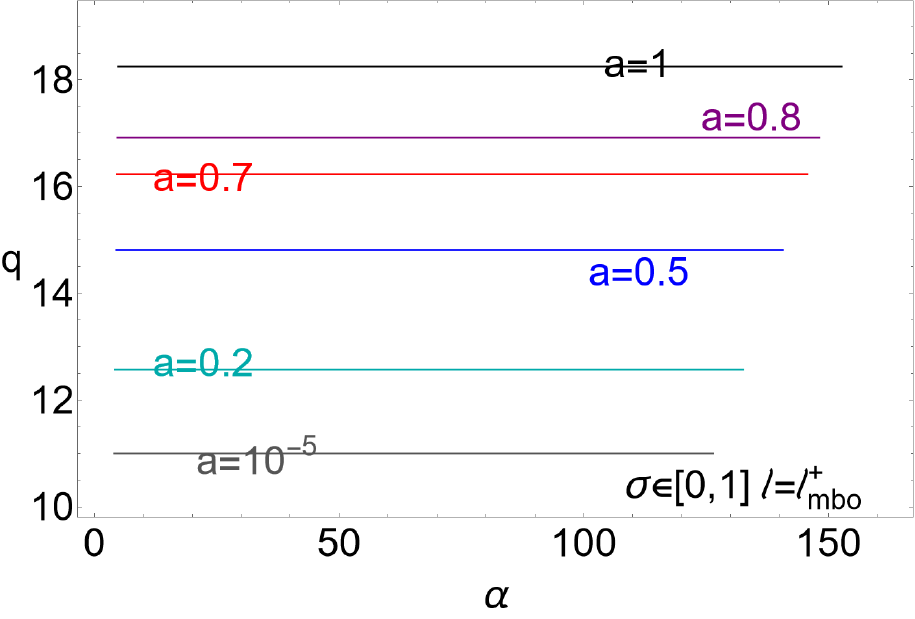}
\caption{Case  $\ell=\ell_{mbo}^+$.
(For further details see Figs\il(\ref{Fig:Plotbetrodon1lmsom}).)} \label{Fig:Plotfondoaglrsigmapb}
\end{figure}
\begin{figure}
\centering
\includegraphics[width=5.6cm]{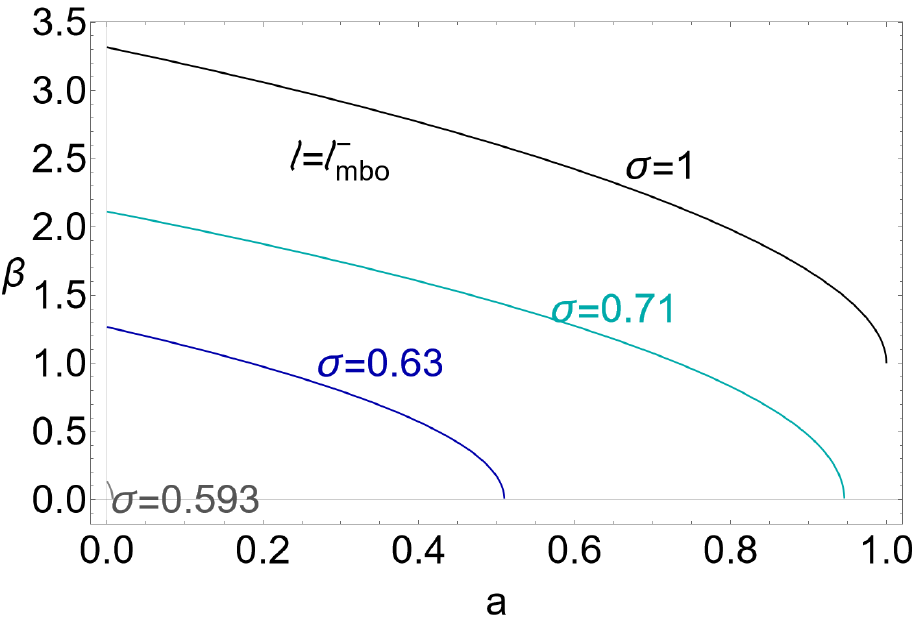}
\includegraphics[width=5.6cm]{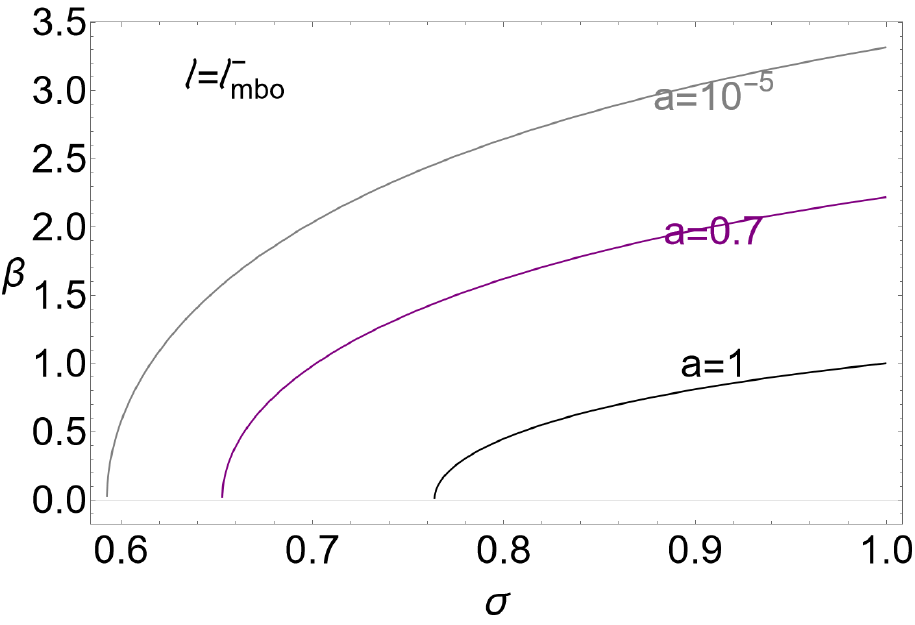}
\includegraphics[width=5.6cm]{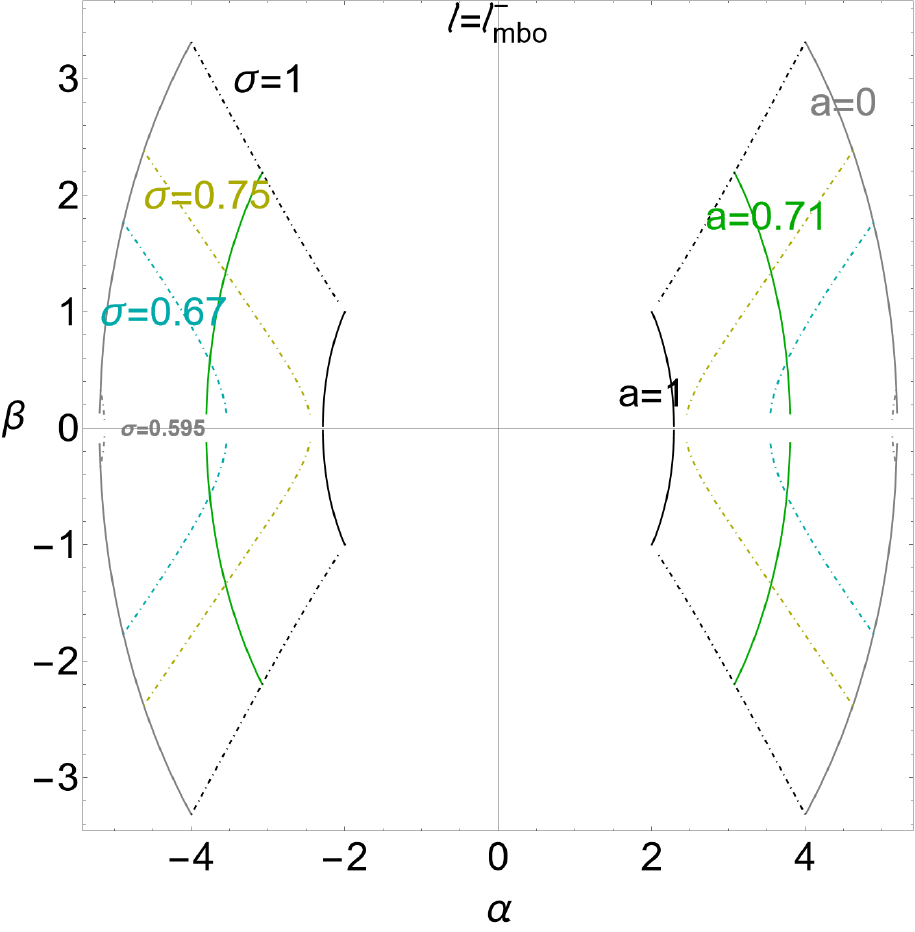}
\includegraphics[width=4.25cm]{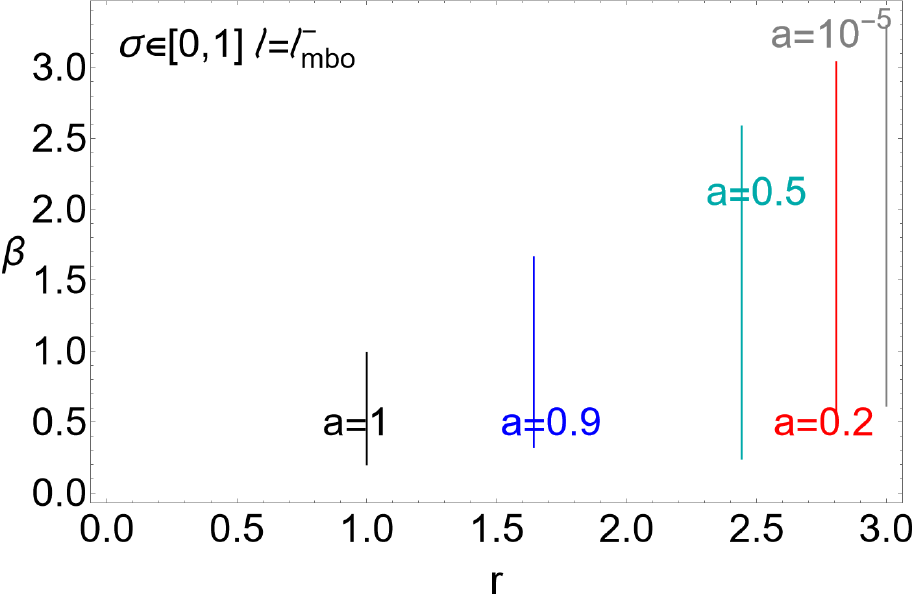}
\includegraphics[width=4.25cm]{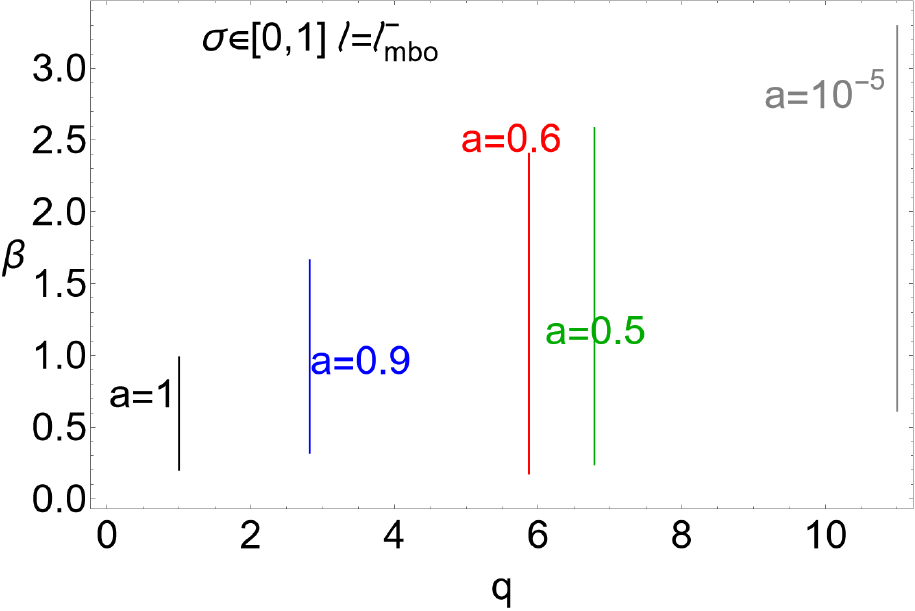}
\includegraphics[width=4.25cm]{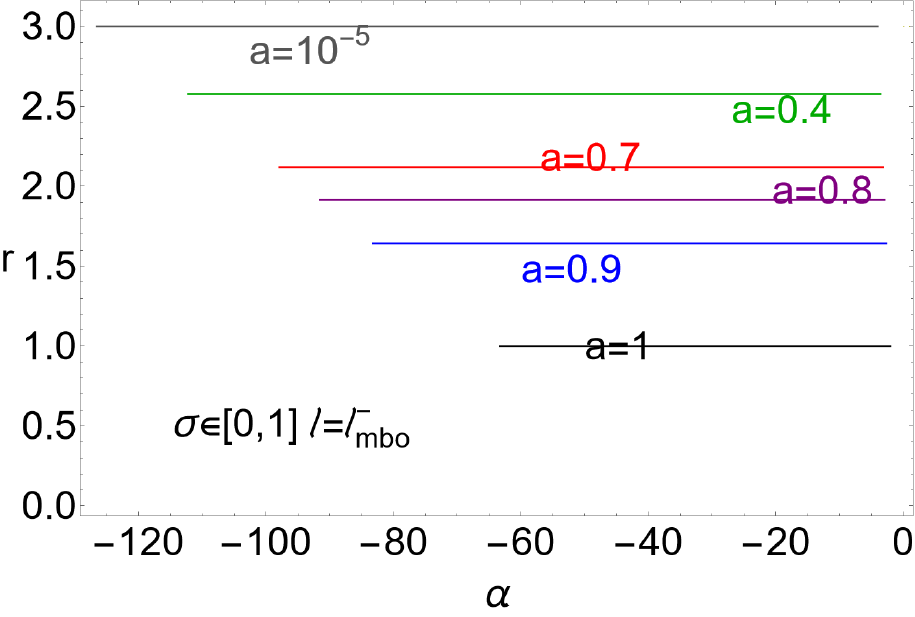}
\includegraphics[width=4.25cm]{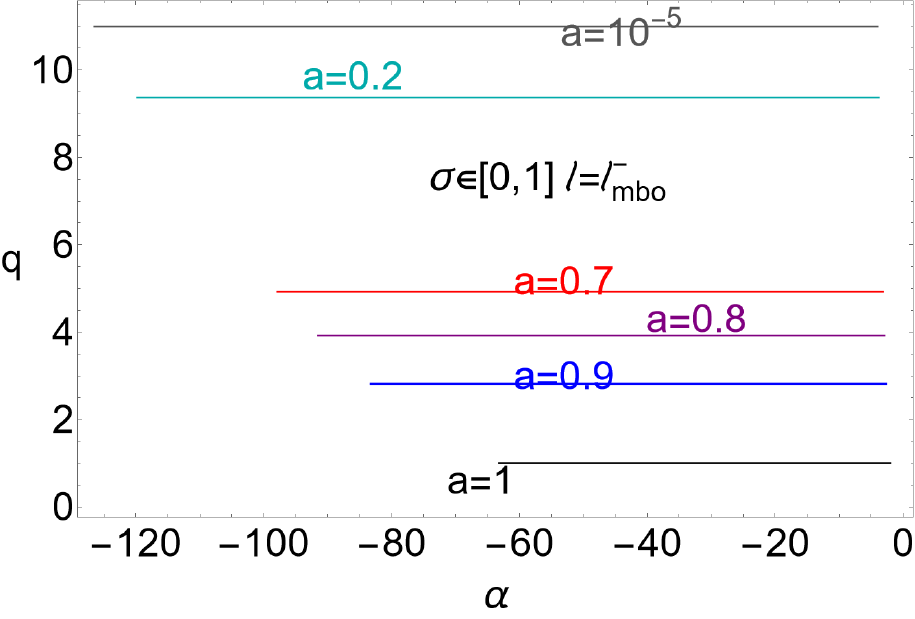}
\caption{Case $\ell=\ell_{mbo}^-$. (For further details see Figs\il(\ref{Fig:Plotbetrodon1lmsom})).}\label{Fig:Plotbetrodon1lmsopbm}
\end{figure}

\medskip

\textbf{{Parameter $\ell=\ell_{mbo}^+$}}

\medskip

Figs\il(\ref{Fig:Plotfondoaglrsigmapb}) present  the results for $\ell=\ell_{mbo}^+$. Coordinate $|\beta|$  increases with the angle $\sigma>0.55$ and the \textbf{BH} spin $a$.  For $\sigma\gg 0.58$ there is $|\beta|>0$.
Smaller values of $(\alpha,\beta)$ (in magnitude), coincident with the  inner regions of the   $\alpha-\beta$ plane, identify the  slowly  spinning \textbf{BHs}.
$|\beta|$ also increases with $(a,r, q)$, according to  Figs\il(\ref{Fig:Plotfondoaglrsigmapb}). Quantities  $r$ and $q$ increase with $a$.

\medskip

\textbf{Parameter $\ell=\ell_{mbo}^-$}

\medskip

The case $\ell=\ell_{mbo}^-$  is in Figs\il(\ref{Fig:Plotbetrodon1lmsopbm}). Celestial coordinate  $|\beta|$ decreases with spin and increases with $\sigma>0.593$. These curves    for fast spinning \textbf{BHs} can be observed in the inner regions of the $\alpha-\beta$ plane.  The coordinate $|\alpha|$, the  range of $\alpha$ values   and $(r,q)$ decrease (in magnitude) with the  spin.

\medskip

{Parameter $\ell=\ell_{\gamma}^\pm$}

\medskip

{For  $\ell=\ell_{\gamma}^\pm $  there are the  photon circular orbits $r_{\gamma}^\pm$,  on the equatorial plane, where $ q=0$ and $\sigma=1$.}

\begin{figure}
\centering
\includegraphics[width=6cm]{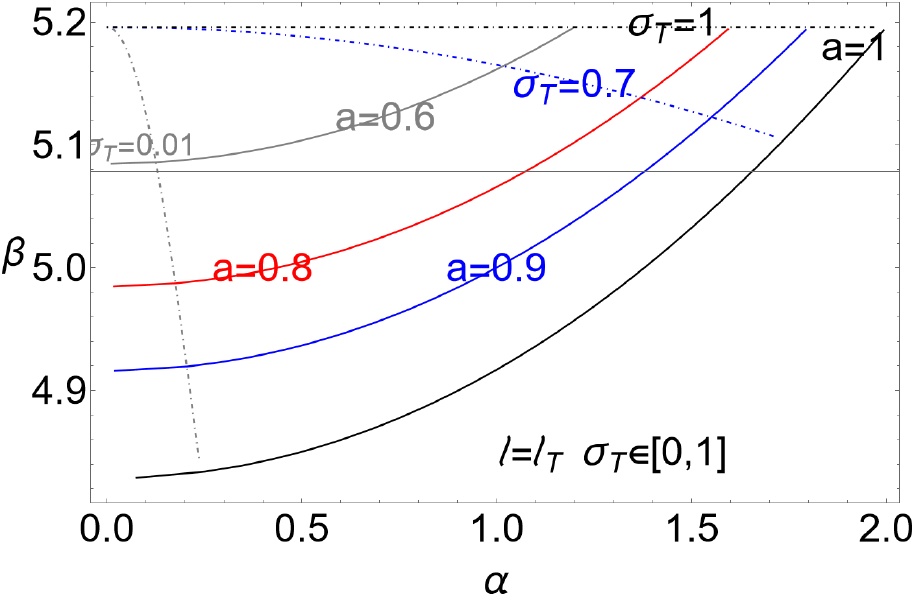}
\includegraphics[width=6cm]{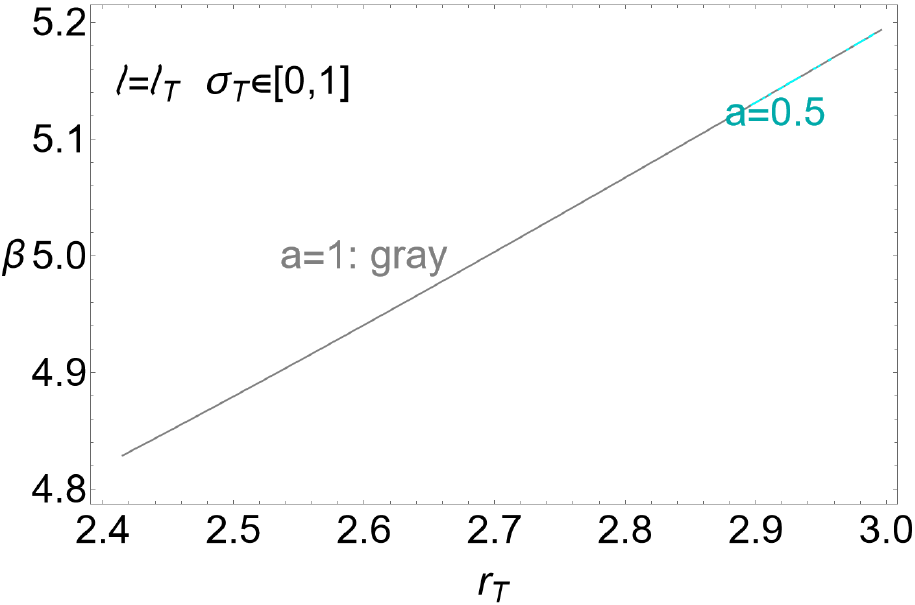}
\includegraphics[width=6cm]{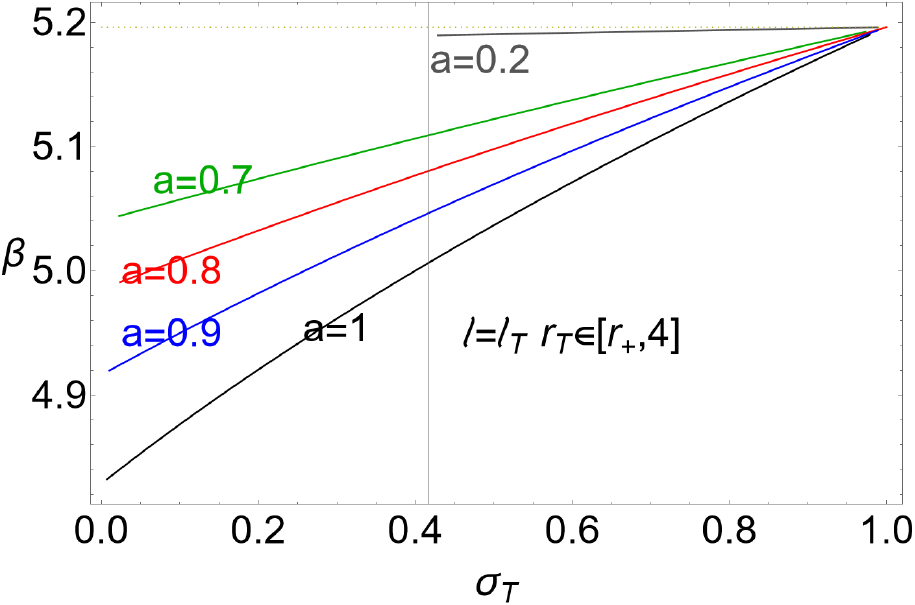}
\includegraphics[width=6cm]{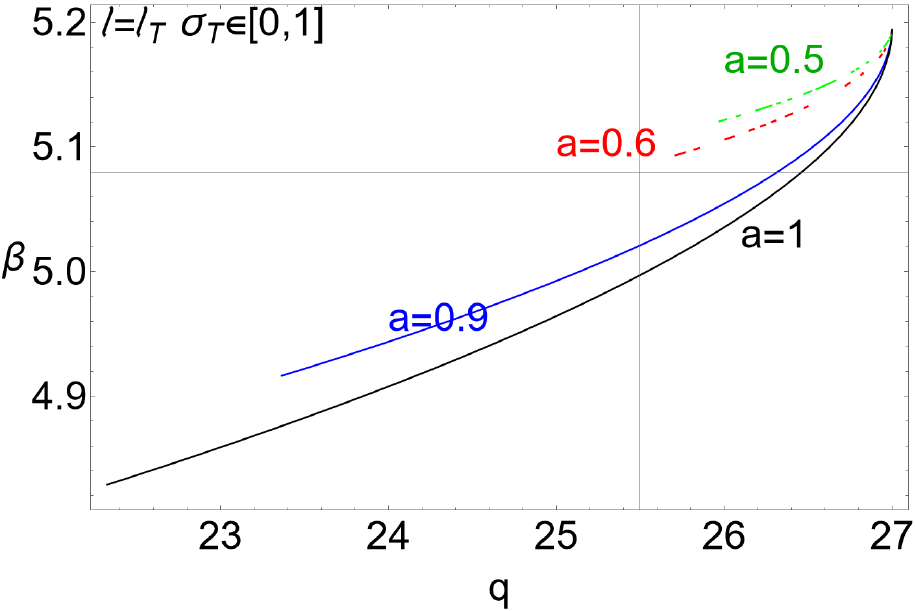}
\caption{Analysis for $\ell=\ell_\Ta$. Upper  panels: each point of the solid curve is for a different angle $\sigma_\Ta$.   Each point of a dotted-dashed  curve is  for a different  $a$. As $\ell_\Ta<0$, there is $\alpha>0$,   the analysis  is for  $(\beta>0,\alpha>0)$.
Bottom left panel: each point of a curve is for a different radius $r_\Ta$. Bottom right panel: each point of each curve is for a different angle $\sigma_\Ta$.}\label{Fig:Plotfarengmbeta}
\end{figure}
\begin{figure}
\centering
\includegraphics[width=5.6cm]{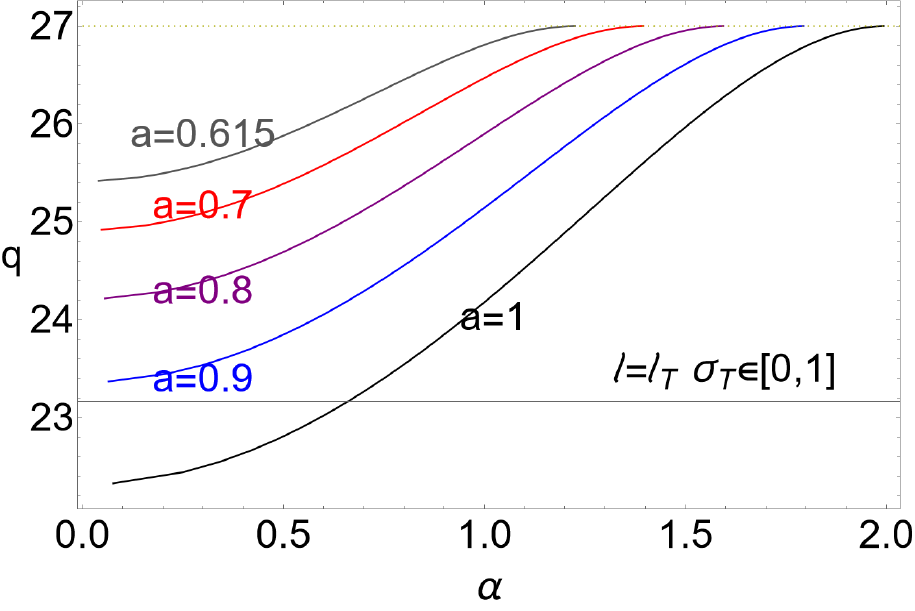}
\includegraphics[width=5.6cm]{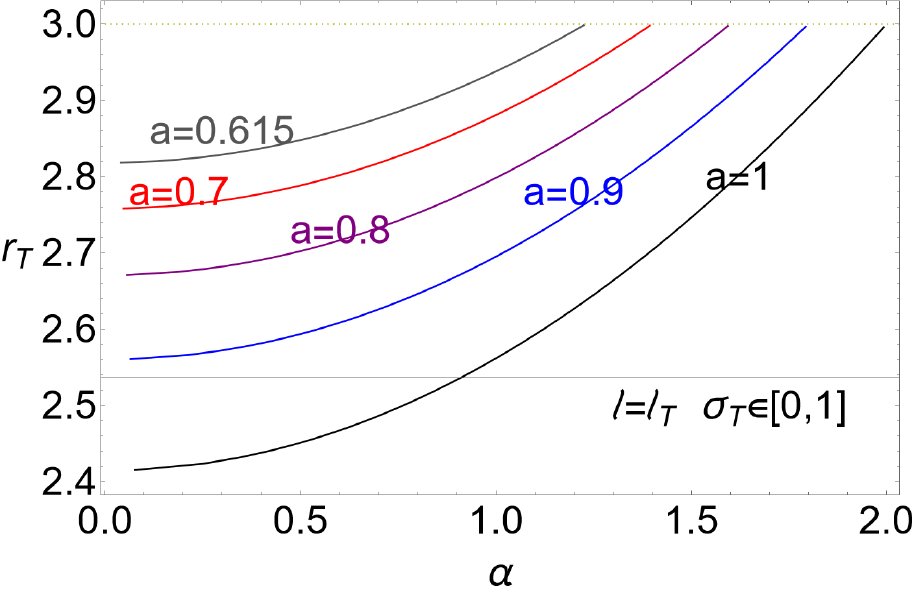}
\includegraphics[width=5.6cm]{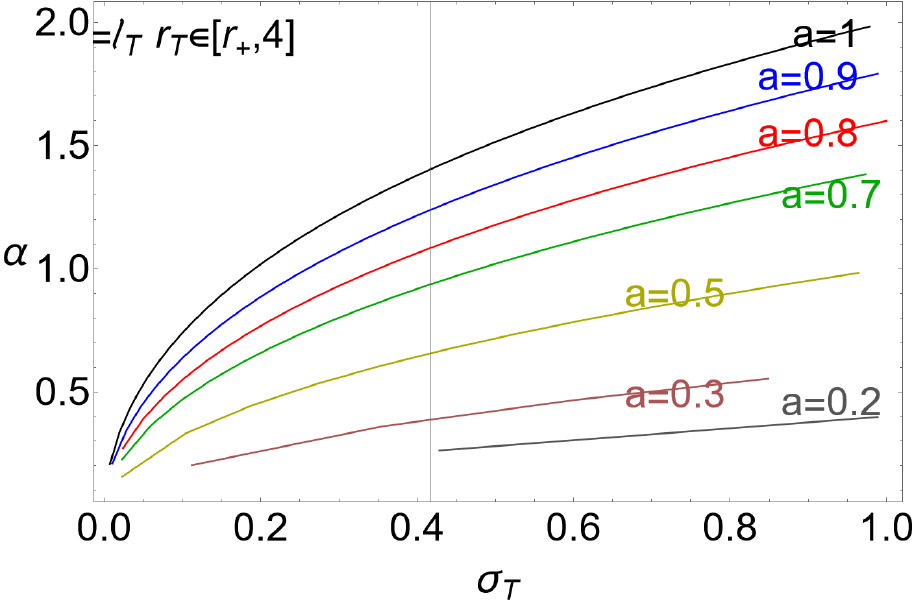}
\caption{Analysis for $\ell=\ell_\Ta$. Left and center panels: each point of a curve is for a different   $\sigma_\Ta$.
Right panel: each point of a curve is for a different radius  $r_\Ta$.}.\label{Fig:Plotfarengmx}
\end{figure}
\begin{figure}
\centering
\includegraphics[width=5.6cm]{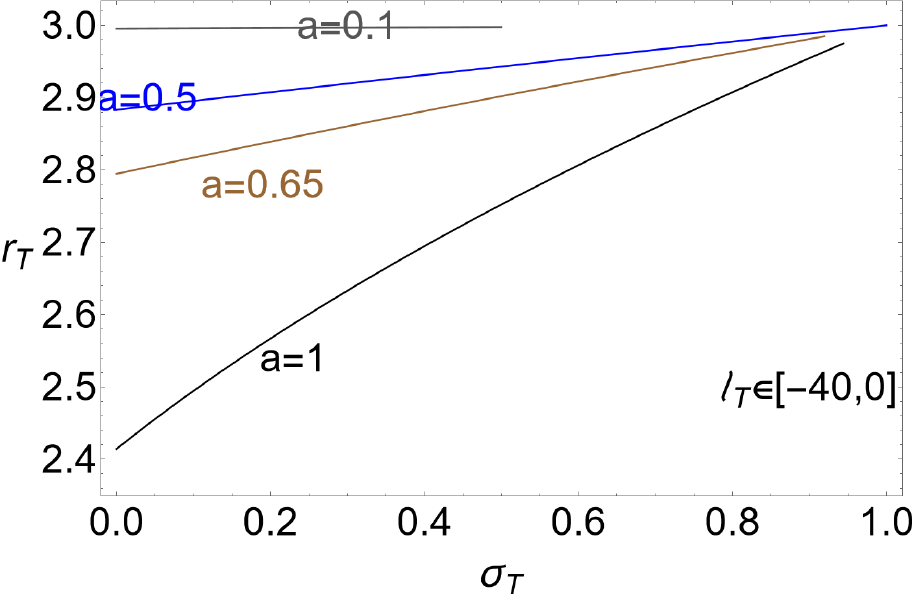}
\includegraphics[width=5.6cm]{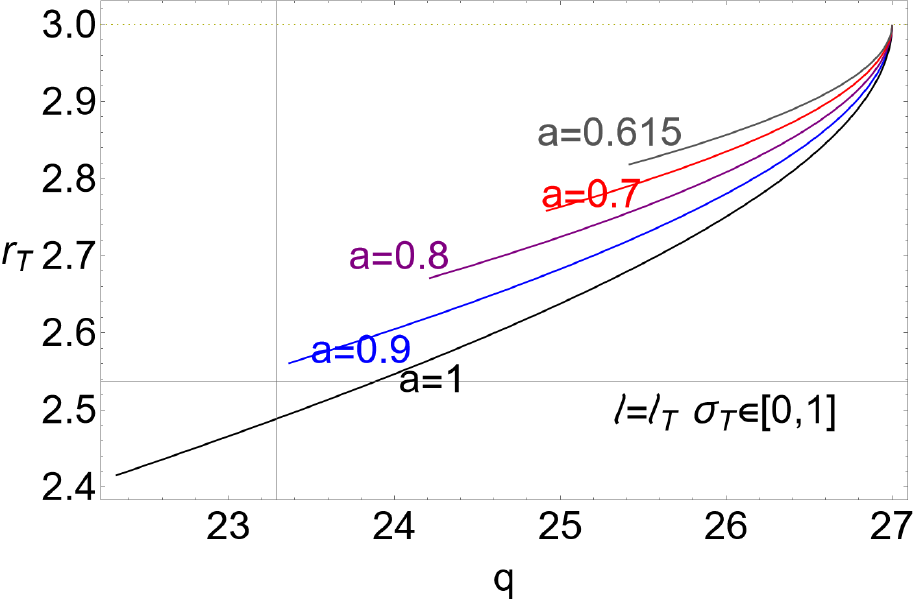}
\includegraphics[width=5.6cm]{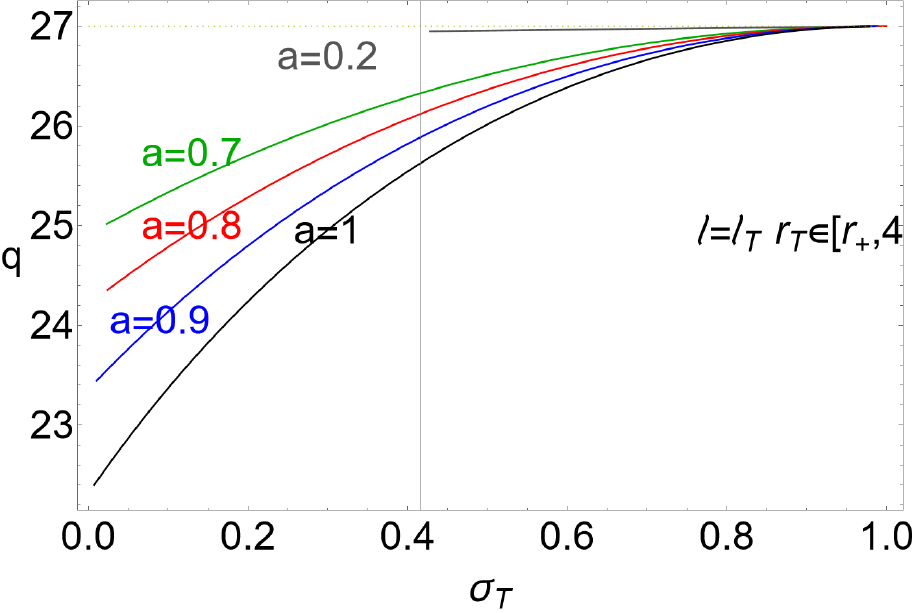}
\caption{Analysis for $\ell=\ell_\Ta$. Left panel:
each point of a curve is for a different   $\ell_\Ta$. Center panel: each point of a curve is a different  $\sigma_\Ta$. Right panel: each point of a curve is a different  radius  $r_\Ta$.  {For further details see also caption of Figs\il(\ref{Fig:Plotfarengmx}).}}.\label{Fig:Plotfarengmy}
\end{figure}
\begin{figure}
\centering
\includegraphics[width=5.6cm]{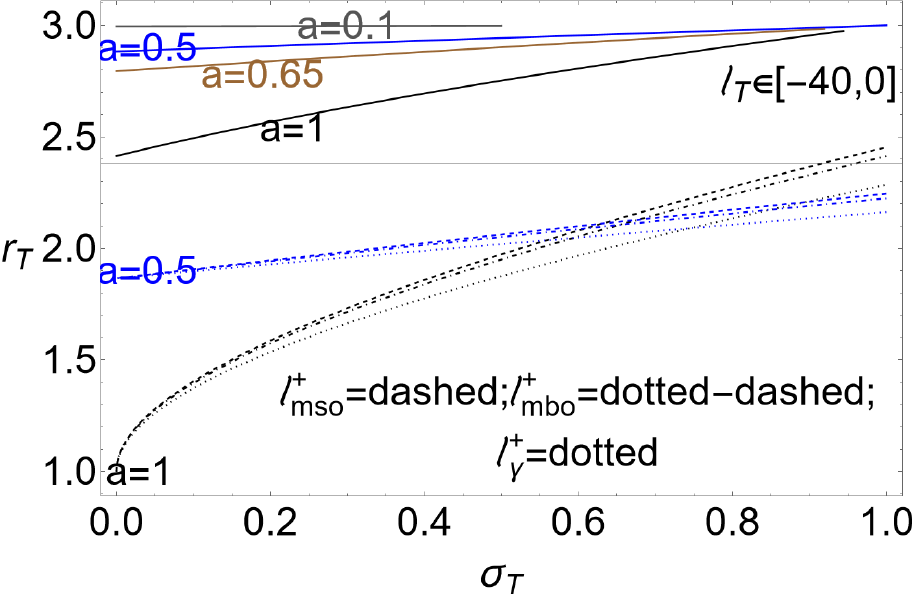}
\includegraphics[width=5.6cm]{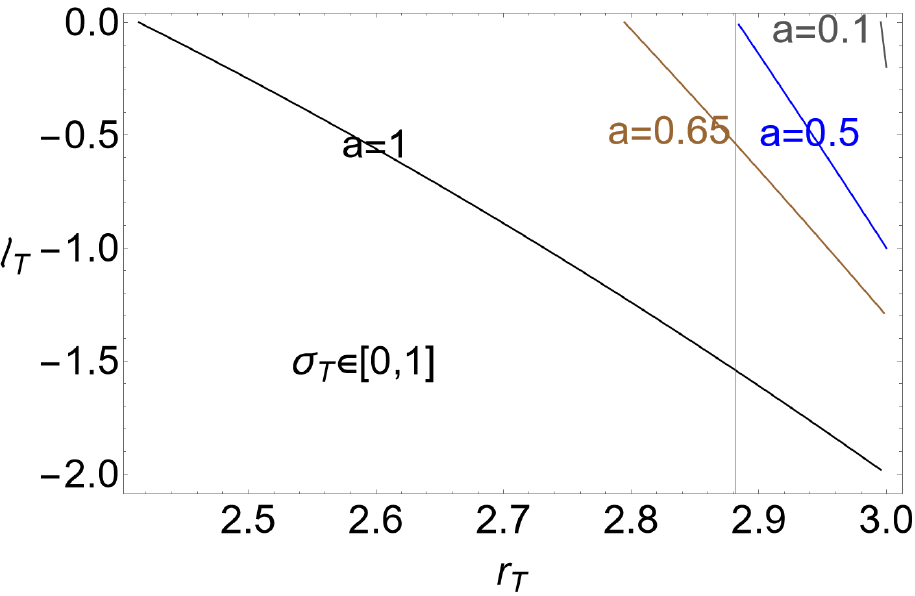}
\includegraphics[width=5.6cm]{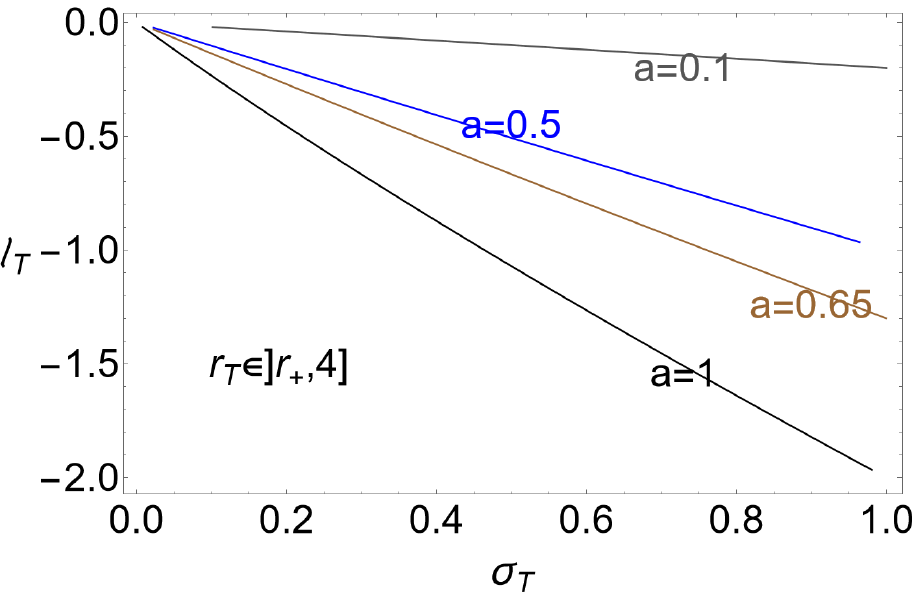}
\caption{Analysis for $\ell=\ell_\Ta$ Left panel:    each point  is for a different   $\ell_\Ta$.
  Middle panel: each point of a curve is for a different angle  $\sigma_\Ta$.  Right panel: each point of a curve is for a different radius  $r_\Ta$.  {For further details see also caption of Figs\il(\ref{Fig:Plotfarengmx}).}}.\label{Fig:Plotfarengmz}
\end{figure}
\begin{figure}
\centering
\includegraphics[width=8cm]{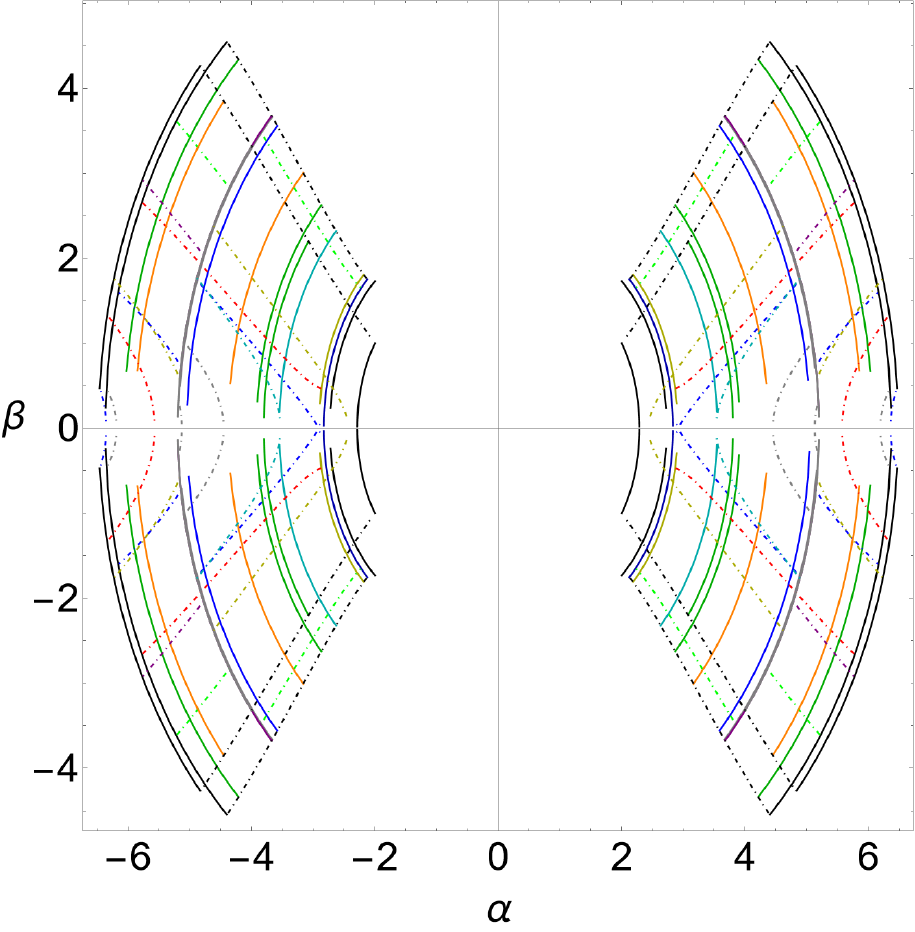}
\includegraphics[width=8cm]{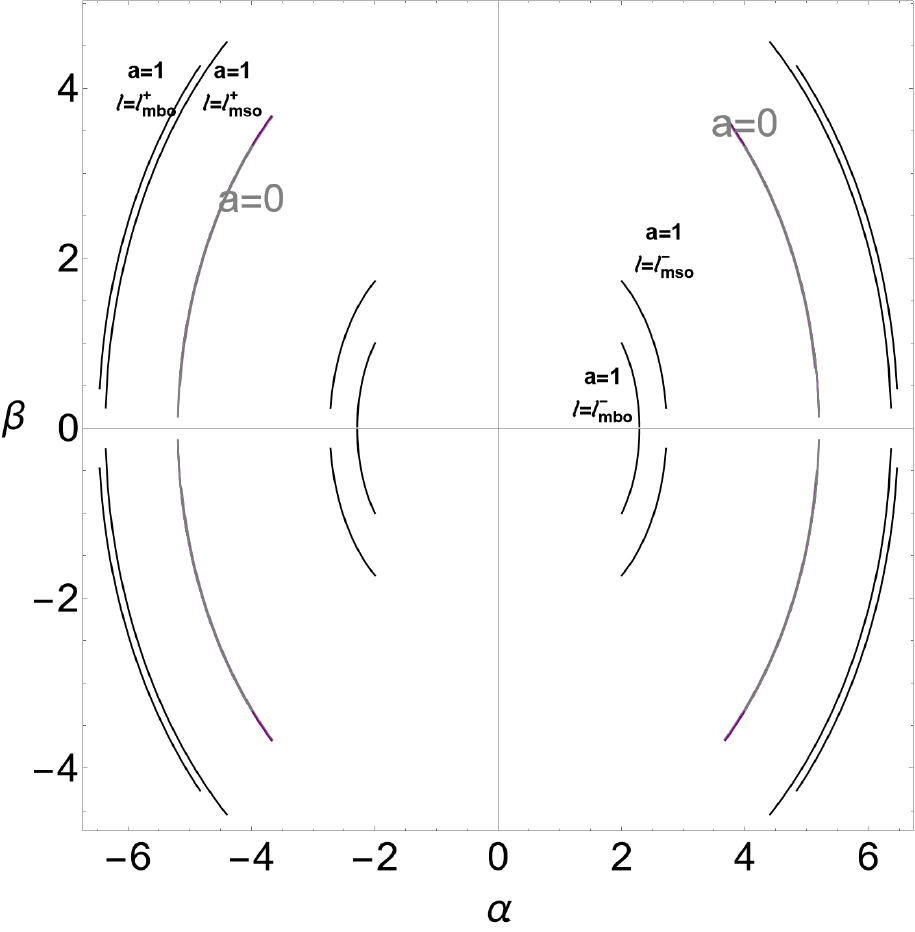}
\caption{Cases $\ell=\{\ell_{mbo}^\pm,\ell_{mso}^\pm\}$.  Left panel: merging of Figs\il(\ref{Fig:Plotbetrodon1lmsopbm},\ref{Fig:Plotfondoaglrsigmapb},\ref{Fig:Plotverinimsop},\ref{Fig:Plotbetrodon1lmsom})--right panels.
Dotted-dashed lines are  for spin $a\in[0,1]$ and for different angles $\sigma$. (Each point of a fixed curve is for a different spin $a$). Solid lines  are for  $\sigma\in[0,1]$ and for different spin. (Each point of each curve is for a different angle $\sigma$). Right panel: each point of each curve is for a different {angle} $\sigma$).}.\label{Fig:Plotvimsounita}
\end{figure}
%

%

\medskip

\textbf{Parameter $\ell=\ell_\Ta$}

\medskip

The case of photons from the inversion surfaces, analyzed assuming  $\ell=\ell_\Ta$,  is shown in Figs\il(\ref{Fig:Plotfarengmbeta}) where $\beta(\alpha)$ for this impact parameter is shown at different $\sigma$ and spin on the inversion surface, and the celestial coordinate $\beta$ is then related to $(r_\Ta, \sigma_\Ta)$. Figs\il(\ref{Fig:Plotfarengmx})   relate $\alpha$ to $(q,r_\Ta,\sigma_\Ta)$, while $(r_\Ta,\sigma_\Ta,q,\ell)$ are analyzed in  Figs\il(\ref{Fig:Plotfarengmy},\ref{Fig:Plotfarengmz}). Within this constraint
we consider $q\geq 0$,
 and  $\dot{\phi}=0$
for the inversion points  i.e.
\bea
\mathcal{F}_\Ta\equiv a^2 (\sigma -1) \ell -2 a r \sigma -(r-2) r \ell=0.
\eea
%
There is  $\alpha>0$ since $\ell_\Ta<0$, and in Figs\il(\ref{Fig:Plotfarengmbeta}) we restricted the analysis  to  $(\beta>0,\alpha>0)$.
Note that for $\ell=\ell_\Ta$  function $\beta(\alpha)$  is strongly distinct from the case $\ell\in\{\ell_{mso}^\pm,\ell_{mbo}^\pm\}$--see also  Figs\il(\ref{Fig:Plotscadenztoaglrl}).

 For fixed $\alpha$, where  $ 4.8\lessapprox |\beta| \lessapprox5.2$, the celestial coordinate $\beta(\alpha)$  decreases (in magnitude) with the \textbf{BH} spin.
  $|\beta|$ increases linearly with $r_\Ta>2.5$, decreases with $a$ for fixed $(\alpha,\sigma_\Ta,q)$, and it  increases with $\sigma_\Ta$ and $q$ (smaller values of  $q$ characterize fast spinning \textbf{BHs}).
Considering Figs\il(\ref{Fig:Plotfarengmx},\ref{Fig:Plotfarengmy},\ref{Fig:Plotfarengmz})
for  $\ell=\ell_\Ta$,  quantity $q$, where   $22\lessapprox q\lessapprox  27$,  increases with $|\alpha|<2$.
 The coordinate $\beta$   is maximum   for $|\beta|\approx 5.196$ where $(r_\Ta=3, q=27)$ for $a=1 $ and  $\sigma_\Ta=1$.
For $ \sigma_\Ta\approx 0$ there is $(r_\Ta=2.41421, |\beta|=4.82843)$, there is $q= 22.31$ for
$\sigma_\Ta\approx 0$ and $a=1$.

Celestial coordinate $|\alpha|$ increases with $q$. Then, $|\alpha|$ (and $\alpha$ range)  increases with  $a$ and $\sigma_\Ta$.
 Quantity $q$  decreases in general  with $a$ and increases with $\sigma_\Ta$.
Radius $r$ increases with $(|\alpha|,\sigma_\Ta,q)$.   However $r_\Ta$ decreases with $a$ only for the limiting value  $\sigma_\Ta\lessapprox 0.63$ (related to the inversion surfaces maximum \citep{retro-inversion}).

\medskip

\textbf{Final remarks on the cases $\ell=\{\ell_{mbo}^\pm,\ell_{mso}^\pm,\ell_\Ta\}$}

\medskip

 Figs\il(\ref{Fig:Plotvimsounita}), merging  of Figs\il(\ref{Fig:Plotbetrodon1lmsom},\ref{Fig:Plotverinimsop},\ref{Fig:Plotfondoaglrsigmapb},\ref{Fig:Plotbetrodon1lmsopbm})--upper right panels, show  cases $\ell=\{\ell_{mbo}^\pm,\ell_{mso}^\pm\}$.
The shadow boundary regions correspondent to the  inversion surfaces are completely distinct from the   cases  correspondent to $\ell=\{\ell_{mbo}^\pm,\ell_{mso}^\pm\}$.
Furthermore, despite  there is $\ell_\Ta<0$, the shadows parts  in the fast spinning \textbf{BHs} correspond to smaller values of $|\beta|$ (and larger range of $|\alpha|$). 

%
%
Comparing  profiles of Figs\il(\ref{Fig:Plotbetrodon1lmsom},\ref{Fig:Plotverinimsop},\ref{Fig:Plotfondoaglrsigmapb},\ref{Fig:Plotbetrodon1lmsopbm})--upper right panels and Figs\il(\ref{Fig:Plotfarengmbeta}, \ref{Fig:Plotvimsounita}),  it can be seen  that in general, orbits for $\ell=
\ell_\Ta$ are located approximately at the central region of the shadow boundary, $\ell_{mbo}^\pm$ is more external (larger values of $\alpha$ in magnitude), $\ell_{mso}^\pm$ is more internal.

Analysis of  $\sigma_\lambda$ in Figs\il(\ref{Fig:Plotsigmaglr})  shows that, for $\ell=\ell_{mso}^-$, the angle  $\sigma_\lambda$  increases with the \textbf{BH} spin $a$ to  the maximum value  at $a=a_\sigma$. For $\ell=\{\ell_{mso}^+,\ell_{mbo}^+\}$, the angle  $\sigma_\lambda$   decreases with  $a$ and, for  $\ell=\ell_{mbo}^-$,  increases with $a$. There is
\bea
\sigma_\lambda(\ell_{mbo}^-)>\sigma_\lambda(\ell_{mbo}^+)>\sigma_\lambda(\ell_{mso}^-)>\sigma_\lambda(\ell_{mso}^+),
 \eea
 and  in general $\sigma_\lambda\gtrapprox0.479$.

From Figs\il(\ref{Fig:Plotfondoaglrsigmapbcasdu})  it is noted  that for $\sigma=1$  coordinate  $\alpha$ decreases (in magnitude) with  the  \textbf{BH} spin for the co-rotating cases and increases with spin for the counter-rotating case;  there is
\bea
-\alpha_\lambda(\ell_{mso}^-)>-\alpha_\lambda(\ell_{mbo}^-)>-\alpha_\lambda(\ell_{\gamma}^-)\quad\mbox{and}\quad
\alpha_\lambda(\ell_{\gamma}^+)>\alpha_\lambda(\ell_{mbo}^+)>\alpha_\lambda(\ell_{mso}^+).
\eea
On the other hand, $q_\lambda$ increases with $a$ for the counter-rotating cases and decreases with $a$ in the co-rotating cases;  there is
\bea&&\nonumber
\mbox{for}\quad a>a_q:\quad
q_\lambda(\ell_{mso}^+)>q_\lambda(\ell_{mbo}^+)>q_\lambda(\ell_{mso}^-)>q_\lambda(\ell_{mbo}^-),\quad\mbox{where}\quad a_q\equiv 0.153054,
\\\label{Eq:aq-definition}
&&\mbox{for}\quad
a\in [0,a_q[:\quad
q_\lambda(\ell_{mso}^+)>q_\lambda(\ell_{mso}^-)>q_\lambda(\ell_{mbo}^+)>q_\lambda(\ell_{mbo}^-),
\eea
see Figs\il(\ref{Fig:Plotfondoaglrsigmapbcasdu}).
From Figs\il(\ref{Fig:Plotscadenztoaglrl}) (and  Figs\il(\ref{Fig:Plotfondoaglrsigmapb})) it is clear that $r_\lambda$ decreases with the \textbf{BH}  spin for the co-rotating cases and it  increases with the \textbf{BH} spin for the counter-rotating case. For $\ell\in\{\ell_{mbo}^+,\ell_{mso}^+\}$ there is $r_\lambda\in [r_{\gamma}^+,r_\Ta(\sigma=1)]$.
However  for $\ell=\ell_{mso}^-$ there is:
\bea
\mbox{for}\quad a<a_{\lambda}^-:\quad r_\lambda\in [r_{\gamma}^-,r_{mbo}^-], \quad\mbox{and for}\quad a>a_{\lambda}^-:\quad r_\lambda\in [r_{mbo}^-,r_{mso}^-].
\eea
From Figs\il(\ref{Fig:Plotvimsounita}), showing the superimposition of the  Figs\il(\ref{Fig:Plotbetrodon1lmsopbm},\ref{Fig:Plotfondoaglrsigmapb},\ref{Fig:Plotverinimsop},\ref{Fig:Plotbetrodon1lmsom})--right panels, it is possible to note that  the  \textbf{BH} spin is capable to distinguish the cases  $\{\ell_{mbo}^\pm,\ell_{mso}^{\pm}\}$.
For the static case ($a=0$) the shadow boundary parts  are very close and practically indistinguishable. The static case  separates the co-rotating case (spread on the more internal and  smaller region of the boundary--smaller values of $(|\alpha|,|\beta|)$) and the counter-rotating case (on a more external --larger values of $(|\alpha|,|\beta|)$) for different angles  $\sigma$.

For an  extreme Kerr \textbf{BH} there is
\bea
 C_\beta(\ell_{mbo}^+)>C_\beta(\ell_{mso}^+)>C_\beta(\ell_{mso}^-)>C_\beta(\ell_{mso}^-),
 \eea
 where $C_\beta(\ell_{u})>C_\beta(\ell_{v})$ indicates that the curve  $C_\beta(\ell_{u})$  for a momentum $\ell_u$  is more external (on the  shadow  boundary as in the  Figs\il(\ref{Fig:Plotvimsounita}) than the curve  $C_\beta(\ell_{v})$  for a momentum $\ell_u$, for different angles  $\sigma$.
Each curve has been evaluated for  $\sigma\in [0,1]$. Celestial coordinate $|\beta|$ increases generally for fixed $\alpha$ with the \textbf{BH} spin for the counter-rotating case and decreases with the  \textbf{BH} spin for co-rotating case; $\alpha$ in magnitude increases for fixed $\beta$ with the  \textbf{BH} spin and the magnitude of $\ell^+$ for the counter-rotating case, viceversa, $\alpha$ in magnitude decreases for fixed $\beta$ with the  \textbf{BH} spin and the $\ell^-$ for the co-rotating case. Whereby the co-rotating and counter-rotating cases are sharply distinct and distinguishable for each angle.

Note, from  Figs\il(\ref{Fig:Plotscadenztoaglrl})
it is clear how in the co-rotating case $
r_\lambda$ can cross the outer ergosurface for fast spinning   \textbf{BHs} (depending on  the angle  $\sigma$). We will focus on this  special case in Sec.\il(\ref{Sec:from-ergo}).

In general, considering $r_\lambda$ as function of $a$, there is   $r_\lambda^+>r_\lambda^-$ for the counter-rotating and co-rotating cases. In the co-rotating case the radius $r_\lambda^-$  decreases with the  \textbf{BH} spin and   there is  $r\approx 3$ at $a\approx 0$. Viceversa  $r_\lambda^+$  increases  with the  \textbf{BH} spin from  $r\approx 3$ at $a\approx 0$ to $r\approx 4$ for $a\approx 1$.
There is $r_\lambda^+(\ell_*^+)>r_\Ta$ for $\ell_*^+\in\{\ell_{mbo}^+,\ell_{mso}^+\}$  (see also Figs\il(\ref{Fig:Plotscadenztoaglrl}))  for any angle $\sigma$. Therefore photons from these inversion surfaces cannot be observed from the \textbf{BH} shadow boundary. However, as   proved in this analysis, photons with  $ -\ell^+=-\ell_{mso}^+$ can be part of the \textbf{BH}  shadow boundary.
Furthermore, there is
\bea
 r_\lambda^+(\ell_{\gamma}^+)>r_\lambda^+(\ell_{mbo}^+)>r_\lambda^+(\ell_{mso}^+)>r_\lambda^-(\ell_{mso}^-)>
 r_\lambda^-(\ell_{mbo}^-)>r_\lambda^-(\ell_{\gamma}^-).
 \eea

In Sec.\il(\ref{Sec:radiii-shadows})  we analyze  in additional details  shadows from
$r_\lambda\in\{r_{mbo}^-,r_{mso}^-\}$.
 \subsection{Assuming $r\in\{r_{mbo}^-,r_{mso}^-\}$}\label{Sec:radiii-shadows}
{In general,   photons spherical orbits, solutions of   system  $\mathfrak{(R)}$   with    $r_\lambda\in\{r_{mso}^\pm,r_{mbo}^\pm\}$,  are  for an impact parameter  $\ell\neq\{\ell_{mso}^\pm,\ell_{mbo}^\pm\}$ and constant $q=q_\lambda\geq 0$.}
{(The radius $r_\lambda$ for $\ell\in\{\ell_{\gamma}^\pm,\ell_{mbo}^\pm,\ell_{mso}^\pm\}$ is compared in Figs\il(\ref{Fig:Plotscadenztoaglrl})   to the radii of the geodesic structures  $r\in\{r_{\gamma}^\pm,r_{mbo}^\pm,r_{mso}^\pm\}$.)}
In this section we assume  $r_\lambda\in[r_{mbo}^\pm,r_{mso}^\pm]$ and $r_\lambda\in[r_{\gamma}^\pm,r_{mbo}^\pm]$.

We will show  that solutions of  $(\mathfrak{R})$ for  $r_\lambda\in[r_{mbo}^\pm,r_{mso}^\pm]$ and $r_\lambda\in[r_{\gamma}^\pm,r_{mbo}^\pm]$ are strongly differentiated. No solutions have been found  for $r\in\{r_{mbo}^+, r_{mso}^+\}$, and  we focus our  study  for     $r\in\{r_{mbo}^-, r_{mso}^-\}$.
\begin{figure}
\centering
\includegraphics[width=5.6cm]{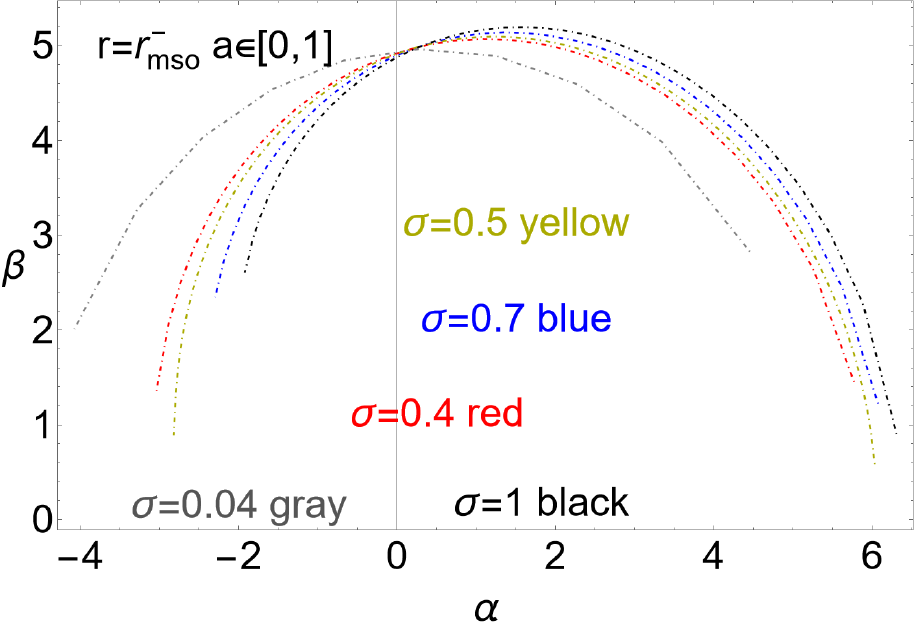}
\includegraphics[width=5.6cm]{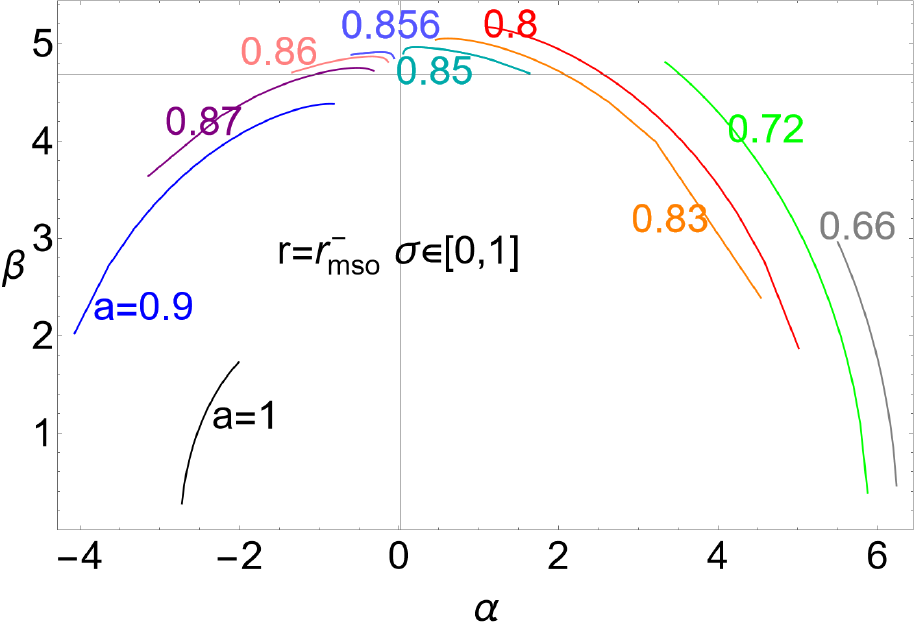}
\includegraphics[width=5.6cm]{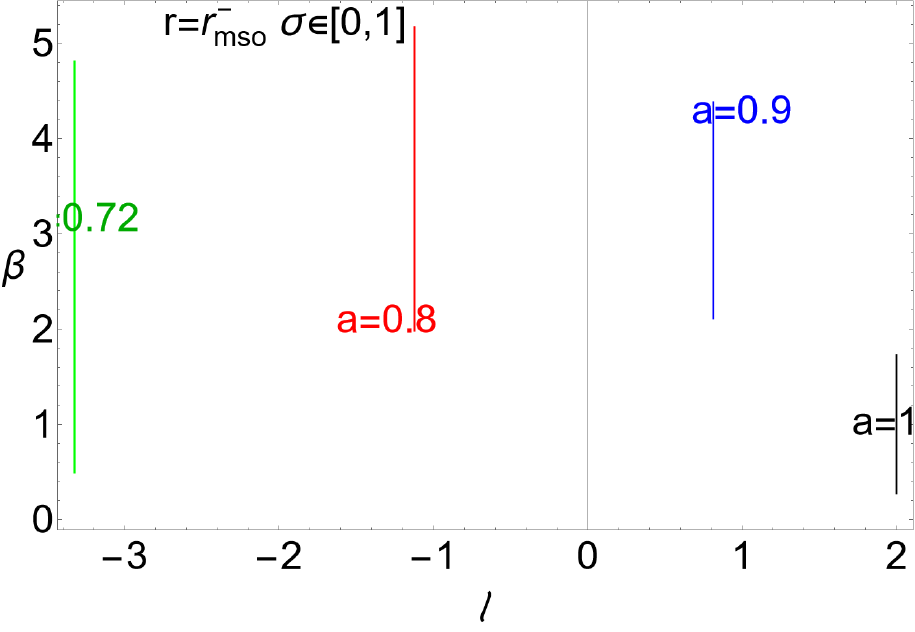}
\includegraphics[width=5.6cm]{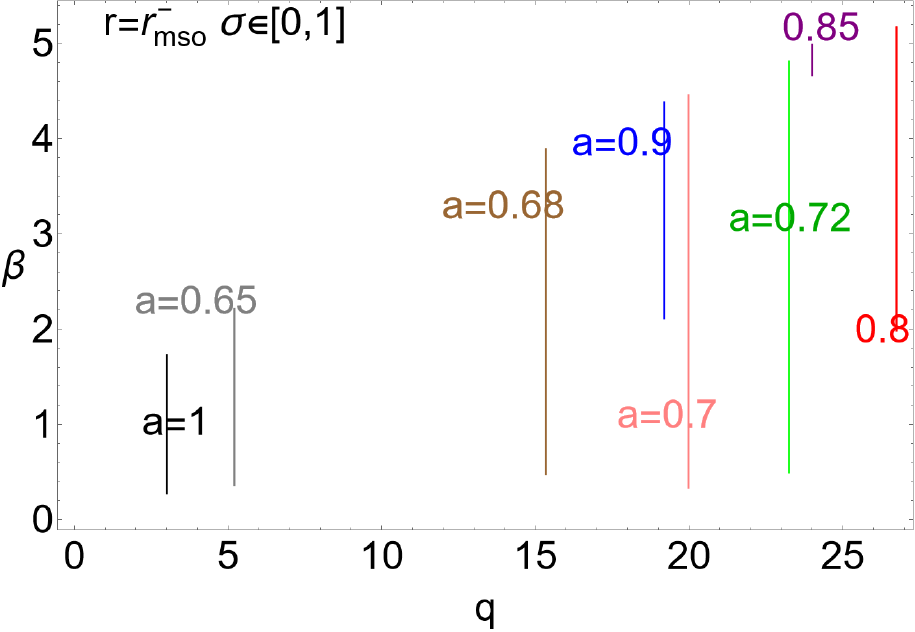}
\includegraphics[width=5.6cm]{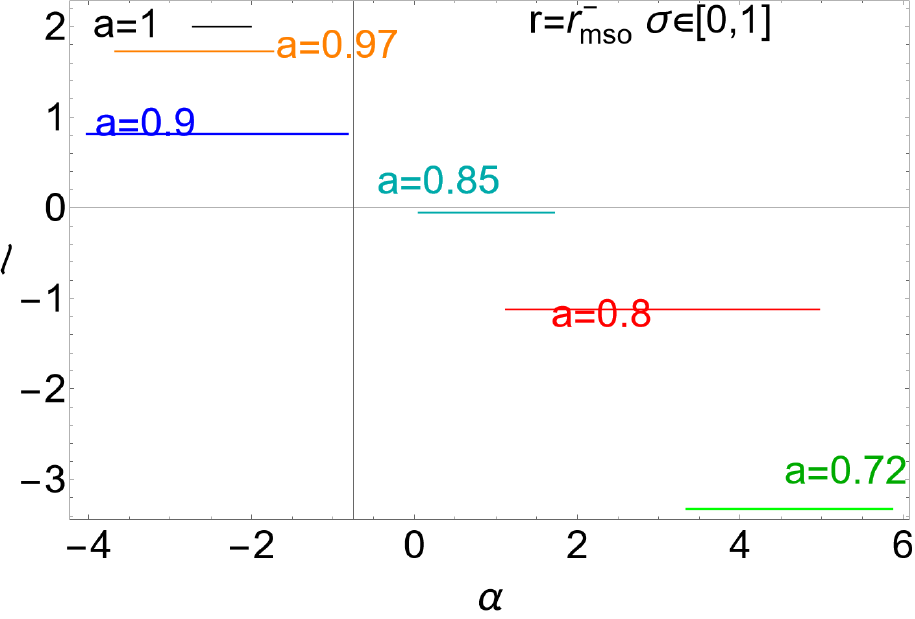}
\includegraphics[width=5.6cm]{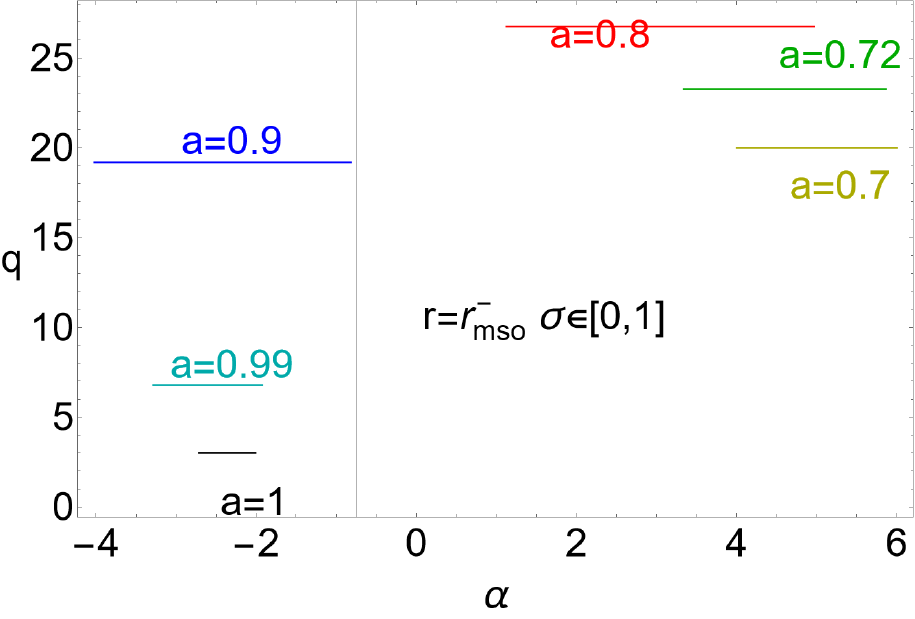}
\caption{Case $r=r_{mso}^-$.  Each point of a curve is for a different   $\sigma$. Upper left panel: each point of a curve is for a different   $a$. }\label{Fig:Plotfarengmrepsrmso}
\end{figure}

\medskip

\textbf{Case $r= r_{mso}^-$}

\medskip

From Figs\il(\ref{Fig:Plotfarengmrepsrmso})
we see the situation for $r= r_{mso}^-$. There are solutions for co-rotating $(\ell>0)$ and counter-rotating $(\ell<0)$ photon orbits. Coordinate
$\alpha$ is larger and positive for the counter-rotating case  for $a\geq 0.66$. Coordinate $|\beta|$ is larger for $a\approx0.85$, and
a minimum occurs for  $a=1$. $|\beta|$ is greater for $\ell$ small in magnitude,  and it increases with $q$.

\medskip

\textbf{Case $r= r_{mbo}^-$}

\medskip

The case
$r=r_{mbo}^-$ is shown in Figs\il(\ref{Fig:Plotfarengm4repslmsomb}) and it is qualitatively  similar to the case $r=r_{mso}^-$.
\begin{figure}
\centering
\includegraphics[width=5.6cm]{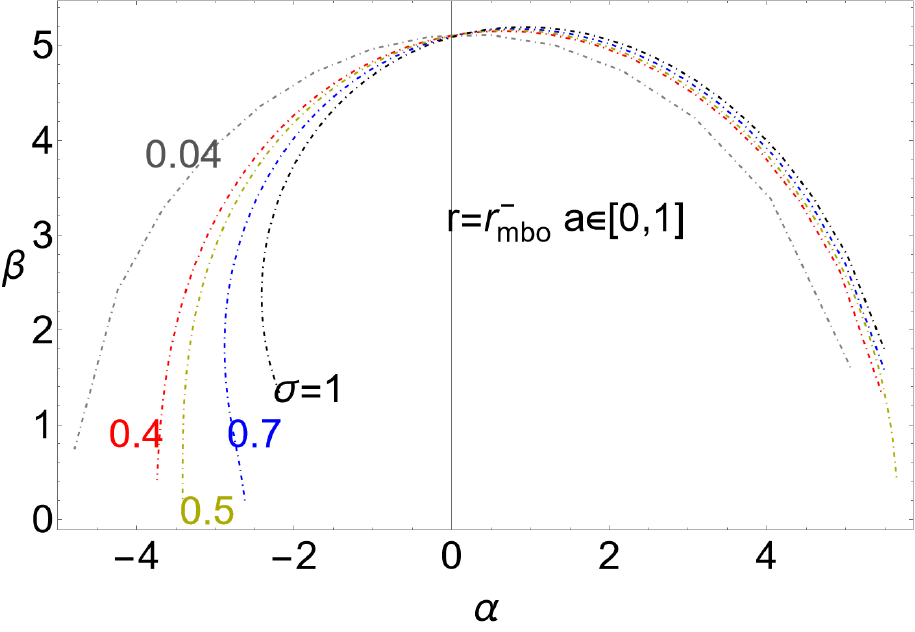}
\includegraphics[width=5.6cm]{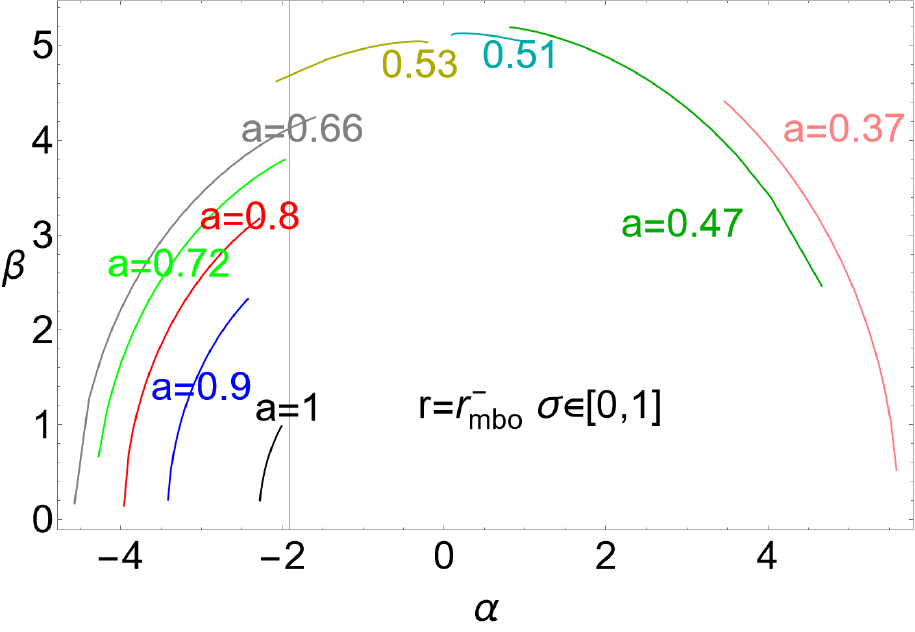}
\includegraphics[width=5.6cm]{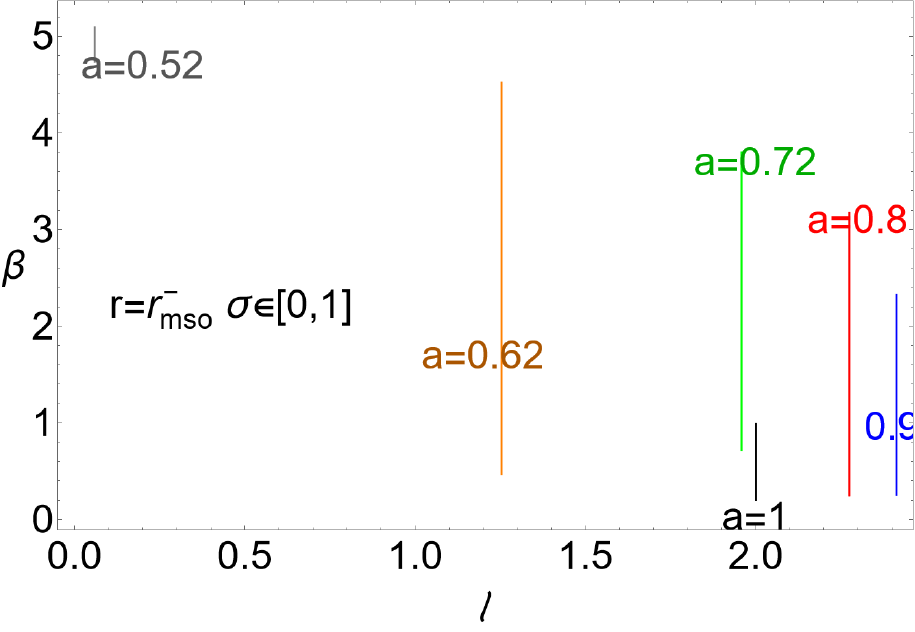}
\includegraphics[width=5.6cm]{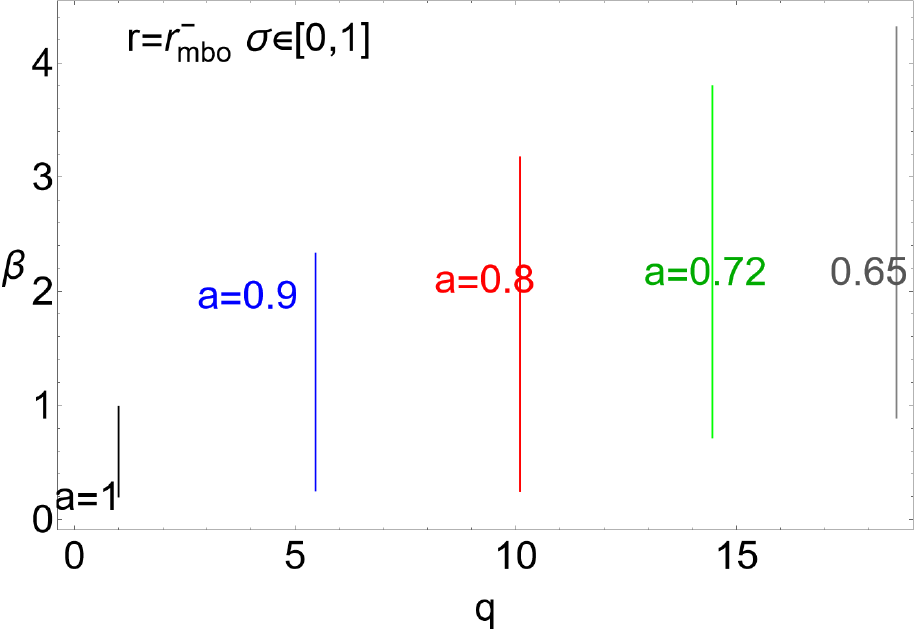}
\includegraphics[width=5.6cm]{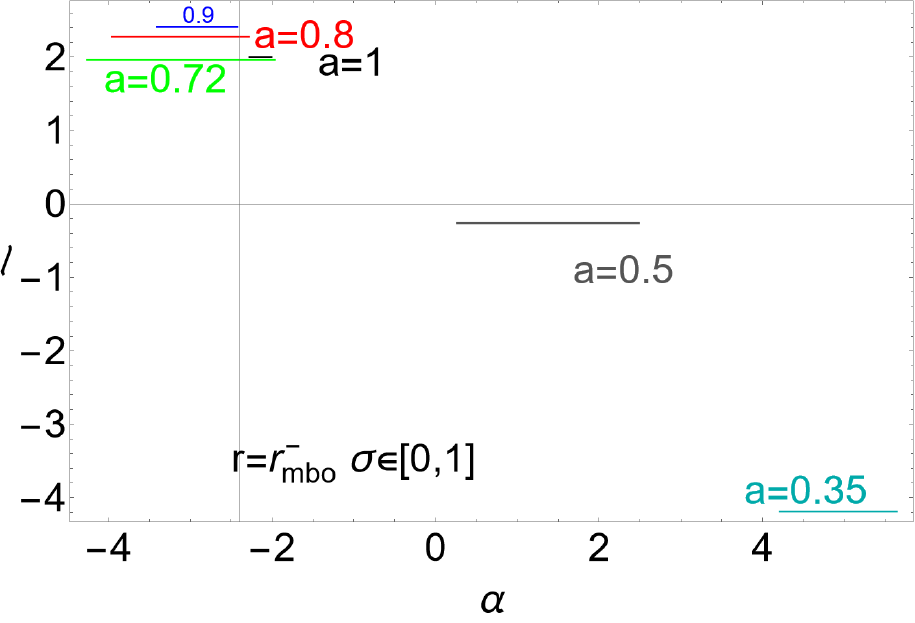}
\includegraphics[width=5.6cm]{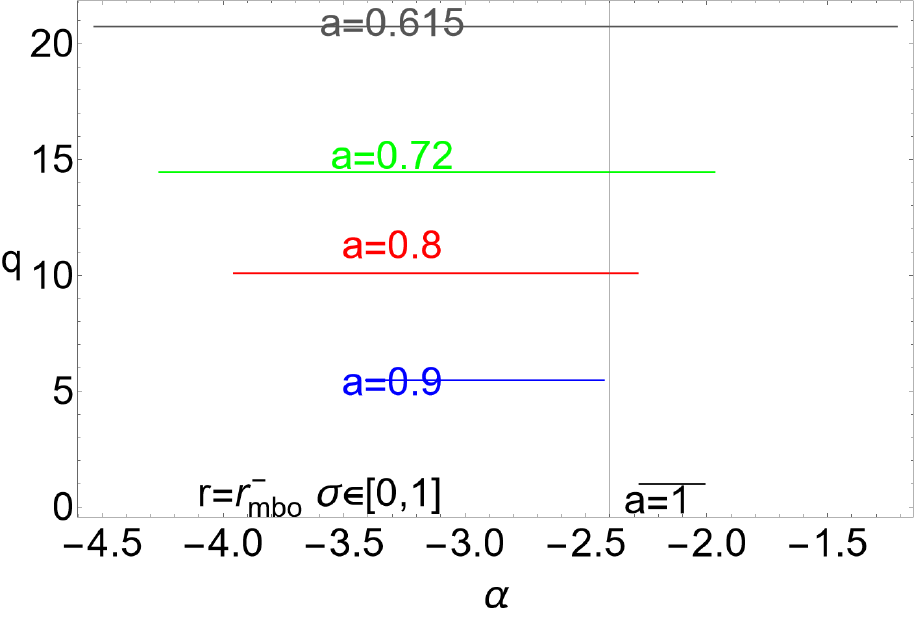}
\caption{Case $r=r_{mbo}^-$. {(For further details see also caption of Figs\il(\ref{Fig:Plotfarengmrepsrmso}).)}
 }.\label{Fig:Plotfarengm4repslmsomb}
\end{figure}

\medskip

\textbf{In the spherical shells $ [r_{mbo}^\pm,r_{mso}^\pm]$}

\medskip

In general, it can be $r_\lambda\in [r_{mbo}^\pm,r_{mso}^\pm]$ for some ranges of values of the impact parameter  $\ell$, and we study the  general solutions: \textbf{(a)} $r_\lambda\in [r_{mbo}^+,r_{mso}^+]$, for  $\ell<0$ and  \textbf{(b)}  $r_\lambda\in [r_{mbo}^-,r_{mso}^-]$ for  $\ell>0$. It is convenient to consider  the  solution   $\ell_\lambda$ of the  equation $r=r_\lambda(\ell)$,  leading to  the  impact parameter
\bea\label{E:llambda-eq-def}
\ell_\lambda\equiv \frac{a^2 (r+1)+(r-3) r^2}{a(1-r)}: r_\lambda(\ell)=r,
\eea
showed in Figs\il(\ref{Fig:Plotuljlp}) for different \textbf{BH} spin, with respect to the Kerr spacetime geodesic structure.
Below we  comment our results focusing  first on the co-rotating case and we close this section  with the analysis of the counter-rotating photons.

\medskip

\textbf{The co-rotating case }

Momentum   $\ell_\lambda>0$ is shown (for $\sigma\in[0,1]$) in   Figs\il(\ref{Fig:Plotuljlp}),
with the  limiting spins
\bea\label{Eq:def-astars}
a_{mso}^*\equiv 0.8528: \ell_\lambda(r_{mso}^-)=0,\quad a_{mbo}^*\equiv 0.516: \ell_\lambda(r_{mbo}^-)=0.
\eea
Assuming $\ell>0$,  there is  $r_\lambda=r_{mso}^-$  for $a>a_{mso}^*$, and $r_\lambda=r_{mbo}^-$   for $a>a_{mbo}^*$, therefore in general $r\in[r_{mbo}^-,r_{mso}^-]$ only for  $a>a_{mbo}^*$ (where $\sigma\in[0,1]$), where further constraints should be applied, considering the full set of equations (\ref{Eq:eqCarter-full}).

From Figs\il(\ref{Fig:Plotscadenztoaglrl})  it can be seen that
$r_{\lambda}(\ell_{mso}^-)\in[r_{mbo}^-,r_{mso}^-]$ for $a>a_\lambda^-$.

More specifically, considering  Figs\il(\ref{Fig:Plotuljlp}) there is
\bea&&
\mbox{for}\quad a\in[0,a_\lambda^-[:\quad
\ell_{\gamma}^-=\ell_\lambda(r_{\gamma}^-)>\ell_{mbo}^->\ell_{mso}^->\ell_\lambda(r_{mbo}^-)>\ell_\lambda(r_{mso}^-),
\\\nonumber&&
\mbox{for}\quad a>a_\lambda^-:\quad
\ell_{\gamma}^-=\ell_\lambda(r_{\gamma}^-)>\ell_{mbo}^->\ell_\lambda(r_{mbo}^-)>\ell_{mso}^->\ell_\lambda(r_{mso}^-),
\eea
for all $\sigma\in[0,1]$.
Hence,  for $\ell>0$, there  is  $r_\lambda\in [r_{mbo}^-,r_{mso}^-]$, with a momentum  $\ell$ smaller then $\ell_{mso}^-$  for $a\in[0,a_\lambda^-[$. For $a>a_\lambda^-$ there is $r_\lambda\in [r_{mbo}^-,r_{mso}^-]$, close to  $r_{mbo}^-$.
(Radius $r_\lambda$ is  located in the ergoregion for sufficiently large spins.).
 Finally, according to the analysis of  Fig.\il(\ref{Fig:PLOTHDIREC31})-left panel, there is
\bea
C_\beta(r_{mbo}^-)>C_\beta(r_{mso}^-),\quad \mbox{for}\quad \sigma=1
\eea
in the plane $\alpha-\beta$,  in the sense that the curve  correspondent to $r=r_{mbo}^-$ is more external then  the curve  for  $r=r_{mso}^-$.
 As  there is $\sigma=1$, and therefore $\beta=\sqrt{q}$, solutions at $q=0$ are only the curves $r=r_{\gamma}^-$.

From Figs\il(\ref{Fig:PLOTHDIREC31})-right panel, we see that
for fixed spin $a=1$, radii $r_{mso}^-$ and $r_{mbo}^-$ are inside the outer ergoregion, and there are no solutions for a spin
$a<a_{min}$ where,  for $r_{mso}^-$, there is $ a_{min}\approx 0.86$ (according to limit  $a=a_{mso}^\lambda\equiv 0.7851: r_{\lambda}(\ell_{mso}^-)=r_{\epsilon}^+$), for $r_{mbo}^-$ there is $ a_{min}=0.66$, for $r_{\epsilon}^+$ there  is $a_{min}=0.73$. Coordinate $|\beta|$ increases with $\sigma$. For large  \textbf{BH} spins,  curves  for $r_{mso}^-$ and $r_{mbo}^-$ cross the ergosurface. (It is interesting to note that the ergosurfaces curves cross at different spins.).
At  fixed $r=r_{mso}^-$,  the $\beta$ variation with $\alpha$   distinguishes different \textbf{BH} spins. (In this analysis we selected the cases
$\ell>0$, therefore focusing on $\alpha<0$).
In  this case, however, there is:
\bea
C_\beta(r_{mbo}^-)<C_\beta(r_{mso}^-),\quad \mbox{for}\quad a\approx 1.
\eea

Figs\il(\ref{Fig:PLOTHDIREC31})-left panel
shows  $\beta(\alpha)$
for all \textbf{BH} (dimensionless) spins  $a\in[0,1]$, for fixed orbits $r\in\{r_{\epsilon}^+,r_{\gamma}^\pm,r_{mso}^-,r_{mbo}^-,r_{\Ta},(r_{\gamma}^+-0.1)\}$
{on the equatorial plane}.
Curves exist only for a spin $a>a_{min}$, in agreement also with the analysis of  Figs\il(\ref{Fig:Plotfarengmrepsrmso}) and  Figs\il(\ref{Fig:Plotfarengm4repslmsomb}), where there is
\bea&&\label{Eq:flous-a-min-aigma1}
\mbox{for}\quad \sigma=1:\quad  a_{\min}(r_{\gamma}^-)\approx 0,\quad
a_{\min}(r_{mbo}^-)=0.52,\quad
a_{\min}(r_{mso}^-)=0.853, \\
&&\nonumber\hspace{3cm}
a_{\min}(r_{\gamma}^+)\approx 0,\quad
a_{\min}(r_{\Ta})\approx 0,\quad
a_{\min}(r_{\epsilon}^+)=0.71,
\eea
--Fig.\il(\ref{Fig:PLOTHDIREC31}). Note,  $a_{\min}(r_{mbo}^-)$ corresponds to  solution $a_{mbo}^*$,  and $ a_{\min}(r_{mso}^-)$ to the limiting spin $a_{mso}^*$ and $a_{\min}(r_{\epsilon}^+)$ to $a_{\gamma}^\epsilon: r_{\gamma}^-=r_{\epsilon}^+(\sigma=1)$.
%
Finally,    solutions with $\ell\in[\ell_{mbo}^-,\ell_{\gamma}^-]$ or  $r_\lambda\in\{r_{mbo}^-,r_{\gamma}^-\}$ are also possible--see Figs\il(\ref{Fig:PLOTHDIREC31}).

In conclusion,  the analysis on the co-rotating case   shows that there are constraints on the spin and on the angle $\sigma$ for the  shadow  boundary.
The second condition  explored   here sees $r_\lambda$ located in the orbital range   for the accretion disk inner edge  (or  for  cusps closer to the \textbf{BH}), addressed   for larger poloidal angles and on the equatorial plane (where $q =0$ only in the case of the photon circular  orbit on the equatorial plane).
In any case, there can be orbits on  the outer ergosurface and in  the outer ergoregion for fast spinning  \textbf{BHs}--see  Figs\il(\ref{Fig:Plotscadenztoaglrl}) and discussion  in Sec.\il(\ref{Sec:from-ergo}).
However, this analysis proves that for any $\sigma\in[0,1]$ there is  (with small impact parameter) a bottom limit, $(a_{min},\sigma_{min})$,  for the \textbf{BH} spin and the $\sigma$ coordinate   according to  Eqs\il(\ref{Eq:def-astars}) and Eq.\il(\ref{Eq:flous-a-min-aigma1}).

This implies that the contribution to the  shadow boundary  from  co-rotating  photons in the ranges considered  here
are constrained,  especially for smaller  $\sigma$ and \textbf{BH} spins.
 Finally, we conclude by  stressing  that these orbital spherical shells can be in the \textbf{BH} ergoregion\footnote{For spin $a\in[a_{mbo}^\epsilon,a_{mso}^\epsilon[$ and will be in the ergoregion for spin $[a_{mso}^\epsilon,1]$, where spins $\{a_{\gamma}^\epsilon,a_{mbo}^\epsilon,a_{mso}^\epsilon\}$ are in Figs\il(\ref{Fig:PlotfDS9s})} and in Sec.\il(\ref{Sec:from-ergo}) we shall focus on shadows boundary from photons orbiting  the outer ergosurface and the outer ergoregion.

\medskip

\textbf{The counter-rotating case }

In the counter-rotating case the situation is strongly different.
As clear from the analysis of
Figs\il(\ref{Fig:Plotfondoaglrsigmapb},\ref{Fig:Plotverinimsop},\ref{Fig:Plotfondoaglrsigmapbcasdu},\ref{Fig:Plotsigmaglr})
there are solutions for
$\ell\in\{\ell_{mso}^+,\ell_{mbo}^+\}$.

Considering Figs\il(\ref{Fig:Plotuljlp}) we find that
\bea
-\ell_\lambda(r_{mso}^+)>-\ell_{\lambda}(r_{mbo}^+)>-\ell_{\lambda}(r_{\gamma}^+)=
-\ell_{\gamma}^+>-\ell_{mbo}^+>-\ell_{mso}^+.
\eea
However, further inspection on the photons trajectories constraints informs that there are \emph{no}  counter--rotating solutions
for $r>r_{\gamma}^+$.
\begin{figure}
\centering
\includegraphics[width=5.5cm]{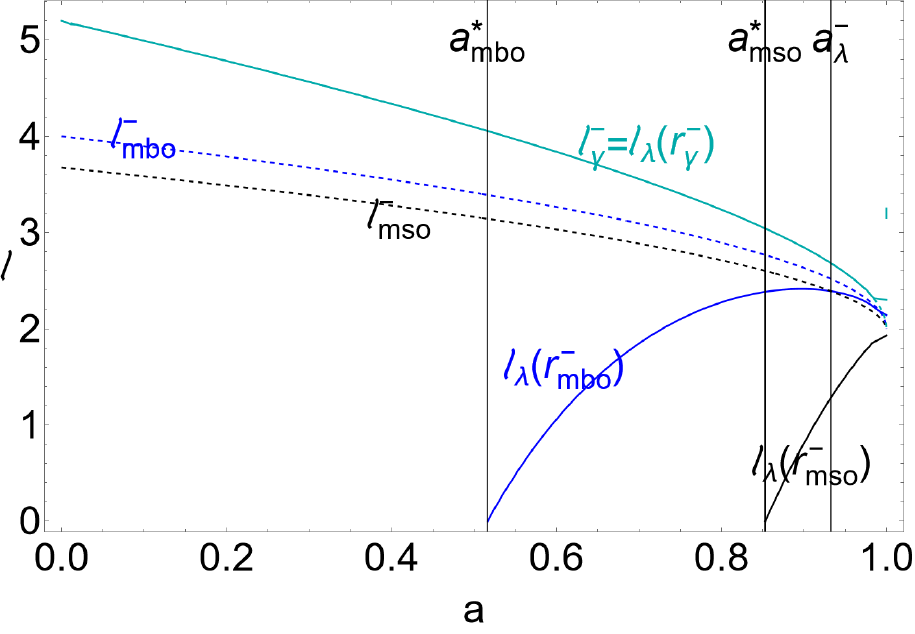}
\includegraphics[width=5.5cm]{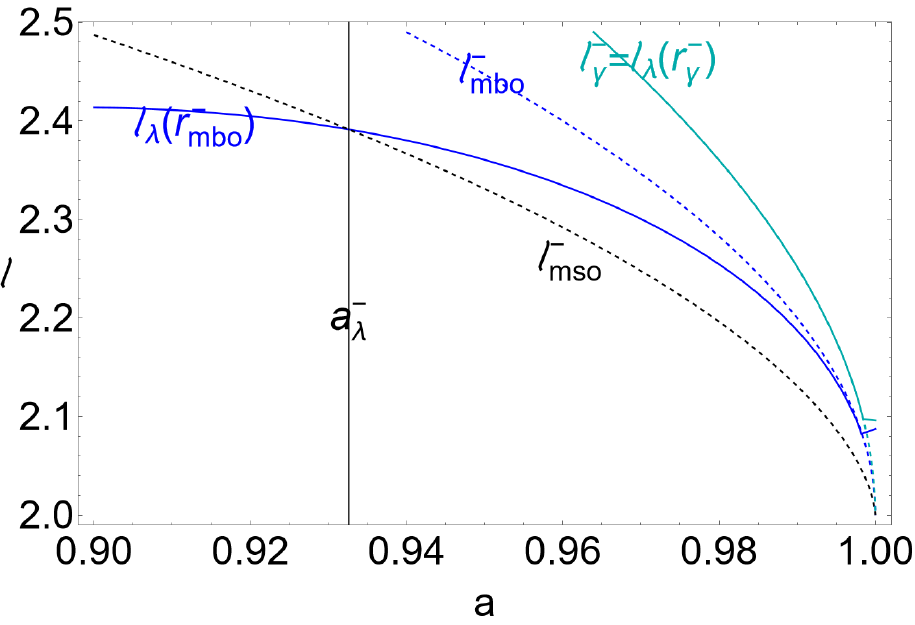}
\includegraphics[width=6.1cm]{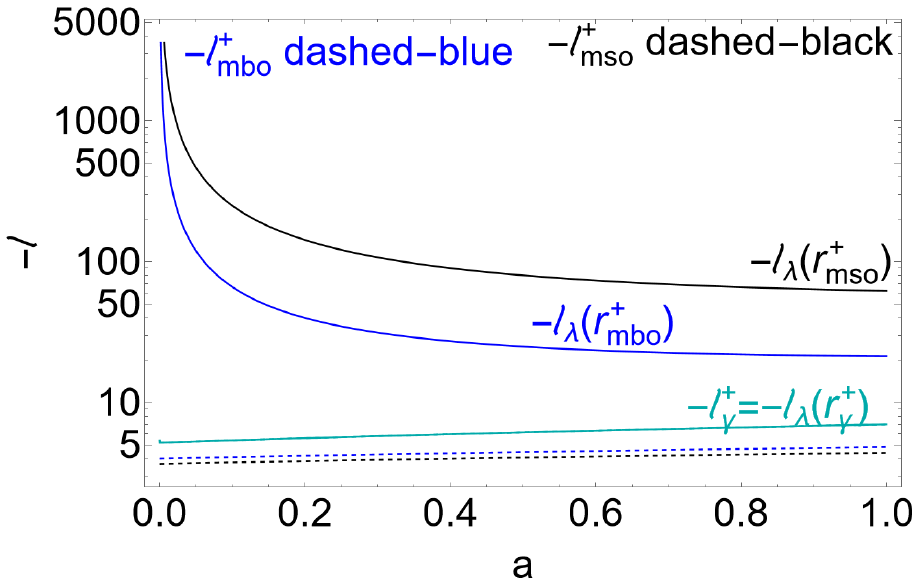}
\caption{Shadows from the inner edge (Sec.\il(\ref{Sec:radiii-shadows})).  $\ell_\lambda$ is  in Eq.\il(\ref{E:llambda-eq-def}).  Left (right) panel shows the situation for the co-rotating (counter-rotating) case, center panel is a close-up view of the left panel.}\label{Fig:Plotuljlp}
\end{figure}
\begin{figure}
\centering
\includegraphics[width=8cm]{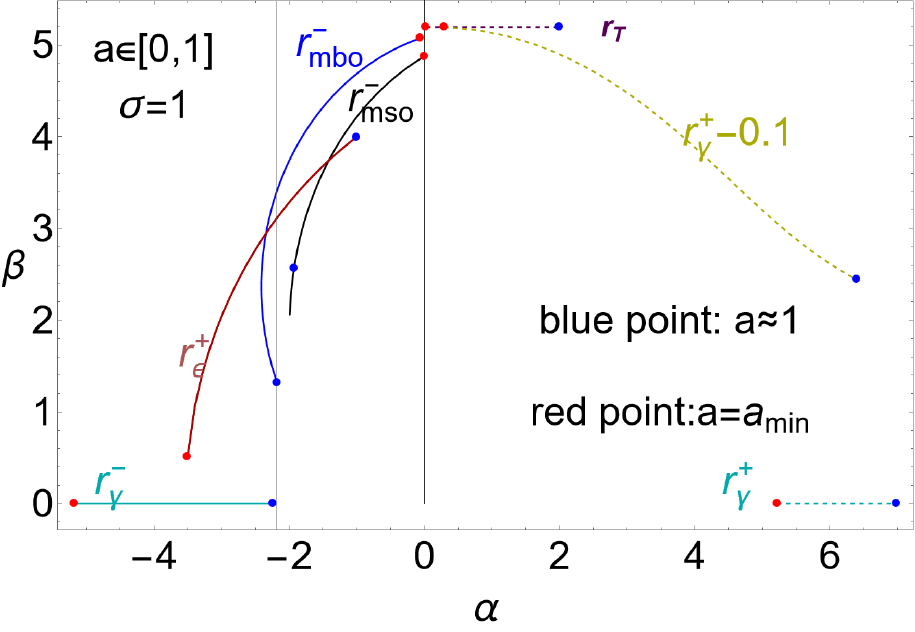}
\includegraphics[width=8cm]{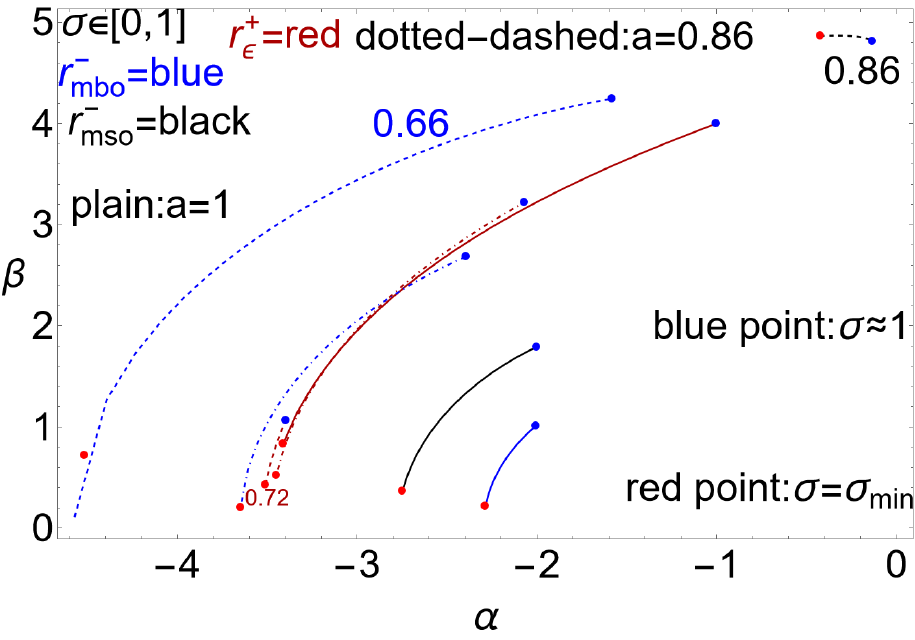}
\caption{Shadows from the inner edge (Sec.\il(\ref{Sec:radiii-shadows})). Left panel:  there are no solutions for $r=\{r_{mso}^+,r_{mbo}^+\}$).  Each point of a curve is for a spin  $a$. $a_{min}$ is the smallest \textbf{BH} spin a solution exists, different for each fixed radius $r$. Right panel:    each point of a curve is for a angle $\sigma$.
 (It is $\ell>0$ ($\alpha<0$) for  $\{r_{mso}^-,r_{mbo}^-,r_{\gamma}^-\}$ and  $\ell<0$  ($\alpha>0$) for $\{r_{mso}^+,r_{mbo}^+,r_{\gamma}^+\}$).   (Solutions $(-\alpha)$ are not represented).}\label{Fig:PLOTHDIREC31}
\end{figure}
More generally, there is \emph{no} solution $r_\lambda>r_{\gamma}^+$  for $\ell<0$ (for \emph{any}  $\sigma$).

On the other hand, there is
$r_\lambda<r_{\gamma}^+$  for $\ell\in\{\ell_{mso}^+,\ell_{mbo}^+\}$  that is, in this range of $\ell$ radius  $r_\lambda$ cannot be in the range for accretion disks inner edge on the equatorial plane.

 Radius  $r_\Ta<r_{\gamma}^+$  is shown, at any  $\sigma$, in Figs\il(\ref{Fig:Plotscadenztoaglrl}).
 Figs\il(\ref{Fig:PLOTHDIREC31})-left panel
shows   $\beta(\alpha)$
for all  $a\in[0,1]$ and  for fixed orbits $r\in\{r_{\epsilon}^+,r_{\gamma}^\pm,r_{mso}^-,r_{mbo}^-,r_{\Ta},(r_{\gamma}^+-0.1)\}$
{on the equatorial plane} (also for the counter-rotating orbits there is a bottom boundary $(a_{min},\sigma_{min})$ for the \textbf{BH} spin and $\sigma$).  If  there is $\sigma=1$, and therefore $\beta=\sqrt{q}$, solutions at $q=0$ are only the orbits  $r=r_{\gamma}^+$.
  Fig.\il(\ref{Fig:PLOTHDIREC31}) shows that  on the equatorial plane the shadow from  the inversion point, i.e.  $r=r_\Ta$, is almost constant in $\alpha$ for different spins $a\in[a_{min} (r_\Ta),1]$.
  It is also clear  how the variation with spin for the curves $\beta(\alpha)$ is opposite for $r_{\gamma}^-$ and  $r_{\gamma}^+$,
 and for  $r_\epsilon^+$  and $\{r_{mso}^-,r_{mbo}^-,r_\Ta,(r_{\gamma}^+-0.1)\}$ as $\beta$ is  greater for the fast spinning \textbf{BHs} and $|\alpha|$ is large for  small spin.   In the counter-rotating case, $|\beta|$ increases with decreasing $r$--Fig.\il(\ref{Fig:PLOTHDIREC31}).

The analysis of  Figs\il(\ref{Fig:Plotuljlp}) confirms  the constrains on the specific angular momentum raising questions on the observability  of  the counter-rotating case as defined in this framework.
\subsection{Shadows from the outer  ergoregion}\label{Sec:from-ergo}
\begin{figure}
\centering
\includegraphics[width=5.8cm]{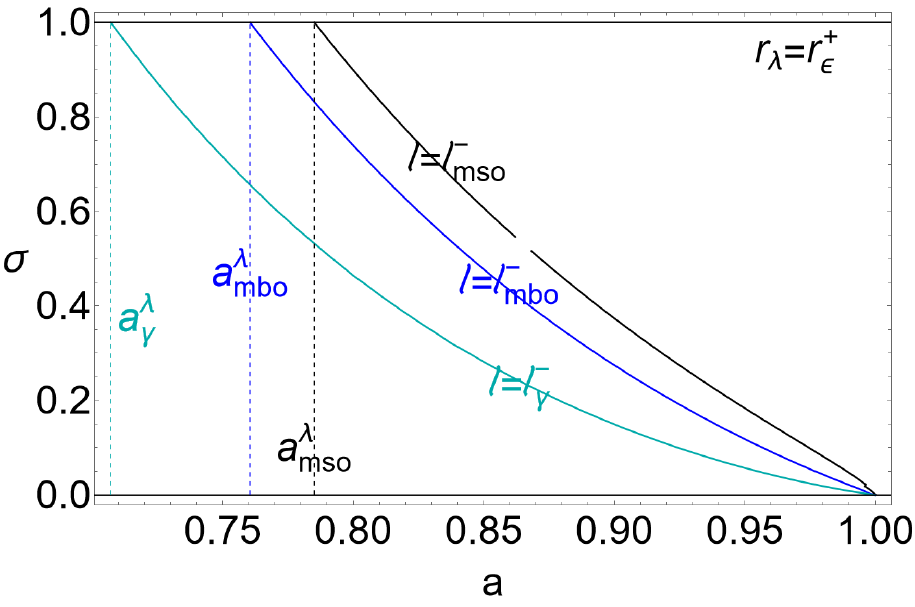}
\includegraphics[width=5.8cm]{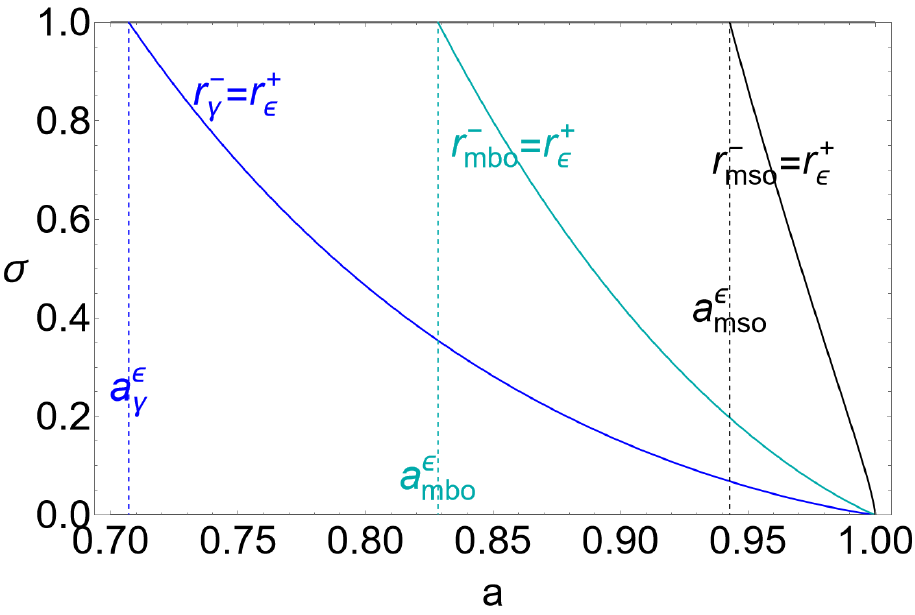}
\includegraphics[width=5.8cm]{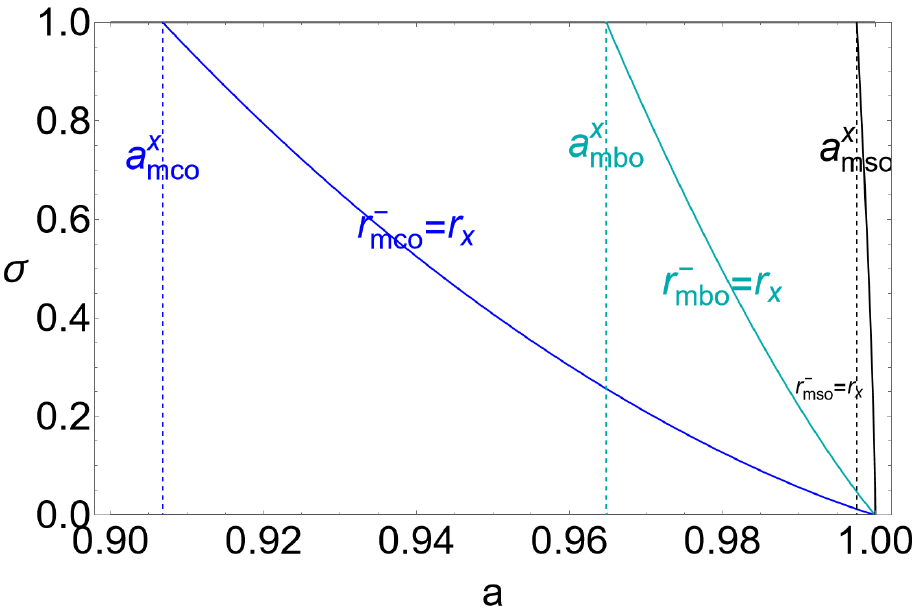}
\caption{Left panel: solutions $(\sigma,a): r_\lambda(\ell_*)=r_{\epsilon}^+$,      $r_\lambda$ is in Eq.\il(\ref{Eq:qlambda})  and  $\ell_*\in\{\ell_{mso}^-,\ell_{mbo}^-,\ell_{\gamma}^-\}$.  For $\sigma=1$  there is $r_\lambda(r_*)=r_\epsilon^+$  in the geometries with  $a_*\in \{a_{mso}^\lambda,a_{mbo}^\lambda,a_{\gamma}^\lambda\}$    in Eqs\il(\ref{Eq:raCub-post}). 	Center (right) panel: solutions $\sigma(a)$ of $r_{\gamma}^-=r_{\epsilon}^+(r_x)$, $r_{mbo}^-=r_{\epsilon}^+(r_x)$, $r_{mso}^-=r_{\epsilon}^+(r_x)$.}\label{Fig:PlotfDS9s}
\end{figure}
\begin{figure}
\centering
\includegraphics[width=6cm]{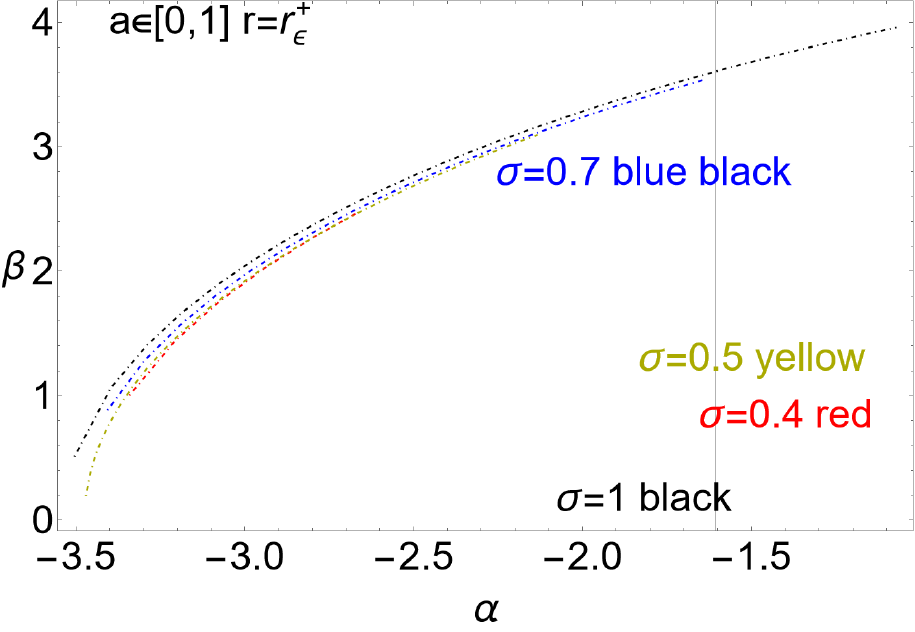}
\includegraphics[width=6cm]{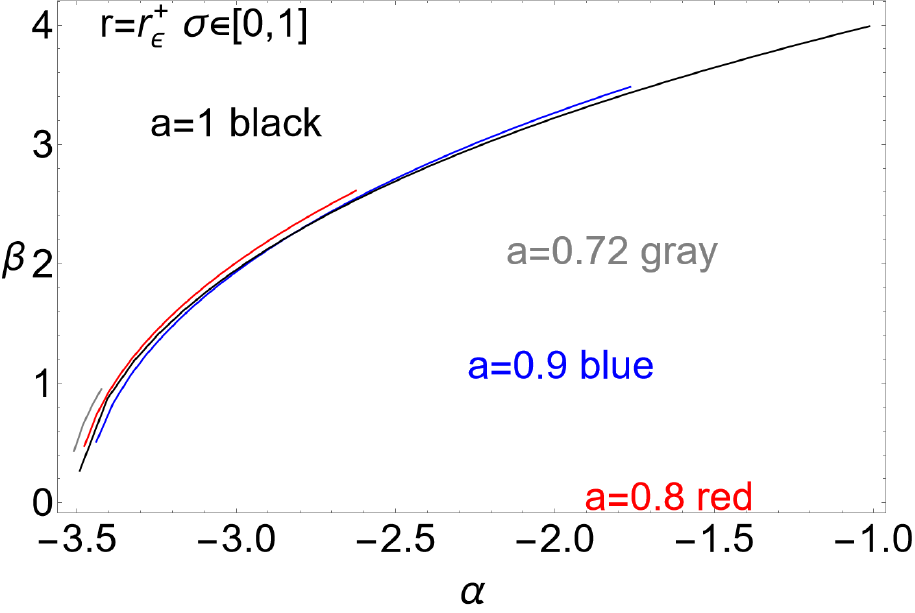}
\includegraphics[width=5.2cm]{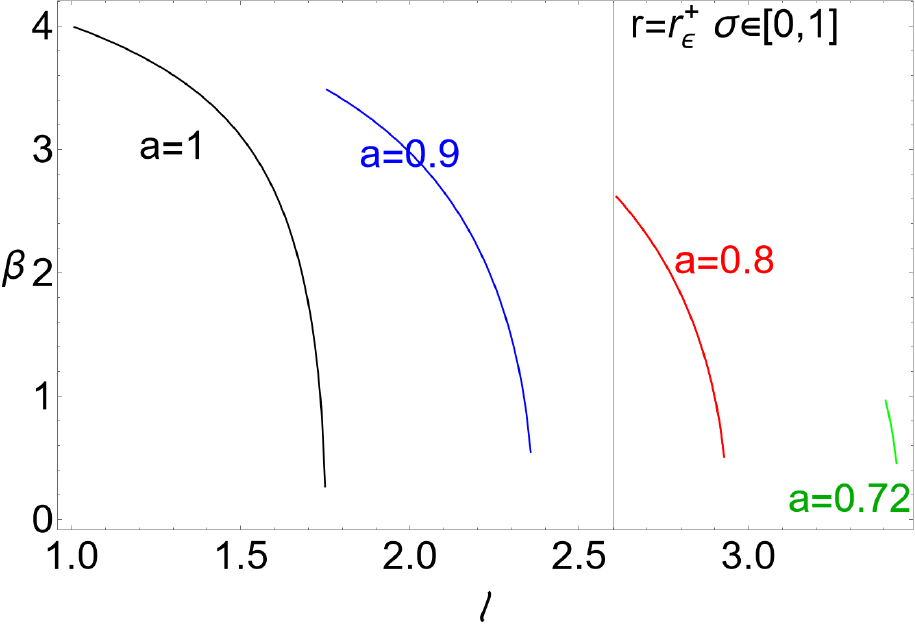}
\includegraphics[width=4.25cm]{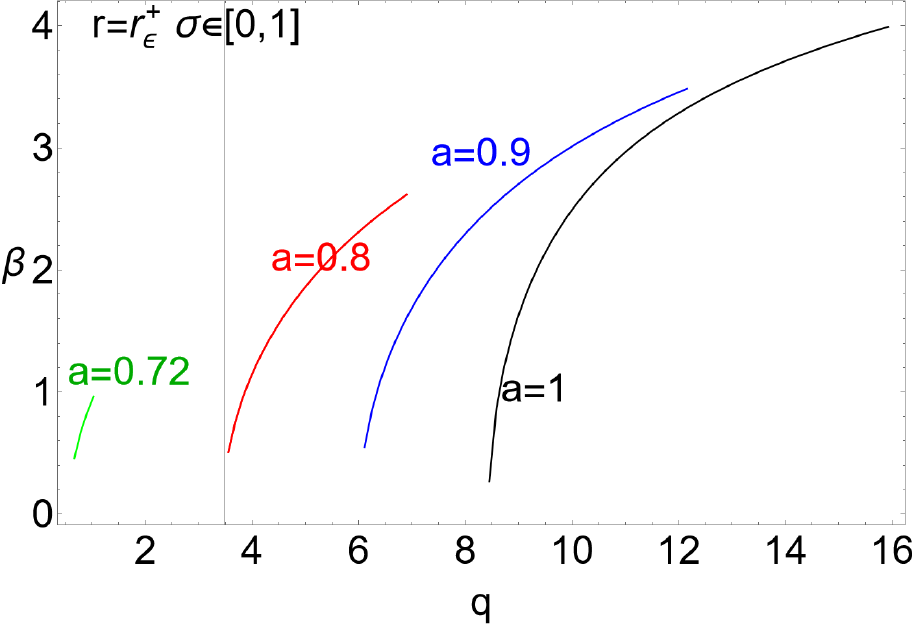}
\includegraphics[width=4.25cm]{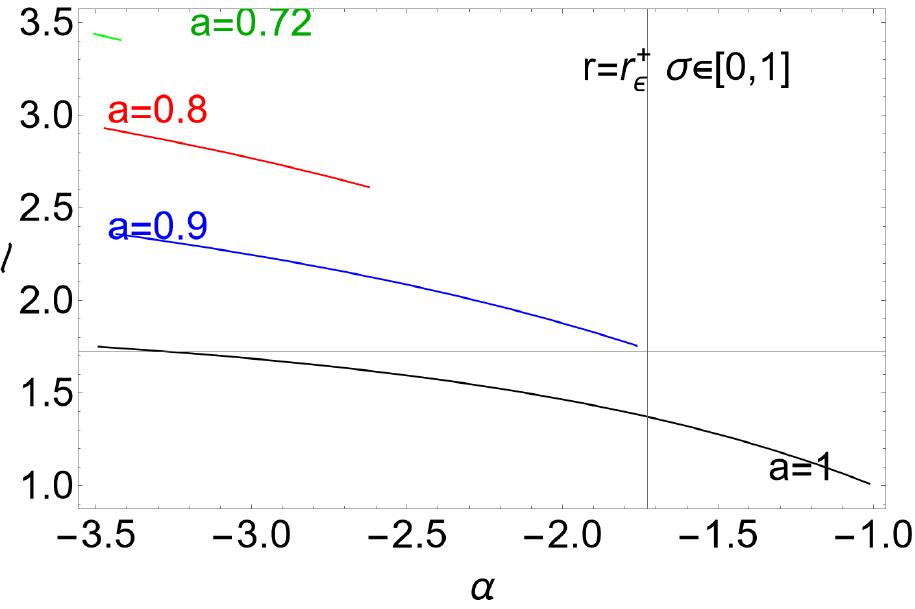}
\includegraphics[width=4.25cm]{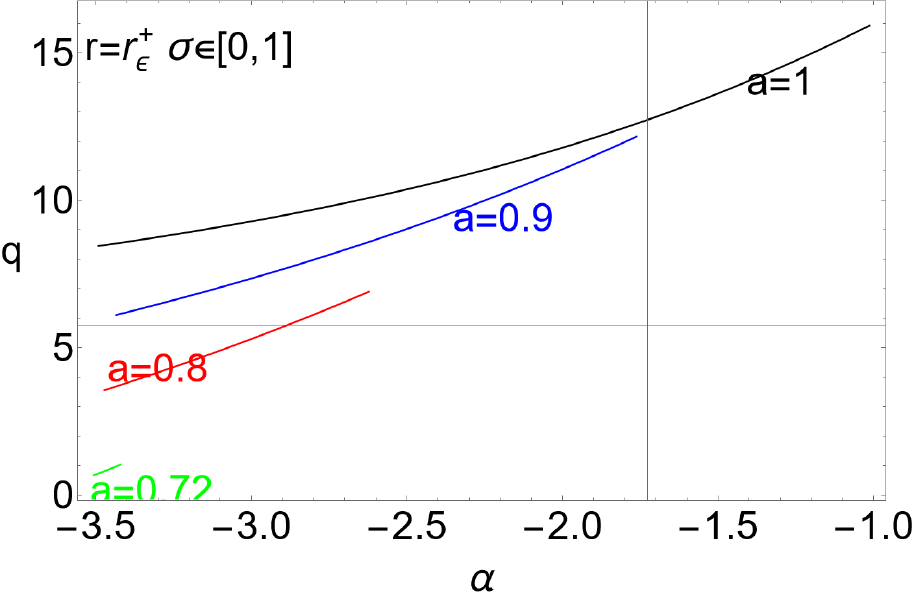}
\includegraphics[width=4.25cm]{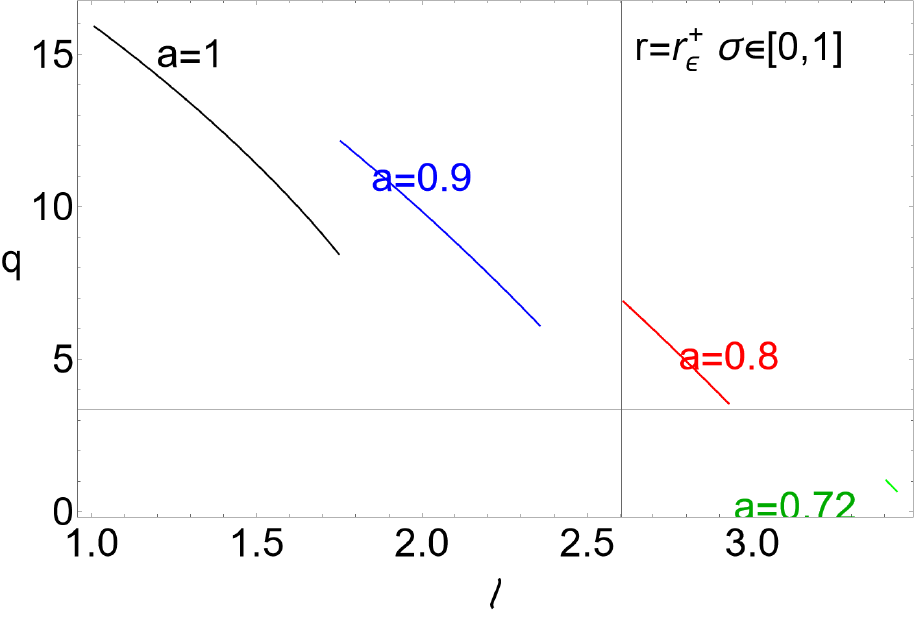}
\caption{Case $r=r_{\epsilon}^+$. Each point of a curve is for a different   $\sigma$.   Upper left panel: each point of a curve is for a different   $a$.}\label{Fig:Plotfarengmreps}
\end{figure}
%
%
%
\begin{figure}
\centering
\includegraphics[width=6cm]{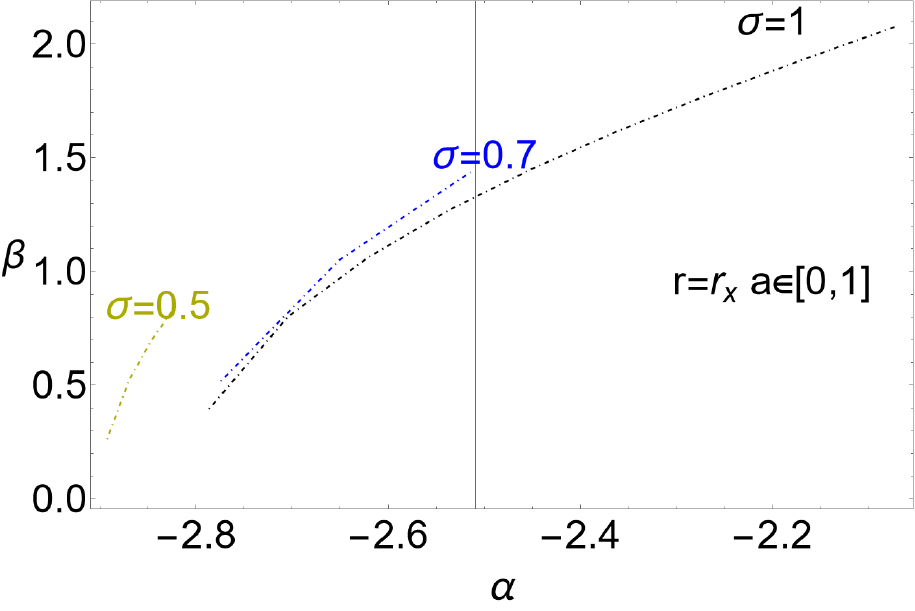}
\includegraphics[width=6cm]{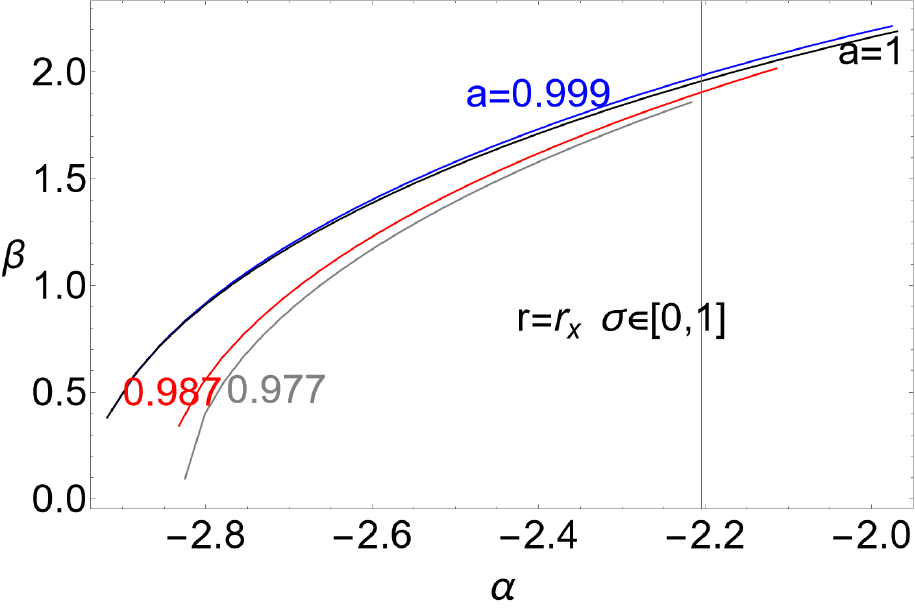}
\includegraphics[width=5cm]{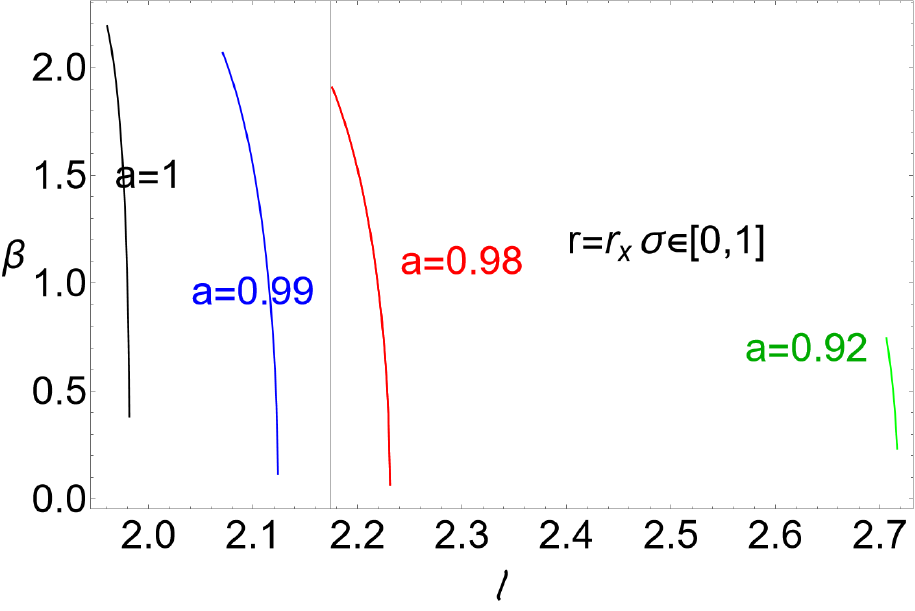}\\
\includegraphics[width=4.25cm]{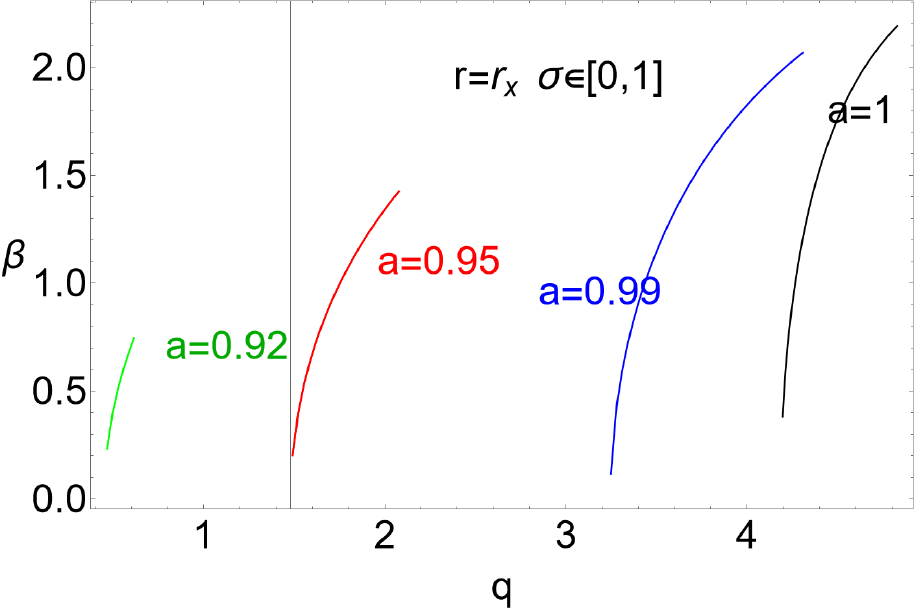}
\includegraphics[width=4.25cm]{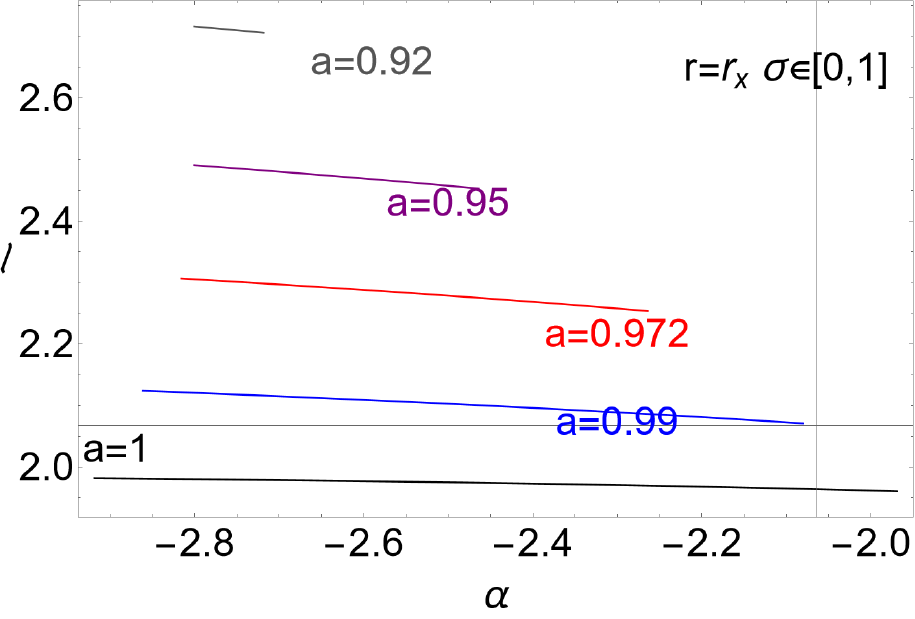}
\includegraphics[width=4.25cm]{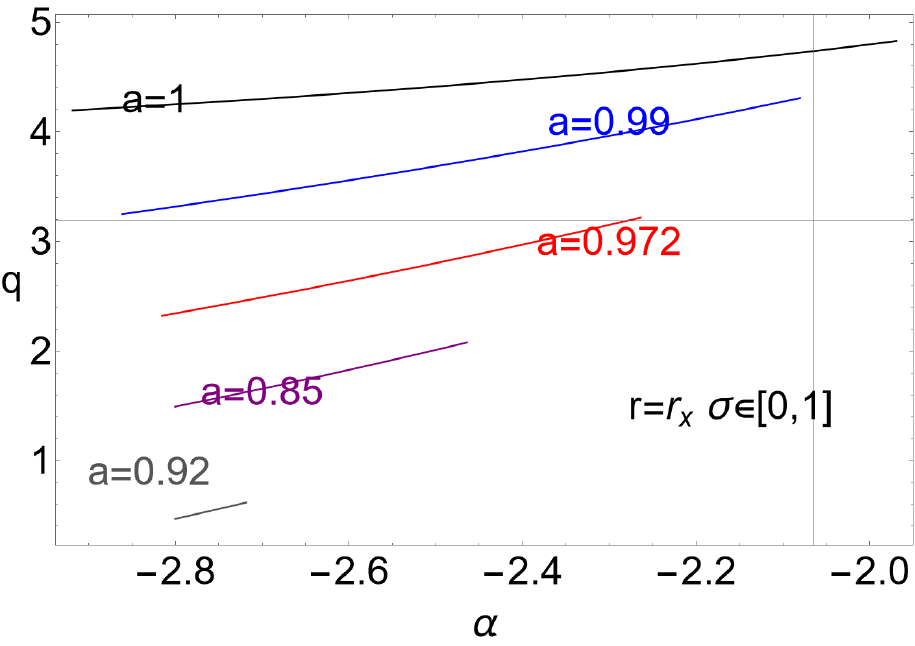}
\includegraphics[width=4.25cm]{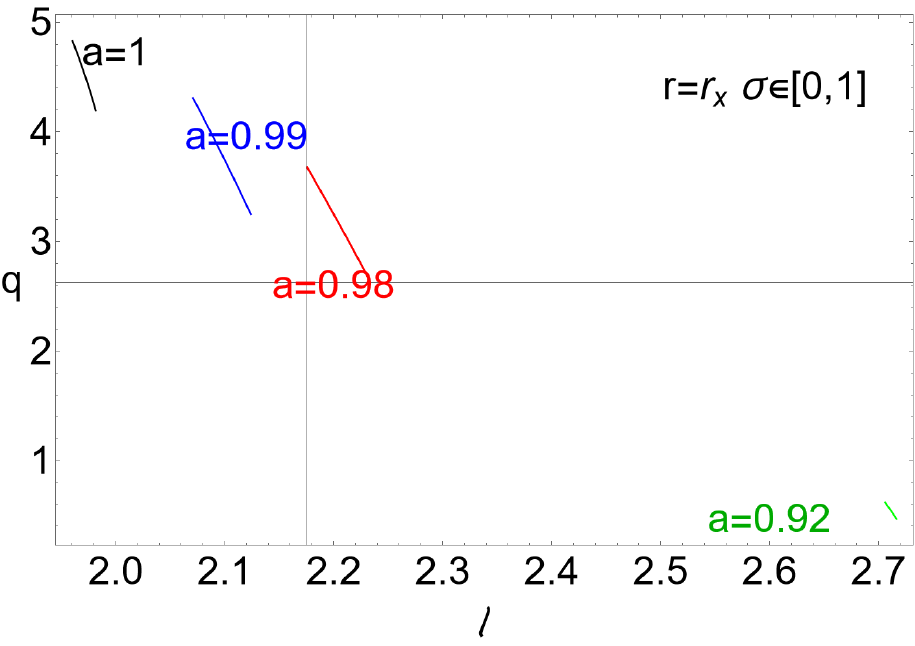}
\caption{Case $r=r_x\in]r_+,r_\epsilon^+]$ of Eq.\il(\ref{Eq:rx-definition}).  {(For further details see also Figs\il(\ref{Fig:Plotfarengmreps}).)}
 }.\label{Fig:Plotfarengmrepsx}
\end{figure}
There can be stable, bound and unstable co-rotating  circular orbits in the outer ergoregion of   the fast spinning  Kerr \textbf{BHs}. On the other hand, the inner edge of a co-rotating  torus,  are possible inside the ergoregion for large \textbf{BHs} spins \citep{dragged}.
In this section we consider the possibility  that the  outer ergoregion  (and the outer ergosurface) will be ``imprinted" in  the shadow boundary, i.e.  here we analyze solutions of the equations $(\mathfrak{R})$ for $r\in]r_+,r_\epsilon^+]$. First,   from  the analysis of   Figs\il(\ref{Fig:Plotscadenztoaglrl})
it is clear, how (depending on  the angle $\sigma$) in the co-rotating case, radius  $
r_\lambda$ can cross the outer ergosurface for large \textbf{BH} spins. 

Fig.\il(\ref{Fig:PLOTHDIREC31}) shows the  shadow profiles  from  fixed orbits $r\in\{r_{\epsilon}^+,r_{\gamma}^\pm,r_{mso}^-,r_{mbo}^-,r_{\Ta},(r_{\gamma}^+-0.1)\}$,  on the equatorial plane $\sigma=1$ for all \textbf{BH}  spins  $a\in[0,1]$.  (No counter-rotating solutions have been found for  $r>r_{\gamma}^+$, and in particular for $r\in[ r_{mso}^+,r_{mbo}^+]$).

The curve $\beta(\alpha)$, correspondent  to the photons  orbit coincident with the outer ergosurface ($r=r_\epsilon^+$), crosses the curve $\beta(\alpha)$ for fixed
$r\in \{r_{mso}^-,r_{mbo}^-\}$ for determined  values of $\alpha$ correspondent  in general to different \textbf{BH} spins (each point of the curve represents a different spin, from a maximum $a=1$, marked with a blue point, to a minimum $a_{min}$, different for each  curve $\beta(\alpha)$ for fixed  orbit $r$ and marked with a red point).
There is $a\in[a_{min} (r_\epsilon^+),1]$ where $a_{min} (r_\epsilon^+)=0.71$.
Therefore, there is $r_{\epsilon}^+\in[r_{mbo}^-,r_{mso}^-]$  only for some ranges of \textbf{BH} spins (in agreement with the  analysis of  Figs\il(\ref{Fig:Plotscadenztoaglrl})).
From Fig.\il(\ref{Fig:PLOTHDIREC31}) it is easy to see that,  as  there is $\sigma=1$ and therefore $\beta=\sqrt{q}$, solutions at $q=0$ are only the curves $r=r_{\gamma}^\pm$.

Therefore we set, at different angles $\sigma\in[0,1]$, the condition $r=r_x\in ]r_+,r_{\epsilon}^+[$  and
 $r=r_\epsilon^+$ respectively,  where
\bea\label{Eq:rx-definition}
r_x\equiv \frac{r_\epsilon^+-r_+}{5}+r_+.
\eea
In Figs\il(\ref{Fig:Plotfarengmreps})  the shadow boundary  is shown  for $r=r_\epsilon^+$, for  different  $a$ and $\sigma$. The plots illustrate the dependence of the  constrained coordinates  $\beta$ and $\alpha$  from parameters  $\ell$ and $q$, and the curves  $\ell(q)$ for $r=r_\epsilon^+$. The analysis of Figs\il(\ref{Fig:Plotfarengmreps})  is reproduced in
Figs\il(\ref{Fig:Plotfarengmrepsx}) for $r=r_x$.
In Figs\il(\ref{Fig:Plotfarengmreps},\ref{Fig:Plotfarengmrepsx}), as $\ell>0$ on $r=r_\epsilon^+$ and $r=r_x$, there is $\alpha<0$, here we restricted the analysis  to  $(\beta>0,\alpha<0)$.

\medskip

\textbf{From the outer ergosurface}

\medskip

Here we   consider  the case $r=r_\epsilon^+$.

From Figs\il(\ref{Fig:PlotfDS9s})--left panel we see the  solutions $(\sigma,a): r_\lambda(\ell_*)=r_{\epsilon}^+$, for  $\sigma\in [0,1]$, and  $\ell_*\in\{\ell_{mso}^-,\ell_{mbo}^-,\ell_{\gamma}^-\}$.
 For $\sigma=1$  there is $r_\lambda(\ell_*)=r_\epsilon^+$  in the geometries with $a\geq a_*$ where $a_*\in \{a_{mso}^\lambda,a_{mbo}^\lambda,a_{\gamma}^\lambda\}$   defined  as follows
\bea\label{Eq:raCub-post}
\mbox{for}\quad   \sigma=1:\quad
a_{mso}^\lambda\equiv 0.7851: r_{\lambda}(\ell_{mso}^-)=r_{\epsilon}^+,\quad a_{mbo}^\lambda\equiv 0.76: r_{\lambda}(\ell_{mbo}^-)=r_{\epsilon}^+,\quad  a_{\gamma}^\lambda\equiv0.7071: r_{\lambda}(\ell_{\gamma}^-)=r_{\epsilon}^+,
\eea
radius $r_\lambda(\ell_*)$ for $\ell_*\in\{\ell_{mso}^-,\ell_{mbo}^-,\ell_{\gamma}^-\}$ is in the outer ergoregion  in the geometries with \textbf{BH} spins $a\geq a_*$ where $a_*\in\{a_{mso}^\lambda,a_{mbo}^\lambda,a_{\gamma}^\lambda\}$  respectively,
see Figs\il(\ref{Fig:PlotfDS9s}).

  In  Figs\il(\ref{Fig:PlotfDS9s})--center and right panel we show the solutions $\sigma(a)$ of the equations $r_{*}=\{r_{\epsilon}^+,r_x\}$, where $r_*\in\{r_{\gamma}^-,r_{mbo}^-,r_{mso}^-\}$.
It is convenient to introduce the spins
\bea\label{Eq:a-rubb-Nobe}
a_{mbo}^\epsilon\equiv 0.828427: r_{mbo}^-=r_{\epsilon}^+(\sigma=1),\quad a_{\gamma}^\epsilon\equiv 0.707107: r_{\gamma}^-=r_{\epsilon}^+(\sigma=1),\quad a_{mso}^\epsilon\equiv 0.942809: r_{mso}^-=r_{\epsilon}^+(\sigma=1),
\eea
similarly spins  $\{a_{\gamma}^x,a_{mbo}^x,a_{mso}^x\}$   are defined by  $r_{*}=r_x(\sigma=1)$--see  Figs\il(\ref{Fig:PlotfDS9s}).
Note, in general $\{a_{mso}^\lambda,a_{mbo}^\lambda,a_{\gamma}^\lambda\}$ do not coincide with spins $\{a_{mso}^\epsilon,a_{mbo}^\epsilon,a_{\gamma}^\epsilon\}$ (both defined for $\sigma=1$) in Eq.\il(\ref{Eq:a-rubb-Nobe}).
However, there is  $a_{\gamma}^\lambda=a_{\gamma}^\epsilon$.

Results of this analysis are shown in Figs\il(\ref{Fig:Plotfarengmreps},\ref{Fig:Plotfarengmrepsx}):
we note that the coordinate
$|\beta|$ decreases with $\ell$ and  $|\alpha|<3.5$, and increases with
$q\lessapprox 16$. At fixed $\alpha$, it  decreases  with $a$ and  increases with $\sigma$. Quantity
$q$ decreases with $\alpha$ in magnitude, $|\alpha|$ increases with  $\ell$ and decreases with $q$.

\medskip
\textbf{From the outer ergoregion}

\medskip

A qualitatively   similar situation occurs for a point $r=r_x$, located  inside the outer ergoregion--Figs\il(\ref{Fig:Plotfarengmrepsx})--but with some notable differences   with respect to the orbits  from the outer ergosurface.
In  Figs\il(\ref{Fig:Plotvimsounitagrwo}) there are the shadows boundaries  for different $(a,\sigma)$ for  $\ell=\{\ell_{mbo}^\pm,\ell_{mso}^\pm,\ell_\Ta\}$ and $r=\{r_\epsilon^+,r_x\}$,  (merging of   Figs\il(\ref{Fig:Plotbetrodon1lmsom},\ref{Fig:Plotverinimsop},\ref{Fig:Plotfondoaglrsigmapb},\ref{Fig:Plotbetrodon1lmsopbm},\ref{Fig:Plotvimsounita})--upper right panles).
The analysis in Figs\il(\ref{Fig:Plotvimsounitagrwo})  allows to clarify how   the situation for   photons orbiting  in  the ergoregion, at $r_\lambda=r_x$, appears  different as compared to  $r_\lambda=r_\epsilon^+$. At $a=1$, the curve $\beta(\alpha)$ at $r_\lambda= r_\epsilon^+$ is more external than the curve at $r_\lambda= r_x$ (in agreement with the situation for the other curves $\beta(\alpha)$).

However,differences appear with the variation of the \textbf{BH} spin at fixed radius $r_\lambda$.  For $
 r_\lambda=r_\epsilon^+$, the curve moves inward  with increasing of the \textbf{BH} spin, viceversa the curve corresponding to the orbits  at
$r_\lambda=r_x$ moves inward  with decreasing of the \textbf{BH} spin in contrast with the other curves for the co-rotating orbits.  (We remind that $r_x$ is a function of $(a,\sigma)$). This divergence appears also from the comparison of  Figs\il(\ref{Fig:Plotfarengmrepsx})  and Figs\il(\ref{Fig:Plotfarengmreps}), also for the curve $\beta(\alpha)$ at different \textbf{BH} spin for a fixed angle.

Therefore,  orbits in  the outer ergosurface and the outer ergoregion are  possible for \textbf{BH} spin $a>a_{min}$ and   $\sigma\in [\sigma_{min},1]$, where the  $\sigma$ are explored in Figs\il(\ref {Fig:PLOTHDIREC31}) and the limits for a general $\ell$ are in Figs\il(\ref{Fig:Plotfarengmreps},\ref{Fig:PlotfDS9s},\ref{Fig:Plotfarengmrepsx}). In general $|\beta|$ is greatest for $\sigma=1$.
\begin{figure}
\centering
\includegraphics[width=8.5cm]{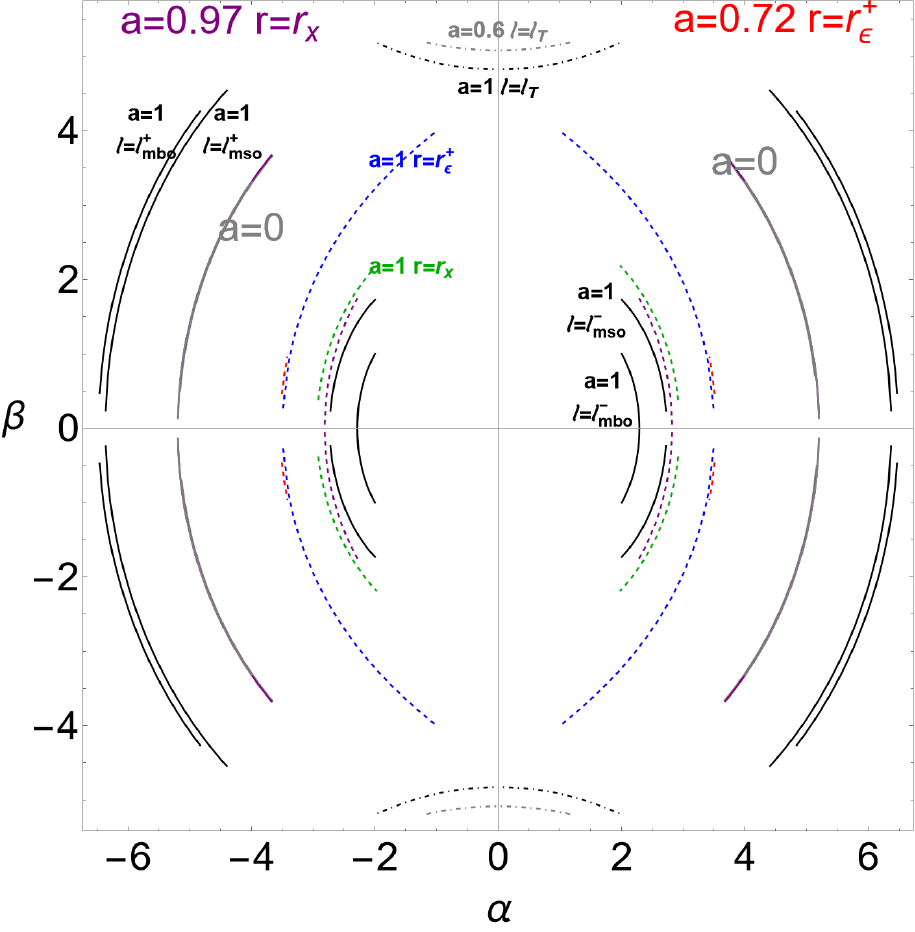}
\caption{Cases $\ell=\{\ell_{mbo}^\pm,\ell_{mso}^\pm,\ell_\Ta\}$ and $r=\{r_\epsilon^+,r_x\}$, merging of Figs\il(\ref{Fig:Plotvimsounita},\ref{Fig:Plotbetrodon1lmsopbm},\ref{Fig:Plotfondoaglrsigmapb},\ref{Fig:Plotverinimsop}).
Each point of  a curve  is for a different value of $\sigma$.} .\label{Fig:Plotvimsounitagrwo}
\end{figure}
Figs\il(\ref{Fig:Plotzing1t171s}) show the location of the photons considered in this analysis (cases $\ell=\{\ell_{mbo}^\pm,\ell_{mso}^\pm,\ell_\Ta\}$ and $r=\{r_\epsilon^+,r_x\}$) on the \textbf{BH} shadow boundary, for  selected values of $(a,\sigma)$, complementing the analysis of Sec.\il(\ref{Sec:allell}).
\begin{figure}
\centering
\includegraphics[width=5.6cm]{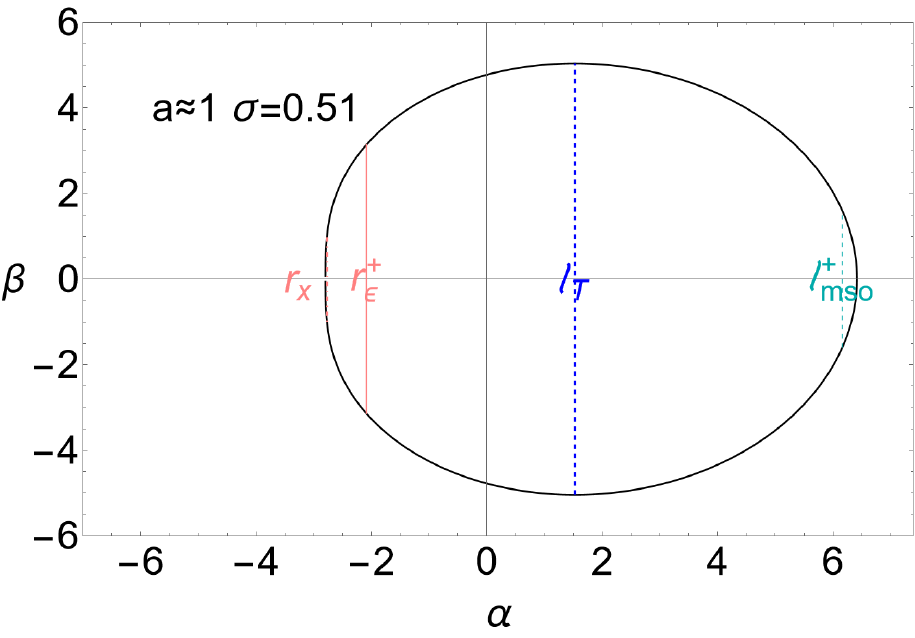}
\includegraphics[width=5.6cm]{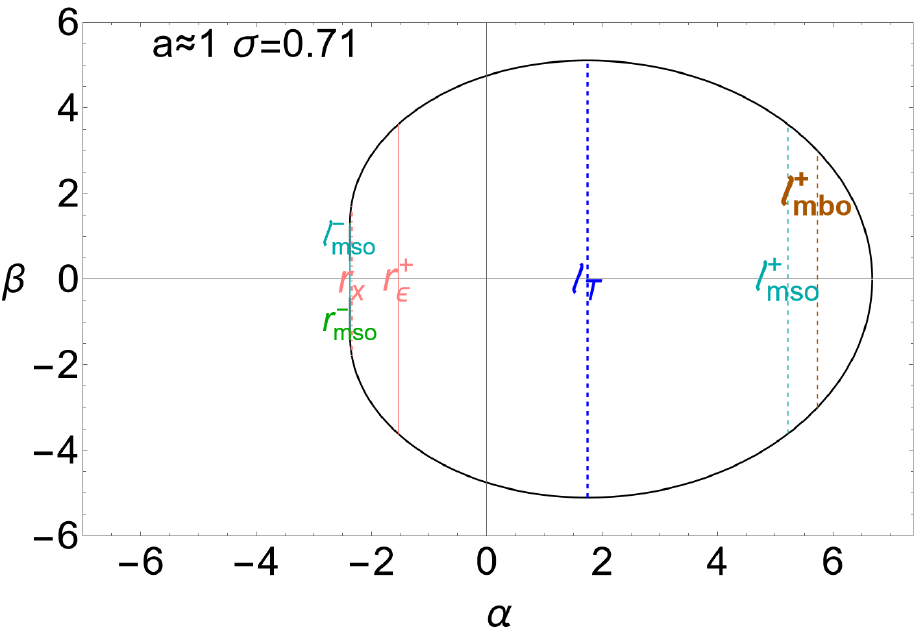}
\includegraphics[width=5.6cm]{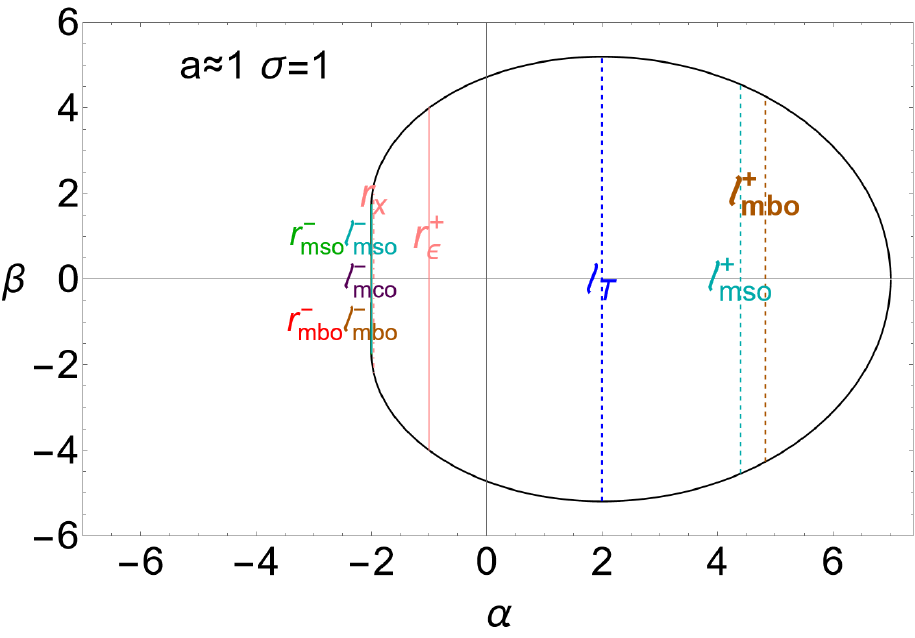}
\includegraphics[width=5.6cm]{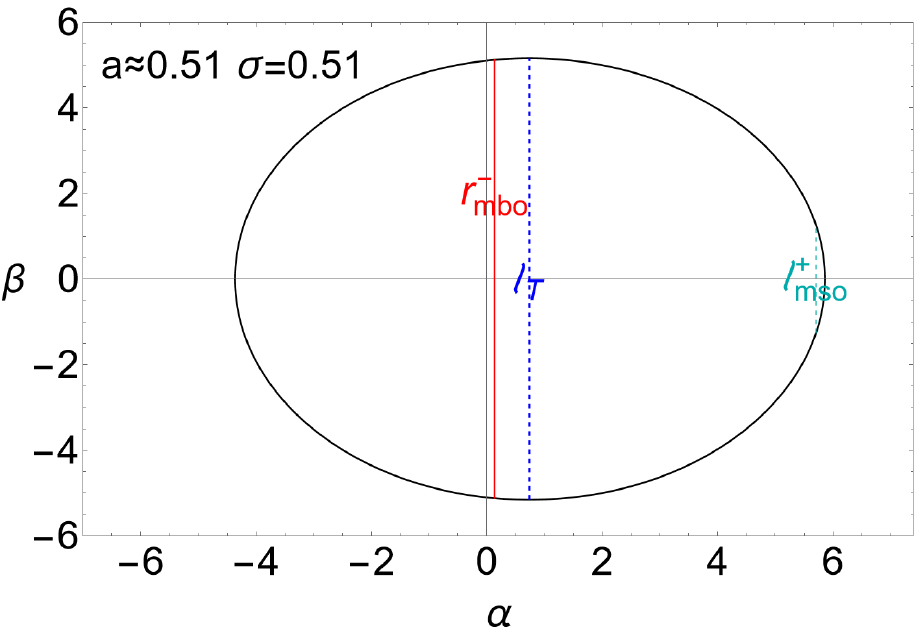}
\includegraphics[width=5.6cm]{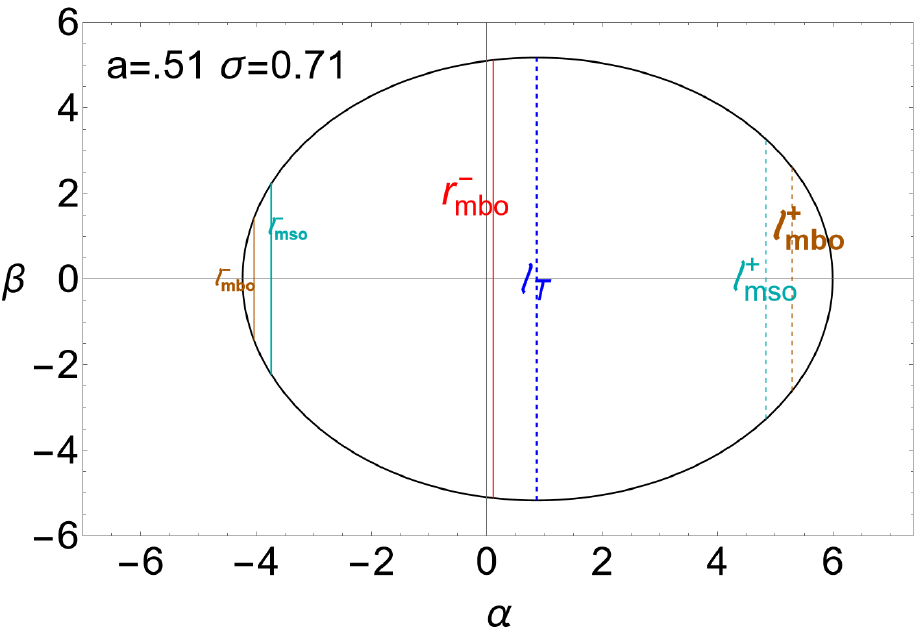}
\includegraphics[width=5.6cm]{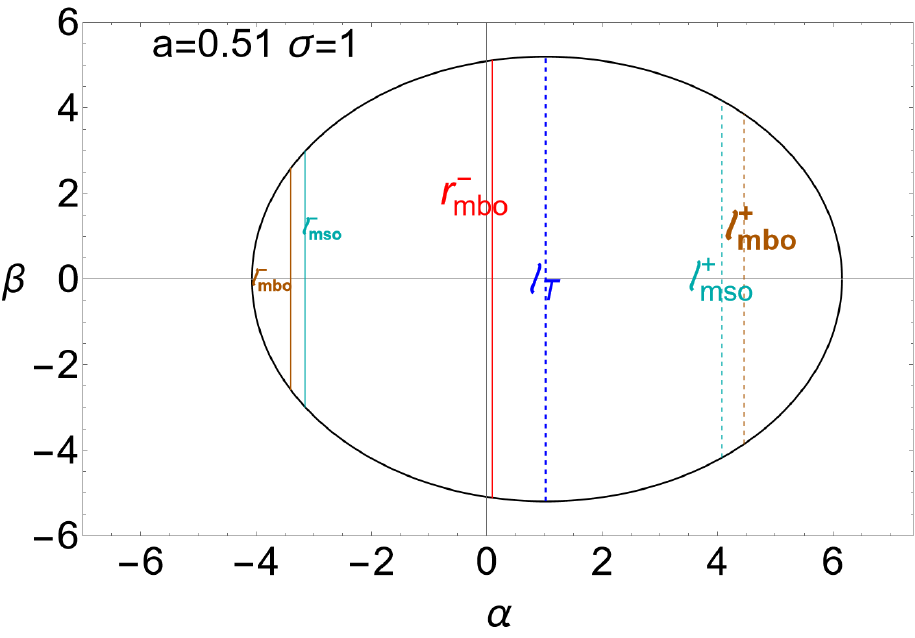}
\includegraphics[width=5.6cm]{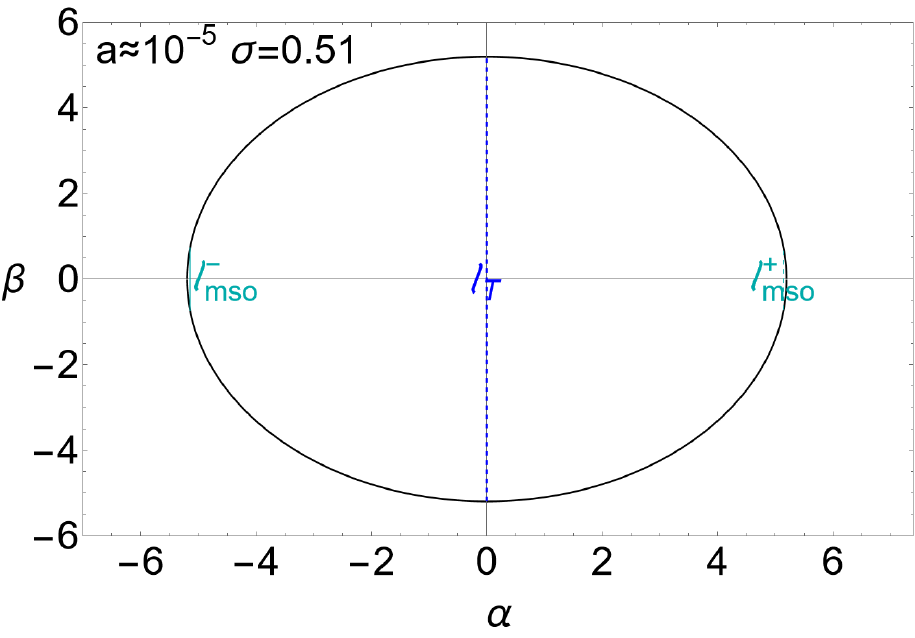}
\includegraphics[width=5.6cm]{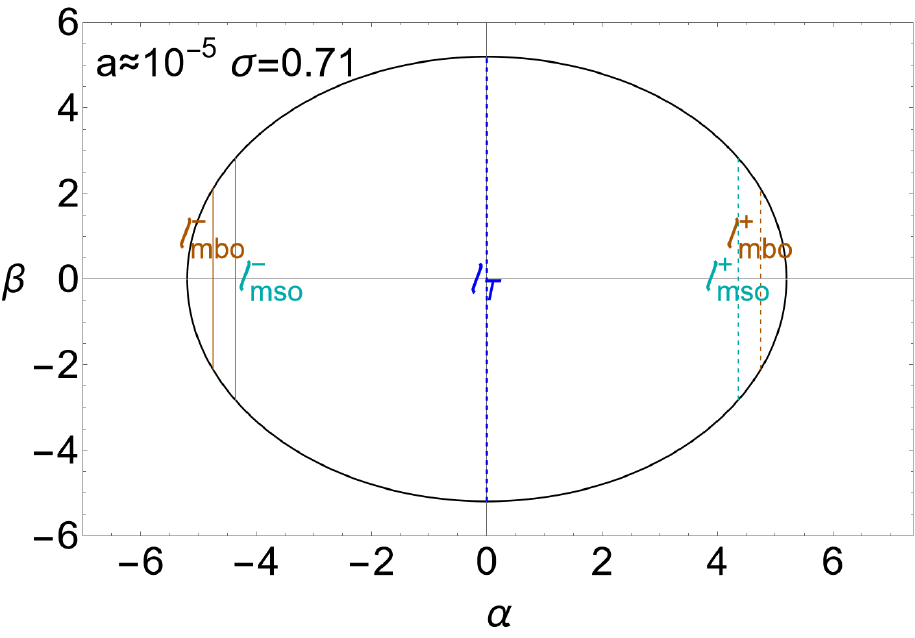}
\includegraphics[width=5.6cm]{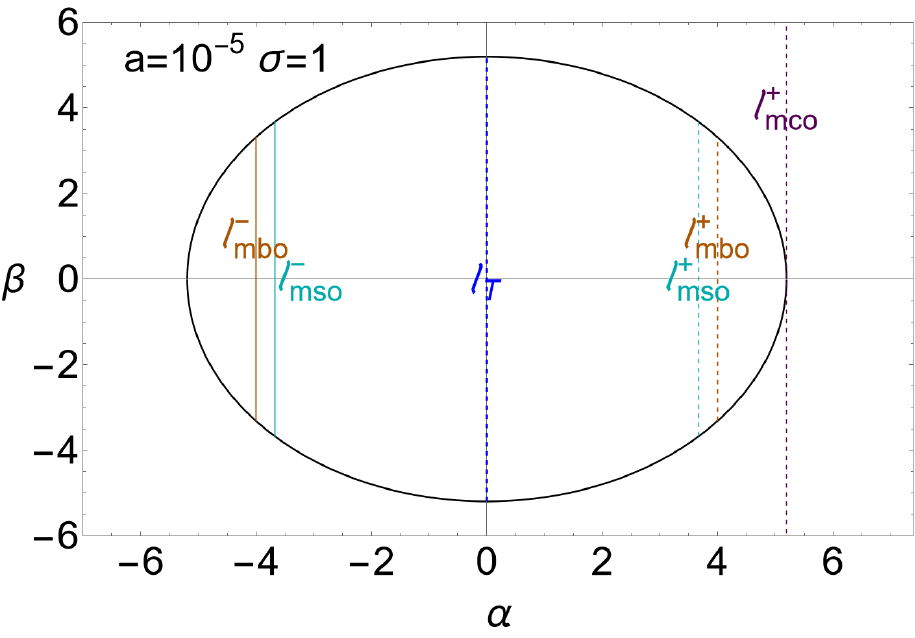}
\includegraphics[width=5.6cm]{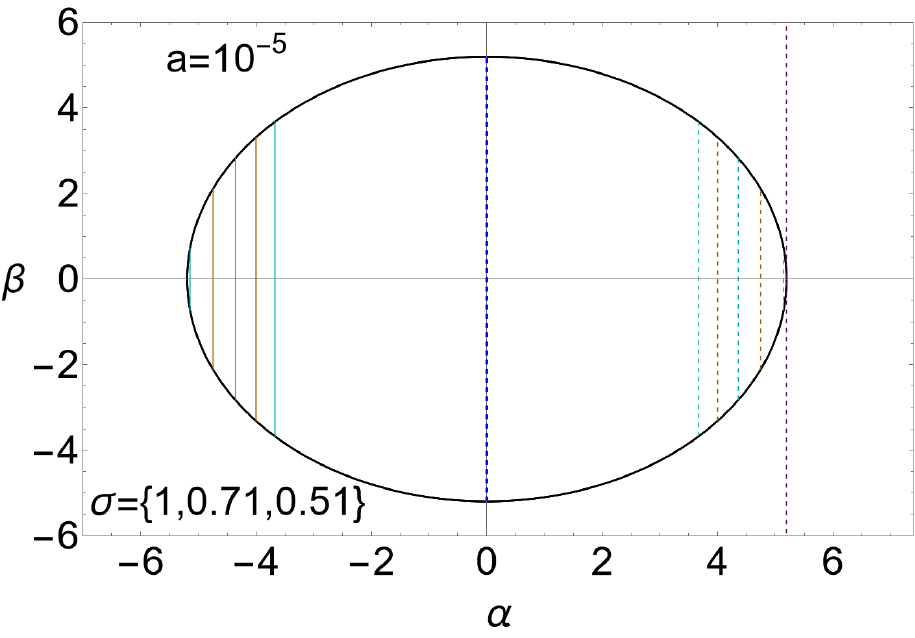}
\includegraphics[width=5.6cm]{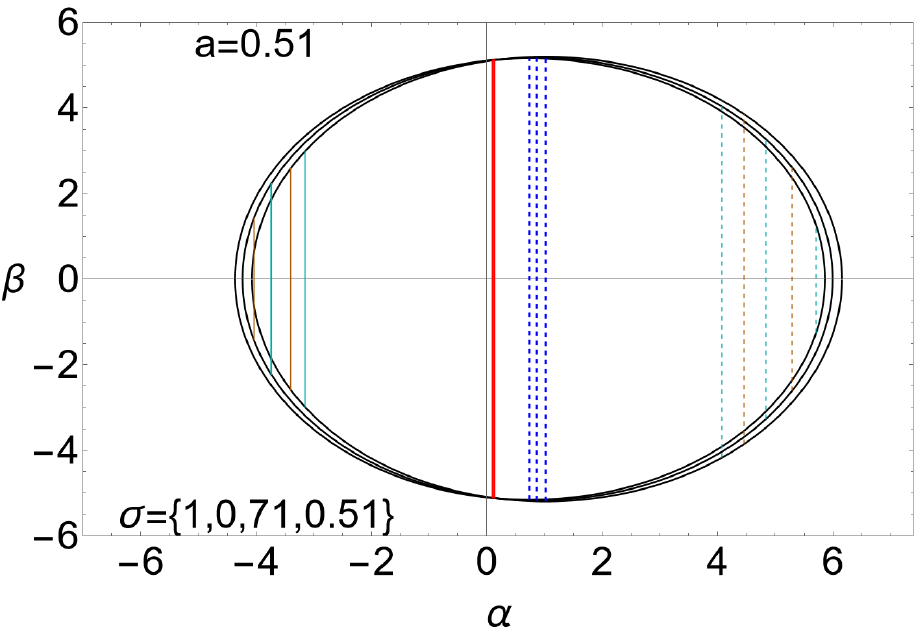}
\includegraphics[width=6cm]{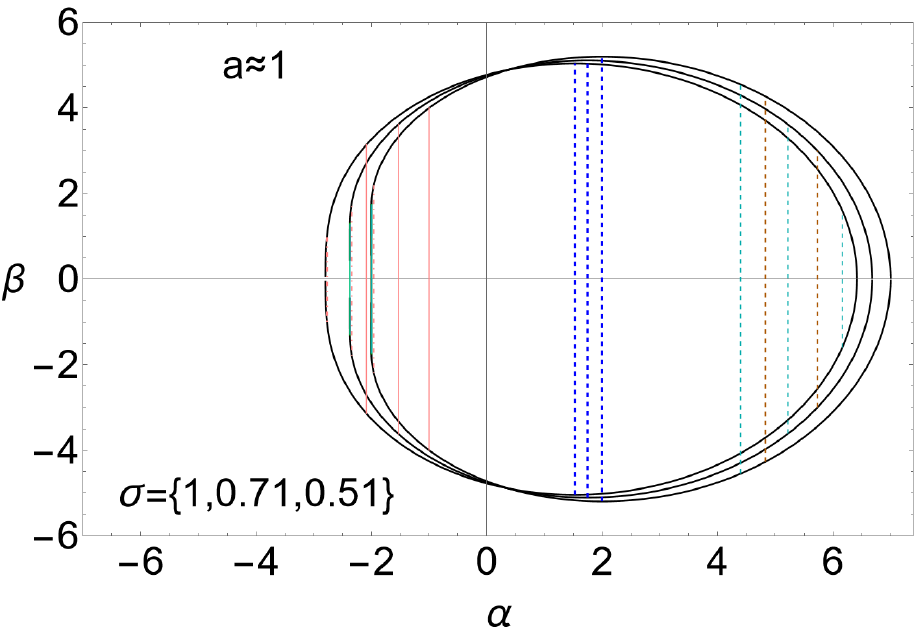}
\caption{Boundary of the \textbf{BH} shadows. Dashed blue vertical lines are $\ell=\ell_\Ta$. Vertical orange, purple and cyan lines  are  $\ell\in \{\ell_{mbo}^\pm,\ell_{\gamma}^\pm,\ell_{mso}^\pm\}$ respectively. (Pink solid,  pink dotted-dashed, red, green) lines  are for  $r=\{r_\epsilon^+,r_x,r_{mbo}^-,r=r_{mso}^-\}$.
($(-\alpha)$ for simplicity  have been  omitted here).}\label{Fig:Plotzing1t171s}
\end{figure}
\begin{figure}
\centering
\includegraphics[width=8cm]{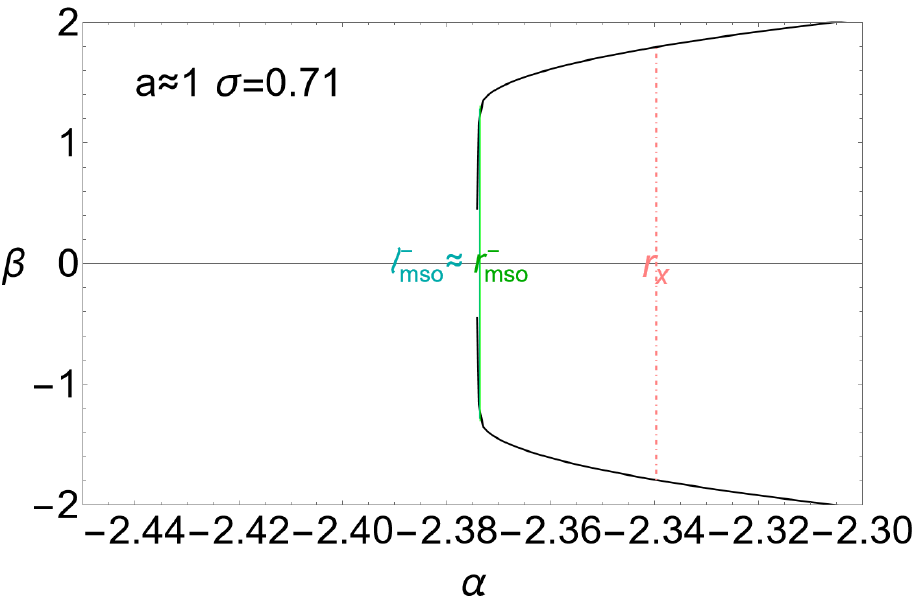}
\caption{A close  view of the case ($a\approx 1$,$\sigma=0.71$)  in Fig.\il(\ref{Fig:Plotzing1t171s}).}\label{Fig:Plotzing1t171z}
\end{figure}
Figs\il(\ref{Fig:Plotvimsounitagrwo})  show cases $\ell=\{\ell_{mbo}^\pm,\ell_{mso}^\pm,\ell_\Ta\}$ and $r=\{r_\epsilon^+,r_x\}$.  Curves at  $\{\ell_{mbo}^-,\ell_{mso}^-\}$  are inside the outer ergoregion.
  There is
  \bea
 C_\beta(a=0.72,r_\epsilon^+)< C_\beta(a=1,r_\epsilon^+)<
 C_\beta(a=1,r_x)<C_\beta(a=0.97,r_x)<C_\beta(a=1,\ell_{mso}^-)<C_\beta(a=1,\ell_{mbo}^-)
  \eea
where $C_\beta(\mathbf{Y})>C_\beta(\mathbf{Z})$ indicates that the curve  $C_\beta(\mathbf{Y})$  for a parameter set $\mathbf{Y}$  is more external (on the shadow boundary  as in the  Figs\il(\ref{Fig:Plotvimsounitagrwo})) than the curve  $C_\beta(\mathbf{Z})$  for the  set of parameters $\mathbf{Z}$, for different  $\sigma\in[0,1]$.

Therefore, at fixed spin, the curve of the plane $\alpha-\beta$, from  orbits  from the outer ergosurface,  is more external than the curve correspondent to photon   orbits  $r_x$ inside the ergoregion.  On the outer ergosurface, as for the other cases with fixed $\ell>0$,  the largest  \textbf{BH}  spins   correspond to innermost curves of the plane $\alpha-\beta$. Viceversa, in agreement with the analysis in Figs\il(\ref{Fig:Plotfarengmrepsx}), inside the ergoregion (on the orbit $r_x$), the  fast spinning  \textbf{BHs}  correspond to outer curves in  the  $\alpha-\beta$ plane.
\section{Discussion and Conclusions}\label{Sec:conclu-Rem}
Co-rotating and counter-rotating null geodesics, solutions of set $(\mathfrak{R})$ in Eqs\il(\ref{Eq:radial-condition})  have been  constrained  by their impact parameter $\ell$ or   radii $r$, and
  related to particular parts  of the shadow  boundary.
  Hence, as results  we  obtained   a  map,  for all $(\sigma,a)$, of regions  on the shadows boundaries,
  correspondent to photon spherical orbits with   $\ell\in\{\ell_\Ta(a),\ell_{mso}^\pm(a),\ell_{mbo}^\pm(a),\ell_{\gamma}^\pm(a)\}$,  or with  radius $r$  in the outer ergoregion, $r\in]r_+,r_\epsilon^+]$,   in the spherical shells  and on  the spherical surfaces defined by the radii  $r=\{r_{mso}^\pm,r_{mbo}^\pm,r_{\gamma}^\pm\}$.
Results are illustrated  in  Figs\il(\ref{Fig:Plotzing1t171s}) (for fixed angle $\sigma$ and spin $a$), in Figs\il(\ref{Fig:Plotvimsounitagrwo}) and Figs\il(\ref{Fig:Plotvimsounita},\ref{Fig:Plotbetrodon1lmsopbm},\ref{Fig:Plotfondoaglrsigmapb},\ref{Fig:Plotverinimsop},\ref{Fig:Plotbetrodon1lmsom})--upper right panels (for all  angles  $\sigma$ and spin $a$).

Note, fixing $(a,\sigma)$, solutions of  $(\mathfrak{R})$ with fixed   $\ell$ or $r$,  are  a set of points  on the shadow boundary   represented by the vertical lines in    the plane$\alpha-\beta$  of   Figs\il(\ref{Fig:Plotzing1t171s}).

On the other hand, solving $(\mathfrak{R})$ with constraints on $\ell$ or $r$, at \emph{fixed} $a$ and for \emph{all}  $\sigma\in[0,1]$,  provides  \emph{curves} of the $\alpha-\beta$ plane, made of points of  the shadow  boundaries  relative to the constrained null spherical geodesics,  for \emph{all} values of $\sigma$. The results of  this analysis are
shown in  Figs\il(\ref{Fig:Plotvimsounitagrwo},\ref{Fig:Plotvimsounita}), and as  solid curves of  Figs\il(\ref{Fig:Plotbetrodon1lmsopbm},\ref{Fig:Plotfondoaglrsigmapb},\ref{Fig:Plotverinimsop},\ref{Fig:Plotbetrodon1lmsom})--upper right panels. In these plots   each solid curve  corresponds to a \emph{fixed}  spin, and  each point of each  solid curve is for \emph{fixed} $\sigma$. Dotted--dashed lines are for a fixed $\sigma$, and each point of a dotted--dashed line is for a different spin.
Then  $(\alpha,\beta)$ have been then  directly related, for each constraint,  to  the quantities
 $(a,\sigma,r,\ell,q)$  in Figs\il(\ref{Fig:Plotfarengmx},\ref{Fig:Plotfarengmy},\ref{Fig:Plotfarengmz}) and  Figs\il(\ref{Fig:Plotvimsounita},\ref{Fig:Plotbetrodon1lmsopbm},\ref{Fig:Plotfondoaglrsigmapb},\ref{Fig:Plotverinimsop},\ref{Fig:Plotbetrodon1lmsom}).

Findings  for photons  spherical orbits in the outer ergoregion are  shown in Figs\il(\ref{Fig:Plotfarengmreps},\ref{Fig:Plotfarengmrepsx},\ref{Fig:Plotvimsounitagrwo}) and discussed in Sec.\il(\ref{Sec:from-ergo}).
The possibility to observe the emission from the outer ergosurface and from the inside the outer ergoregion is limited  for \textbf{BH} spin $a>a_{min}$ as  from a angle $\sigma\in [\sigma_{min},1]$ (in general $|\beta|$ is greater for $\sigma=1$) where $(a_{min},\sigma_{min})$ have been discussed in  Sec.\il(\ref{Sec:from-ergo})--see also
Figs\il(\ref{Fig:Plotzing1t171s}).

In general there are solutions for parameters $\ell\in\{\ell_{mso}^+,\ell_{mbo}^+,\ell_{\gamma}^+,\ell_{\Ta}\}$--see for example Figs\il(\ref{Fig:Plotverinimsop},\ref{Fig:Plotfondoaglrsigmapb})--,
while no solutions have been found  for $r\in\{r_{mbo}^+, r_{mso}^+\}$--Sec.\il(\ref{Sec:radiii-shadows}).

Figs\il(\ref{Fig:Plotvimsounita}) and Figs\il(\ref{Fig:Plotvimsounitagrwo})  summarize results showing   the difference between the co-rotating and counter-rotating photon orbits considered within the different constraints, at different  $(a,\sigma)$.

 Results for   $\ell=\ell_{mso}^-$ are  in   Figs\il(\ref{Fig:Plotbetrodon1lmsom}), solutions are  $\sigma\gtrapprox0.53$ and $\beta$ (in magnitude) increases with $\sigma$ and decreases with $a$ for $\sigma<\sigma_\sigma$ and $a<a_\sigma$--Figs\il(\ref{Fig:Plotsigmaglr})).
 Differently  for  $\sigma>\sigma_\sigma$,   and   $\sigma\in [0.53,\sigma_\sigma]$.
With  $(\sigma>\sigma_\sigma, a>a_\sigma)$,  solutions   appear for small values of $(\beta,\alpha)$ in magnitude, distinguishing   slower from faster  spinning \textbf{BHs}, and the smaller $(\sigma\approx 0.56)$ from larger $\sigma$.

Photons with  $\ell=\ell_{mso}^+$ are considered in  Figs\il(\ref{Fig:Plotverinimsop}).  Inner regions of the $\alpha-\beta$ plane,  characterize  slowly spinning  \textbf{BHs}.    $|\beta|$ increases with the \textbf{BH} spin and  the angle $\sigma>0.479$. For $\sigma>0.55$ there is $|\beta|>0$.
The case $\ell=\ell_{mbo}^-$  is in Figs\il(\ref{Fig:Plotbetrodon1lmsopbm}) and   the results for $\ell=\ell_{mbo}^+$ are for
Figs\il(\ref{Fig:Plotfondoaglrsigmapb}).
($\ell=\ell_{\gamma}^\pm $  correspond to   photon circular orbits $r_{\gamma}^\pm$,  on the equatorial plane($ q=0$)).
The case of photons from the inversion surfaces, is shown in Figs\il(\ref{Fig:Plotfarengmbeta},\ref{Fig:Plotfarengmx},\ref{Fig:Plotfarengmy},\ref{Fig:Plotfarengmz}).
Solutions of  $(\mathfrak{R})$ for  $r_\lambda\in[r_{mbo}^\pm,r_{mso}^\pm]$ and $r_\lambda\in[r_{\gamma}^\pm,r_{mbo}^\pm]$ are strongly differentiated and there are no solutions have been found  for $r\in\{r_{mbo}^+, r_{mso}^+\}$. The case
$r=r_{mbo}^-$ is shown in Figs\il(\ref{Fig:Plotfarengm4repslmsomb}). Case
$r= r_{mso}^-$ is in Figs\il(\ref{Fig:Plotfarengmrepsrmso}  for co-rotating  and counter-rotating  photon orbits.
Orbits in the shell  $r_\lambda\in [r_{mbo}^+,r_{mso}^+]$ are studied for  $\ell<0$ and    $\ell>0$  for $r_\lambda\in [r_{mbo}^-,r_{mso}^-]$  only for  $a>a_{mbo}^*$, in fact
we found solutions  on   $r_\lambda=r_{mso}^-$  for $a>a_{mso}^*$, and on $r_\lambda=r_{mbo}^-$   for $a>a_{mbo}^*$. On the other hand,
from Figs\il(\ref{Fig:Plotscadenztoaglrl})  it can be seen that
$r_{\lambda}(\ell_{mso}^-)\in[r_{mbo}^-,r_{mso}^-]$ for $a>a_\lambda^-$.
As clear from the analysis of
Figs\il(\ref{Fig:Plotfondoaglrsigmapb},\ref{Fig:Plotverinimsop},\ref{Fig:Plotfondoaglrsigmapbcasdu},\ref{Fig:Plotsigmaglr}),
there are solutions for
$\ell\in\{\ell_{mso}^+,\ell_{mbo}^+\}$.
There are \emph{no}  counter--rotating solutions
for $r>r_{\gamma}^+$.

\medskip

The poles $(\sigma\approx 0)$ and the ergoregion of a Kerr \textbf{BH},  explored in this work,  are   regions  where it would be possible to trace important information on the  spacetime  structure for these  \textbf{BHs}  and the   processes involving  fields and matter constituting  the astrophysical \textbf{BH}   embedding  environment.
We stress  that  realistic   \textbf{BH} shadows can depend on properties   of the region of the  light
distribution  and its  source, as
the accretion disks, with  photons    interacting  with  the accreting  plasma. \textbf{BH} shadow  may also be affected by   physical processes involving  the  accreting disk inner edge\footnote{Disk inner edge location is not fixed in time but can move inward, towards the central  \textbf{BH}, or  outward as, for  example,  in the runaway instability\citep{Abra83,Abramowicz:1997sg,Font:2002bi,Hamersky:2013cza,Lot2013}, or  due to  establishment of successive, interrupted, accretion phases   modelling    different phases of super-Eddington accretion-see for example  \cite{apite1,apite2,apite3,Li:2012ts,Oka2017,Kawa,Allen:2006mh}.
 The inner regions of the accretion disks are also characterized by oscillations and modes, as  quasi-periodic oscillations.
A further interesting aspect  to consider in this frame would be to  \textbf{BH} shadows following a  spin variation (precession) process, for  tori  not located in the equatorial plane of the central  \textbf{BH} \citep{King:2018mgw,Aly:2015vqa,NixonKing(2012b),Nelson:2000mw,Fragile:2007dk,1996MNRAS.282..291S,King:2005mv,liska,2015MNRAS.449.1251D,2006MNRAS.368.1196L,2009MNRAS.400..383M,King:2005mv,Nealon:2015jya,bib:spso-sa,King:2008au,Nixon:2013qfa}.}.

  We expect that the regions highlighted here will be distinctly recognizable in future observational  enhancement.
 \textbf{EHT} Collaboration has already  compared  numerical torus models
directly to observations in several comprehensive analyses, performing  \textbf{GRMHD} models fit with the observations\footnote{Various \textbf{GRMHD} analyses have been implemented to   \textbf{EHT} images interpretation in  \cite{EHT2d,EHT2f} and especially in \cite{EHT2e}--see also \cite{Porth,Gralla19,Gralla20,Johnson,Narayan19}, investigating the origin of the photons forming the \textbf{EHT}
image  at  base of the \textbf{M87} jet and disk--\citep{EHT2e,Porth}.
  An analytic disk model  has been also  developed in  \cite{Vincent21} for the \textbf{M87*} accretion flow,
  computing the  synchrotron emission from the disk model   assuming different spacetimes,  and numerical fits  to the
\textbf{EHT} data.
\textbf{GRMHD}
simulations  with jet  ejection and    accretion flows  simulated from first principles were in \cite{Janssen}  with the \textbf{EHT}  observations of the jet
launching and collimation in Centaurus \textbf{A} (see \cite{Chatterjee19}).
In \cite{Chatterjee20etaletal}
\textbf{EHT} images were analysed and interpreted  by \textbf{GRMHD}
models of tilted  accretion discs,
finding  that \textbf{M87} may feature a tilted disc/jet systems--see also \cite{EHT2,White20}.
In \cite{Anantua23}
there are \textbf{GRMHD} simulations of  novel
models for   high-energy particles in systems of  jet/accretion
flow   with emission of  synchrotron radiation.
Vast simulations  of different accretion models were provided  in  \cite{EHT22e,EHT2e}
and  simulations of both thick torii and thin disks and different implementations of the
interaction between radiation and the  plasma fluid are  in \cite{Chatterjee20etaletal,Curd23}.
 \cite{EHT21b} focused on
 description of   the  polarimetric  observations and
the relativistic jet, using
 a large library of
simulated polarimetric images from \textbf{GRMHD} simulations. (Magnetically arrested accretion
disks were considered as consistent \textbf{GRMHD} models.).
 Numerical calculations of the polarization configuration  were generated by an orbiting toroidal source giving raise to a
phenomenological model of a torus--see also \cite{Vincent19}.
In \cite{EHT2e}  it has been shown how \textbf{GRMHD} simulations
 produce images consistent with the  \textbf{M87*}  observation.   \cite{EHT21b,EHT21a} found that
only a few simulation images fit the polarimetric data (magnetically arrested disk).
Semi-analytical models of
the \textbf{M87} spectra  were also considered in \cite{Lucchini19}.
In \cite{Emami21}
 examples of semi-analytic \textbf{GRMHD}  jet and accretion flow  models
were discussed to simulate the emission from \textbf{M87*} and
\textbf{SgrA*}.
(The emission originates in a geometrically thick
equatorial accretion flow, radiatively inefficient
accretion flow and \textbf{ADAF}  for \textbf{SgrA*} can fit the observed
spectral energy density  where different emitting regions could
contribute to  different regions of  \textbf{M87*} and  \textbf{SgrA*}).
Disk/jet/\textbf{BH} systems  are also modelled to   fit observations  in particular between the co-rotating or counter-rotating (and more in general tilted) models, both with respect to  the jet  and for the disk components and  with respect to the rotation of the central  \textbf{BH}--see\cite{EHT22e,EHT21b,Janssen,Chatterjee20etaletal,Emami21}.
In \cite{EHT22e} several astrophysical models were tested for the
 \textbf{SgrA*}    image.
In  \cite{EHT22f}
the first comprehensive  interpretation of
the \textbf{EHT} 2017 \textbf{SgrA*} data  was provided
using  a library of models based on time-dependent \textbf{GRMHD}  simulations, in particular  for  aligned and tilted configurations. Physical and numerical limitations of the models were thoroughly discussed.}, for example in \cite{EHT2e,EHT22e}. For this reason  there is no fixing accretion or  accretion disk model. In this sense our  analysis can be   adaptable  to and complement  the constraints imposed by the specific numerical or analytical model of accretion disks.

\section*{Data availability}
There are no new data associated with this article.
No new data were generated or analysed in support of this research.

\appendix

\section{Co-rotating and counter-rotating geodesic structure}\label{Sec:co-rota}
The marginally stable co-rotating $(-)$ and counter--rotating $(+)$   radius is
\bea
 r_{mso}^\mp \equiv Z_2+3 \mp\sqrt{(3-Z_1) (Z_1+2 Z_2+3)},
 \eea
 where
 \bea
Z_1\equiv \left(\sqrt[3]{1-a}+\sqrt[3]{a+1}\right) \sqrt[3]{1-a^2}+1,\quad Z_2\equiv \sqrt{3 a^2+\left(\left(\sqrt[3]{1-a}+\sqrt[3]{a+1}\right) \sqrt[3]{1-a^2}+1\right)^2}.
 \eea
 The  marginally bounded radius is
 \bea
r_{mbo}^\pm \equiv \pm a+2 \sqrt{1\pm a}+2.
\eea
The photons  circular  counter-rotating and co-rotating orbits on the equatorial plane are
\bea
r_{\gamma}^+ \equiv 4 \cos ^2\left(\frac{1}{6} \arccos\left(2 a^2-1\right)\right),\quad  r_{\gamma}^- \equiv 2 \left[\sin \left(\frac{1}{3} \arcsin\left(1-2 a^2\right)\right)+1\right].
\eea
%

There is
$\ell_{mso}^\pm\equiv \ell^\pm(r_{mso}^\pm)$,
$\ell_{mbo}^\pm\equiv \ell^\pm(r_{mbo}^\pm)$ and
$\ell_{\gamma}^\pm\equiv \ell^\pm(r_{\gamma}^\pm)$ respectively, where, considering Eq.\il(\ref{Eq:flo-adding}), there is
\bea
\ell^\mp(r)=\frac{a^3\mp\sqrt{r^3 \left(a^2+(r-2) r\right)^2}-a (4-3 r) r}{a^2-(r-2)^2 r}.
\eea
The location of the tori  centers, $r_{center}^\pm$, is constrained by the fluids specific angular momentum    $\ell$,  according   to    the radii  $r_{(mbo)}^{\pm}$ and $r_{(\gamma)}^{\pm}$, defined by the  relations:
\bea&&\nonumber
r_{\mathrm{(mbo)}}^{\pm}:\;\ell^{\pm}(r_{\mathrm{mbo}}^{\pm})=
 \ell^{\pm}(r_{\mathrm{(mbo)}}^{\pm})\equiv \mathbf{\ell_{\mathrm{mbo}}^{\pm}},\quad
  r_{(\gamma)}^{\pm}: \ell^{\pm}(r_{\gamma}^{\pm})=
  \ell^{\pm}(r_{(\gamma)}^{\pm})\equiv \ell_{\gamma}^{\pm} \quad \mbox{where} 
  \\&&\label{Eq:def-nota-ell}
r_{\gamma}^{\pm}<r_{\mathrm{mbo}}^{\pm}<r_{\mathrm{mso}}^{\pm}<
 r_{\mathrm{(mbo)}}^{\pm}<
  r_{(\gamma)}^{\pm}
  \eea
  respectively.
On the equatorial plane, accretion tori    cusps are  in   $\in]r^{\pm}_{mbo},r^{\pm}_{mso}]$, with   center in  $r^{\pm}_{center}\in]r^{\pm}_{mso},r^{\pm}_{(mbo)}]$.
Toroids  with $\mp\ell^{\pm}\equiv[\mp \ell_{mbo}^{\pm},\mp\ell_{\gamma}^{\pm}[ $ (the cusp is  located in  $]r_{\gamma}^{\pm},r_{mbo}^{\pm}]$), have  center  in  $r_{center}^{\pm}\in]r_{(mbo)}^{\pm},r_{(\gamma)}^{\pm}]$. Toroids with   $\mp \ell^{\pm}\geq\mp\ell_{\gamma}^{\pm}$,  have centers  $r^{\pm}_{center}>r_{(\gamma)}^{\pm}$--\cite{dsystem,ringed}.
\section{On solutions of $(\mathfrak{R})$}\label{App:allell}
Here we consider again  solutions of $(\mathfrak{R})$ for  $\ell\in\{\ell_{mso}^\pm,\ell_{mbo}^\pm,\ell_{\gamma}^\pm\}$.
In order to clarify aspects of the shadow boundaries in dependence on the parameter $\ell$,  spin $a$ and coordinates $(r,\sigma)$  it is convenient to introduce the following quantities:
\bea\label{Eqs:amsom}
a_{mso}^-\equiv0.839825,\quad a_\lambda^- \equiv 0.93259, \quad
a_{\sigma }\equiv 0.98217,\quad
\sigma _{\sigma }\equiv 0.563773,\quad \sigma_0\equiv 0.148148,
\eea
(see Figs\il(\ref{Fig:Plotsigmaglr}) )
and   spin function $a_C$:
\bea
\label{Eq:ag-lalpha}
&&
a_C\equiv \frac{3 \left(\lambda_C^{2/3}-1\right)}{\sqrt[3]{\lambda_C}}-\ell,\quad \mbox{where}\quad \lambda_C\equiv \ell+\sqrt{\ell ^2+1},
\eea
 \begin{figure}
\centering
\includegraphics[width=5.51cm]{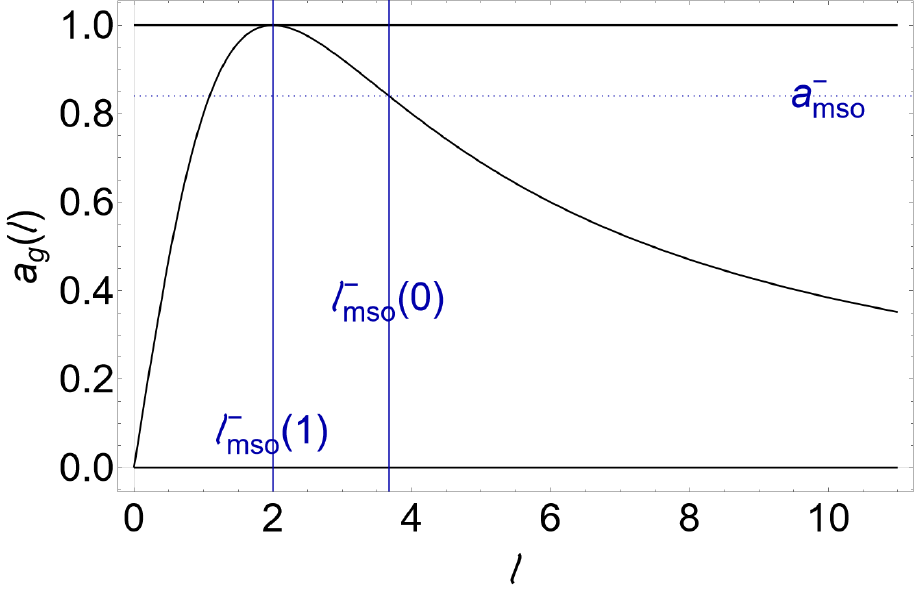}
\includegraphics[width=5.51cm]{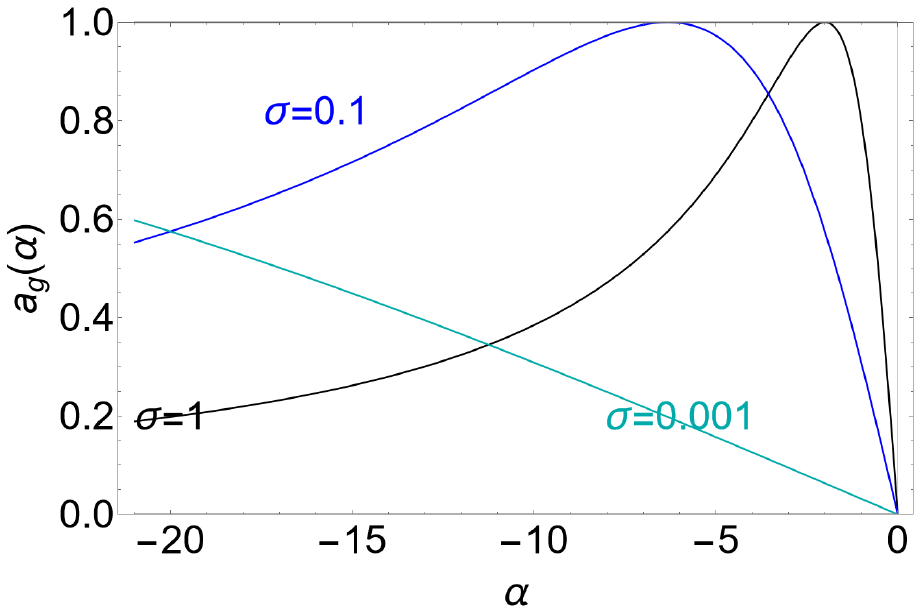}
\includegraphics[width=5.51cm]{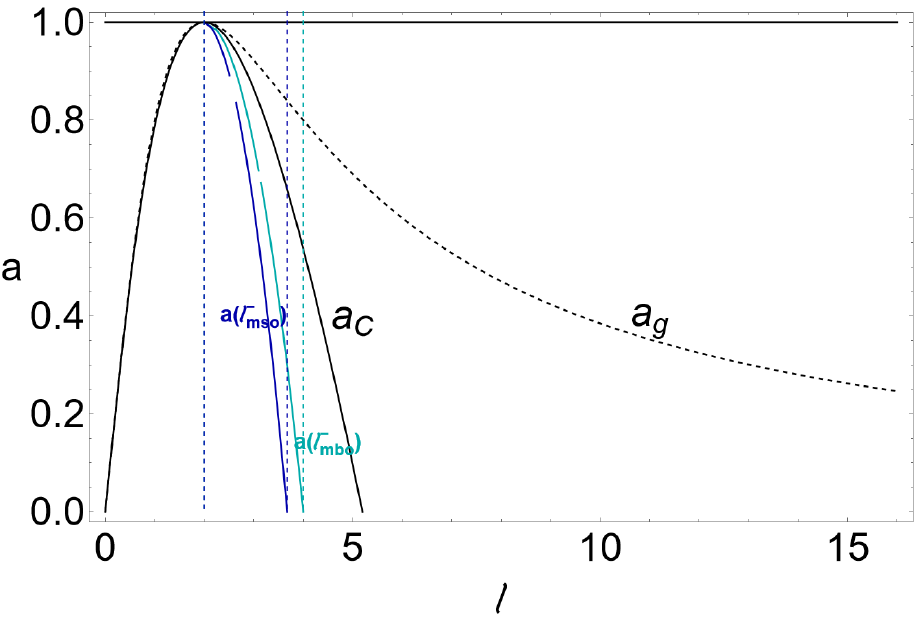}
\caption{Spin $a_g(\ell)$ of Eq.\il(\ref{Eq:ag-lalpha}) expression of the \textbf{BH} horizons as function of the \textbf{BH} horizons angular momentum.
Spin function $a_C$ is in Eq.\il(\ref{Eq:ag-lalpha}).}
\label{Fig:Plotfondoagl}
\end{figure}
 see Figs\il(\ref{Fig:Plotfondoagl}).
 Quantities in Eqs\il(\ref{Eqs:amsom}) and
 Eqs\il(\ref{Eq:ag-lalpha}), bound the constrained  solutions of equations $(\mathfrak{R})$ we discuss below.

\textbf{Case: $\ell=\ell_{mso}^-$}

\medskip

We consider   $\ell\in [\ell_{mso}^-(a=1),\ell_{mso}^-(a=0)]$\footnote{Where $\ell_{mso}^-(a=1)=2$ and  $\ell_{mso}^-(a=0)\equiv 3.674$. This case is illustrated in the Figs\il(\ref{Fig:Plotbetrodon1lmsom}) and Figs\il(\ref{Fig:Plotverinimsop},\ref{Fig:Plotfondoaglrsigmapbcasdu},\ref{Fig:Plotsigmaglr}).}. There is then  $ r=3$
 for $a=0$.

Whereas, for $a\in]0,a_g(\ell)[$,
   there is
$r=r_\lambda$ and  $q= q_\lambda$, with $q_\lambda\in]q_\lambda(a=1),q_\lambda(a=0)[$\footnote{Where $q_\lambda(a=1)=2.999$ and  $q_\lambda(a=0)=13.5$.}--see Figs\il(\ref{Fig:Plotfondoaglrsigmapbcasdu}).
There is   $a_g(\ell_{mso}^-)\in [a_{mso}^-,1]$, constraining  the photons orbit  depending on the central \textbf{BH} spin--mass ratio --Figs\il(\ref{Fig:Plotfondoagl}).
Therefore, there is
\bea&&
\mbox{for}\quad a=0:\quad\sigma\in [ \sigma_{\ell},1],\quad \mbox{where}\quad \sigma_{\ell}(\ell_{mso}^-) \in [0.5,1],
\\\nonumber
&&\mbox{for}\quad
a\in ]0,1]:
\sigma \in [\sigma_\lambda,1], \quad\mbox{where}\quad \sigma_\lambda\in [\sigma_\lambda(a=0),\sigma_\sigma],  \quad\mbox{and}\quad \sigma_\lambda(a=0)=0.5,
 \eea
where $\sigma_\sigma$ is a maximum at  $a=a_\sigma$--see Figs\il(\ref{Fig:Plotfondoaglrsigmapbcasdu}).
There is\footnote{However, note that $r_C(\ell_\bullet)<r_+$ for $\ell_\bullet\in\{\ell_{\gamma}^-,\ell_{mbo}^-,\ell_{mso}^-\}$.
In fact,   considering the condition $T\geq0$,
the orbit  $r=3$  is for  $a=0$ on the angle $\sigma=\sigma_0$ ($\sigma_0\neq1$) with $q>0$, i.e.
\bea
(\sigma\in ]\sigma_0,0.5], \ell\in [\ell_{mso}^-(a=1), \ell_s]),\quad (\sigma\in]0.5,1], \ell\in [\ell_{mso}^-(a=1),\ell_{mso}^-(a=0)]),\quad \mbox{where},\quad \ell_s\equiv 5.19615 \sqrt{\sigma }.
\eea}
%
 \bea
 q=q_\lambda:\quad \ell\in [\ell_{mso}^-(a=1),\ell_{mso}^-(a=0)],\quad(a\in]0,a_g[, r=r_\lambda);\quad  (a\in ]a_g,1], r=r_C),\quad\mbox{where}\quad  r_C\equiv \sqrt{a (\ell- a)}.
 \eea
Similarly, using the constraint  $T\geq0$,  for   $a\in ]0,1]$ the solution can be written as
%
%
\bea&&
r=r_{\lambda},\quad q=q_{\lambda}: \quad (a=a_C, \sigma =1);\quad (a\in ]0,a_C[, \sigma\in[\sigma_C,1]);
\eea
where  $\sigma_C$ is  a zero of the polynomial $\sum_{i=0}^6 x_i \sigma_C^i=0$, with coefficients
\bea
\\&& \nonumber
x_0\equiv \left(a^2-1\right) \ell ^6,\\&& \nonumber
x_1\equiv \ell ^4 \left[27-a^2 \left(a^2-4 a \ell +\ell ^2+30\right)\right],
\\&& \nonumber
x_2\equiv a \ell ^2 \left[-4 \left(a^4+27\right) \ell +96 a^3-4 a^2 \ell ^3+\left(7 a^2+33\right) a \ell ^2\right],
\\&& \nonumber x_3\equiv 2a^2[a^2(-32 a^2+4 a \ell ^3-3  \ell ^4)-3 \left(a^4+22 a^2-27\right)  \ell ^2],\\&&
\nonumber x_4=\frac{a^2}{\ell ^2}x_2, 
\quad x_5=\frac{a^4}{\ell^4}x_1,
\quad x_6=\frac{a^6}{\ell^6}x_0,
 \eea
There is $\sigma_C(\ell_\bullet)=
\sigma_\lambda(\ell_\bullet,q_\lambda(\ell_\bullet))$ where $\ell_\bullet=\{\ell_{mso}^-,\ell_{mbo}^-\}$ (note for $\ell_\bullet=\ell_{\gamma}^\pm$ there is
$\sigma_C=\sigma_\lambda=1$).

\medskip

\textbf{Case: $\ell=\ell_{mso}^+$}

For $\ell=\ell_{mso}^+$ there is   $r=r_\lambda$ and $q=q_\lambda$ for $a\in [0,1]$.
There is  $\ell\in [\ell_{mso}^+(a=1),\ell_{mso}^+(a=0) ]$\footnote{Where $\ell_{mso}^+(a=1)=-4.4$ and $\ell_{mso}^+(a=0)= -3.674$. This case is considered in  Figs\il(\ref{Fig:Plotverinimsop},\ref{Fig:Plotfondoaglrsigmapbcasdu},\ref{Fig:Plotsigmaglr}).},  and
$q_\lambda\in [q_\lambda(a=0),q_\lambda(a=1)]$\footnote{Where $q_\lambda(a=0)=13.5$ and $q_\lambda(a=1)= 20.679$.}, where
\bea
&&\mbox{for}\quad a=0:\quad \sigma\in [\sigma_\ell,1];
\quad \mbox{and for}\quad  a\in ]0,1]: \sigma\in [\sigma_\lambda,1],\quad \mbox{where}\quad \sigma_\lambda\in [\sigma_\lambda(a=1),\sigma_\lambda(a=0)],
\eea
with $\sigma_\lambda(a=0)=0.5$ and $ \sigma_\lambda(a=1)=0.477$--Figs\il(\ref{Fig:Plotfondoaglrsigmapbcasdu}).

\medskip

\textbf{Case: $\ell=\ell_{mbo}^+$}

\medskip

For $\ell\in[\ell_{mbo}^+(a=0),\ell_{mbo}^+(a=1)]$\footnote{Where $\ell_{mbo}^+(a=0)=-4$ and  $\ell_{mbo}^+(a=1)= -4.828$. This case is shown in Figs\il(\ref{Fig:Plotfondoaglrsigmapb}) and Figs\il(\ref{Fig:Plotverinimsop},\ref{Fig:Plotfondoaglrsigmapbcasdu},\ref{Fig:Plotsigmaglr}).}, for
 $a\in [0,1]$, there is $r=r_\lambda$ and $q=q_\lambda$, where
$q_\lambda \in[q_\lambda(a=0),q_\lambda(a=1)]$\footnote{With $q_\lambda(a=0)=11$ and $q_\lambda(a=1)=18.2482$.}.

Condition $T\geq0$  implies that:
\bea
&&
\mbox{for}\quad a=0:\quad \sigma\in [\sigma_\ell,1],
\\\nonumber
&&
\mbox{for}\quad  a\in ]0,1]:\quad \sigma\in [\sigma_\lambda,1],\quad\mbox{ where}\quad
\sigma_\lambda\in[\sigma_\lambda(a=1),\sigma_\lambda(a=0)],
\eea
with $\{\sigma_\lambda(a=0)=0.591797, \sigma_\lambda(a=1)=0.554998\}$--Figs\il(\ref{Fig:Plotfondoaglrsigmapbcasdu}).
\medskip

\textbf{Case: $\ell=\ell_{mbo}^- $}

 \medskip

We consider the case  $\ell=\ell_{mbo}^- \in [\ell_{mbo}^- (a=1),\ell_{mbo}^- (a=0)]$\footnote{Where $\ell_{mbo}^- (a=1)=2$  and $\ell_{mbo}^- (a=0)=4$. This case is considered in Figs\il(\ref{Fig:Plotbetrodon1lmsopbm}) and Figs\il(\ref{Fig:Plotverinimsop},\ref{Fig:Plotfondoaglrsigmapbcasdu},\ref{Fig:Plotsigmaglr})} for  $a\in [0,1]$.
There is $q=q_\lambda$ and $r=r_\lambda$  where
$
q\in[q_\lambda(a=1),q_\lambda(a=0)]$\footnote{With $ q_\lambda(a=1)=1$ and $q_\lambda(a=0)=11$.}.

 Condition $T\geq0$  implies that
 \bea&&
\mbox{for}\quad a=0:\quad \sigma\in [\sigma_\ell,1],
\\\nonumber
&&
\mbox{for}\quad  a\in ]0,1]:\quad \sigma\in [\sigma_\lambda,1],\quad\mbox{ where }\quad \sigma_\lambda \in [\sigma_\lambda (a=0),\sigma_\lambda, (a=1)]
  \eea
  with $\{ \sigma_\lambda (a=0)=0.59256, \sigma_\lambda (a=1)=0.76391\}$.

There is
$q_\lambda\in [q_\lambda(a=1),q_\lambda(a=0)]$, where $\{q_\lambda(a=1)=1,q_\lambda(a=0)=11\}$.

More in details there is
\bea&&
  q=q_{\lambda}\quad \mbox{for}
  \\
 &&\nonumber
\ell =\ell_{mbo}^-(a=1): \quad (a=0, \sigma\in[\sigma_s, 1], r=3); \quad (a\in]0,1[, \sigma\in[\sigma_C,1], r=r_{\lambda })); \quad \mbox{where}\quad \sigma_s\equiv \frac{4}{27},
\\\nonumber
&&\ell\in]\ell_{mbo}^-(a=1),\ell_{mbo}^-(a=0)]:\quad  (a=0,\sigma\in[ \sigma_f, 1], r=3);
\quad  (a\in ]0,a_C[, \sigma\in[\sigma_C,1], r=r_{\lambda });\\
&&\nonumber \hspace{5cm} (a=a_C, \sigma =1, r=r_{\lambda }),\quad\mbox{where}\quad \sigma_f\equiv \frac{\ell ^2}{27}.
\eea

\medskip

\textbf{Case $\ell=\ell_{\gamma}^\pm$}

For  $\ell=\ell_{\gamma}^\pm $  there are the  photon circular orbits $r_{\gamma}^\pm$, where $ q=0$ and $\sigma=1$.

\end{document}